\documentclass[useAMS,usenatbib]{mn2e}
\usepackage{amsmath}
\usepackage{graphicx}
\usepackage[normalem]{ulem}
\pdfoutput=1

\title[The mass evolution of the first galaxies]{The mass evolution of the first galaxies: stellar mass functions and star formation rates at $4 < z < 7$ in the CANDELS GOODS-South field}
\author[K Duncan et al.]{K. Duncan $^{1}$\thanks{E-mail:
ppxkd@nottingham.ac.uk}, C. J. Conselice$^{1}$, A Mortlock$^{1,2}$, W. G. Hartley $^{1,3}$, Y. Guo$^{4,5}$,  \newauthor  H. C. Ferguson$^{6}$, R. Dav\'{e}$^{7}$,
Y. Lu$^{8}$, J. Ownsworth$^{1}$, M. L. N. Ashby$^{9}$, \newauthor A. Dekel$^{10}$, M. Dickinson$^{11}$, 
S. Faber$^{4}$, M. Giavalisco$^{5}$, N. Grogin$^{8}$, \newauthor D. Kocevski$^{12}$, A. Koekemoer$^{6}$, R. S. Somerville$^{13}$, C. E. White$^{14}$
\\
$^{1}$University of Nottingham, School of Physics \& Astronomy, Nottingham NG7 2RD \\
$^{2}$SUPA, Institute for Astronomy, The University of Edinburgh, Royal Observatory, Edinburgh, EH9 3HJ\\
$^{3}$Department of Physics, ETH Zurich, CH-8093 Zurich, Switzerland \\
$^{4}$UCO/Lick Observatory, Department of Astronomy and Astrophysics, University of California, Santa Cruz, CA, USA \\
$^{5}$Department of Astronomy, University of Massachusetts, Amherst, MA, USA \\
$^{6}$Space Telescope Science Institute, 3700 San Martin Drive, Baltimore, MD 21218, USA\\
$^{7}$University of the Western Cape, Bellville, Cape Town 7535, South Africa\\
$^{8}$Kavli Institute for Particle Astrophysics and Cosmology, Stanford, CA 94309, USA\\
$^{9}$Harvard-Smithsonian Center for Astrophysics, 60 Garden St., Cambridge, MA 02138, USA\\
$^{10}$Racah Institute of Physics, The Hebrew University, Jerusalem 91904, Israel\\
$^{11}$National Optical Astronomy Observatory, Tucson, AZ, USA\\
$^{12}$Department of Physics and Astronomy, University of Kentucky, Lexington, KY, USA\\
$^{13}$Department of Physics and Astronomy, Rutgers University, 136 Frelinghuysen Road, Piscataway, NJ 08854, USA\\
$^{14}$Department of Physics and Astronomy, The Johns Hopkins University, 3400 North Charles Street, Baltimore, MD 21218, USA
}
\begin{document}

\date{Accepted 2014 August 7. Received 2014 July 16; in original form 2013 November 13}

\pagerange{\pageref{firstpage}--\pageref{lastpage}} \pubyear{2013}

\maketitle

\label{firstpage}

\begin{abstract}
We measure new estimates for the galaxy stellar mass function and star formation rates for samples of galaxies at $z \sim 4,~5,~6~\&~7$ using data in the CANDELS GOODS South field. The deep near-infrared observations allow us to construct the stellar mass function at $z \geq 6$ directly for the first time. We estimate stellar masses for our sample by fitting the observed spectral energy distributions with synthetic stellar populations, including nebular line and continuum emission. The observed UV luminosity functions for the samples are consistent with previous observations, however we find that the observed $M_{UV}$ - M$_{*}$ relation has a shallow slope more consistent with a constant mass to light ratio and a normalisation which evolves with redshift. Our stellar mass functions have steep low-mass slopes ($\alpha \approx -1.9$), steeper than previously observed at these redshifts and closer to that of the UV luminosity function. Integrating our new mass functions, we find the observed stellar mass density evolves from $\log_{10} \rho_{*} = 6.64^{+0.58}_{-0.89}$ at $z \sim 7$ to  $7.36\pm0.06$ $\text{M}_{\odot} \text{Mpc}^{-3}$ at $z \sim 4$. Finally, combining the measured UV continuum slopes ($\beta$) with their rest-frame UV luminosities, we calculate dust corrected star-formation rates (SFR) for our sample. We find the specific star-formation rate for a fixed stellar mass increases with redshift whilst the global SFR density falls rapidly over this period. Our new SFR density estimates are higher than previously observed at this redshift.
\end{abstract}


\section{Introduction}
Thanks to the unprecedented sensitivity of the latest extragalactic surveys, the last decade has seen a revolution in the observations of galaxies in the high-redshift universe. It is now possible to study the beginnings of the mechanisms and processes which formed the diverse array of galaxies we find in the local universe today. Since the first successful detections through the Lyman break technique, via the characteristic `break' induced by blanketing hydrogen absorption of the UV continuum \citep{1990ApJ...357L...9G,1992AJ....104..941S}, the study of high-redshift galaxies has progressed rapidly. With the introduction of the Wide-field Camera 3 (WFC3) in 2009 and the unprecedented depth in the near-infrared it provides, the study of galaxies out to redshifts of $z > 6$ has become commonplace. 

The numerous measurements of the UV luminosity function of high-redshift galaxies spanning the redshift range $4 \leq z \leq 9$ \citep{2007ApJ...670..928B,2009MNRAS.395.2196M,Oesch:2009ew,Bouwens:2010dk,2011A&A...532A..33G,Lorenzoni:2011iz,McLure:2013hh,Schenker:2013cl} are not only giving an insight into the processes of galaxy formation, they are also helping us to understand the role those galaxies played in the ionization of the intergalactic medium during the epoch of reionization (EoR). These surveys have put strong constraints on the contribution of star-forming galaxies to reionization, requiring a significant contribution from faint galaxies below the current detection limits to complete reionization within the observed redshift.

Because they represent the time integral of all past star-formation, the stellar masses of galaxies provide additional independent constraints on their contribution to reionization through the observed stellar mass density \citep{2010Natur.468...49R}. Successful models of galaxy evolution and reionization must therefore be able to reconcile both the star-formation observed directly, and the record of past star-formation contained in the observed stellar masses. The galaxy stellar mass function (SMF) and its integral the stellar mass density (SMD), are therefore  important tools in the study of galaxy evolution.

However, accurately measuring the stellar masses of galaxies at high-redshift is very difficult for a number of fundamental reasons. These reasons stem from the fact that the rest-frame wavelengths probed by optical/near-infrared surveys extend only to the UV continuum, requiring mid-infrared observations to extend past $\lambda_{rest} = 4000$\AA. Even when rest-frame optical measurements are available through deep \emph{Spitzer} 3.6 and 4.5 $\mu m$ observations, e.g.  \citet{2009ApJ...697.1493S,Labbe:2010ho,Gonzalez:2011dn,Yan:2012ky}, the degeneracies between dust extinction, age and metallicity are large (see \citeauthor{Dunlop:2012hs}~\citeyear{Dunlop:2012hs} for a detailed discussion).

More recently it has also been shown that the spectral energy distributions (SEDs) of high-redshift galaxies, for both photometrically selected \citep{2009A&A...502..423S,2010A&A...515A..73S,Ono:2010ed,2011MNRAS.418.2074M,Lorenzoni:2011iz,2012ApJ...755..148G} and smaller spectroscopically confirmed samples \citep{Shim:2011cw,2013MNRAS.429..302C,Stark:2013ix}, can exhibit colours which are best fit by the inclusion of nebular emission lines when measuring galaxy properties such as mass and age. The inclusion of these lines results in fitted ages which are significantly younger compared to fits without nebular emission, as well as being of lower stellar mass. Another consequence of the degeneracies in SED fitting at high-redshift is the increased importance in the assumed star formation history (SFH) of the models being fit. Observational studies of the mass growth of galaxies at a constant number density (\citeauthor{2011MNRAS.412.1123P}~\citeyear{2011MNRAS.412.1123P}, \citeauthor{Salmon:2014tm}~\citeyear{Salmon:2014tm}) and hydrodynamical simulations \citep{2011MNRAS.410.1703F} imply that the star formation histories for galaxies at $z > 3$ are smoothly rising. This is in contrast to the smoothly falling or constant SFH commonly used at low-redshift (see \citeauthor{Conroy:2013dk}~\citeyear{Conroy:2013dk} for a review). In \citet{Maraston:2010dl}, it is also shown that exponentially rising star-formation histories can provide improved fits to both simulated and observed SEDs of high redshift galaxies.

\citet{Gonzalez:2011dn} were amongst the first to make use of the capabilities offered by the WFC3/IR data, using data from the Early Release Science (ERS; \citeauthor{2011ApJS..193...27W}~\citeyear{2011ApJS..193...27W}) to measure the stellar mass to UV luminosity ratio for a sample of dropout galaxies and applying it to the observed luminosity functions to measure the stellar mass function out to $z \sim 7$. In contrast to the steep faint-end slope of the UV luminosity function ($\alpha = -2~\text{to}~-1.7$) at high-redshifts, this work observed a notably shallower mass function ($\alpha = -1.6 ~\text{to}~-1.4$). Subsequent work by \citet{2012ApJ...752...66L} with much greater sample sizes from ground-based near-infrared found a similarly shallow slope at $z \sim 4$ and 5. In contrast, observations by \citet{2011MNRAS.413..162C} and \citet{Santini:2012jq} observe a significantly steeper low-mass slope at $z > 3$.
 
In \citet{Gonzalez:2011dn}, the shallow low-mass slope arises due to the observed evolution of the mass-to-light ratio with UV luminosity. Similarly, \citet{2012ApJ...752...66L} infer an evolving mass-to-light in order to reconcile the luminosity and mass function slopes observed. The primary physical explanation for this evolving mass-to-light ratio is luminosity dependent dust-extinction. However, observations of the stellar populations of high-redshift galaxies have produced conflicting results on the existence and strength of any luminosity dependence. When measuring the UV continuum slope, $\beta$ \citep{1994ApJ...429..582C}, for samples of high-redshift galaxies, \citet{Wilkins:2011fs}, \citet{2012ApJ...754...83B} and \citet{Bouwens:2013vf} find evidence for a strong UV luminosity dependence across all redshifts at $z > 3$. In contrast, similar studies by \citet{Dunlop:2011jl}, \citet{2012ApJ...756..164F} and \citet{2013MNRAS.429.2456R} find no clear evidence for a luminosity dependence on $\beta$. Several of these studies outline the importance of the selection of high redshift galaxies (through either Lyman break or photometric redshift selection) and the treatment of their biases. To this end, more recent analyses \citep{Bouwens:2013vf,Rogers:2014bn} which increase sample sizes and minimise biases in the sample selection and $\beta$ measurements are in good agreement, with both studies finding a clear luminosity dependence.

The deep near-infrared observations of the GOODS South field made as part of the Cosmic Assembly Near-infrared Deep Extragalactic Survey (CANDELS; Co-PIs: Faber \& Ferguson; \citeauthor{2011ApJS..197...35G}~\citeyear{2011ApJS..197...35G}; \citeauthor{Koekemoer:2011br}~\citeyear{Koekemoer:2011br}), combined with the extensive existing optical observations make it a data set ideally suited to the study of galaxy evolution at the so-called `cosmic dawn'. Covering an area approximately 200\% larger than the WFC3 ERS observations alone \citep{2011ApJS..193...27W}, and incorporating the even deeper UDF observations, the CANDELS data combines the high sensitivity of the WFC3 observations with high-redshift samples large enough to attempt the first direct derivation of the stellar mass function at $z > 5$. In this paper, we make use of this comprehensive data set to study galaxy stellar masses across the redshift range $z \sim 4$ to $z \sim 7$. In particular, we aim to estimate stellar masses for a large and robust sample of high-redshift galaxies, investigating how the inclusion of nebular emission and increasing star-formation histories affect the observed stellar mass - UV luminosity relation and the shape of the stellar mass function. For this same sample, we also aim to measure the dust-corrected star-formation rates which will combine to make a detailed census of the stellar mass growth of high-redshift galaxies. 

The structure of this paper is as follows. In Section~\ref{sec:data}, we describe the properties of the optical and near-IR data sets used in this study as well as the methods of photometry extraction used. In Section~\ref{sec:redshift}, we describe the photometric redshift analysis and the selection criteria used to construct the high-redshift samples used in our analysis. The SED fitting method used to estimate the stellar masses for our sample is outlined in Section~\ref{sec:masses}. Section~\ref{sec:simulations} then describes the simulations undertaken to account for the effects of completeness and sample selection. In Section~\ref{sec:results} we present the results of this work and our analyses. Finally, Section~\ref{sec:summary} presents our summary and conclusions of the work in this paper.
Throughout this paper all magnitudes are quoted in the AB system \citep{1983ApJ...266..713O} and we assume a $\Lambda$CDM cosmology with $H_{0} = 70$ kms$^{-1}$Mpc$^{-1}$, $\Omega_{m}=0.3$ and $\Omega_{\Lambda}=0.7$. Quoted observables are expressed as actual values assuming this cosmology. Note that luminosities and luminosity based properties such as observed stellar masses and star-formation rates scale as $h^{-2}$, whilst densities scale as $h^{3}$.

\section{The Data}\label{sec:data}
The photometry used throughout this work is taken from the catalog of \citet{Guo:2013ig}, a UV to mid-infrared multi-wavelength catalog in the CANDELS GOODS South field based on the CANDELS WFC3/IR observations combined with existing public data.

\subsection{Imaging Data}
The near-infrared WFC3/IR data combines observations from the CANDELS survey \citep{2011ApJS..197...35G,Koekemoer:2011br} with the WFC3 Early Release Science (ERS; \citeauthor{2011ApJS..193...27W}~\citeyear{2011ApJS..193...27W}) and Hubble Ultra Deep Field (HUDF; PI Illingworth; \citeauthor{Bouwens:2010dk}~\citeyear{Bouwens:2010dk}) surveys. The southern two thirds of the field (incorporating the CANDELS `DEEP' and `WIDE' regions and the UDF) were observed in the F105W, F125W and F160W bands. The northern-most third, comprising the ERS region, was observed in F098M, F125W and F160W. In addition to the initial CANDELS observations, the GOODS South field was also observed in the alternative J band filter, F140W, as part of the 3D-HST survey (Brammer et al. 2012).

The optical HST images from the Advanced Camera for Surveys (ACS) images are version v3.0 of the mosaicked images from the GOODS HST/ACS Treasury Program, combining the data of \citet{2004ApJ...600L..93G} with the subsequent observations obtained by \citet{2006AJ....132.1729B} and \citep{Koekemoer:2011br}. The field was observed in the F435W, F606W, F775W, F814W and F850LP bands. Throughout the paper, we will refer to the HST filters F435W, F606W, F775W, F814W, F850LP, F098M, F105W, F125W, F160W as $B_{435}$, $V_{606}$, $i_{775}$, $I_{814}$, $z_{850}$, $Y_{098}$, $Y_{105}$, $J_{125}$, $H_{160}$ respectively. 

The \emph{Spitzer}/IRAC \citep{Fazio:2004eb} 3.6 and 4.5$\mu m$ images were taken from the Spitzer Extended Deep Survey (PI: G. Fazio, \citeauthor{Ashby:2013cc}~\citeyear{Ashby:2013cc}) incorporating the pre-existing cryogenic observations from the GOODS Spitzer Legacy project (PI: M. Dickinson). Complementary to the space based imaging of HST and Spitzer is the ground-based imaging of the CTIO U band, VLT/VIMOS U band \citep{Nonino:2009hf}, VLT/ISAAC $K_{s}$ \citep{Retzlaff:2010co} and VLT/HAWK-I $K_{s}$ (Fontana et al. \emph{in prep.}) bands.

\subsection{Source photometry and deconfusion}
The full details on how the source photometry was obtained are outlined in \citet{Guo:2013ig}, however we provide a brief summary of the method used for reference here. Photometry for the HST bands was done using SExtractor's dual image mode, using the WFC3 H band mosaic as the detection image and the respective ACS/WFC3 mosaics as the measurement image after matching of the point-spread function (PSF). 

For the ground-based (VIMOS and CTIO U band and ISAAC and Hawk-I Ks) and Spitzer IRAC bands, deconvolution and photometry was done using template fitting photometry (TFIT). We refer the reader to \citet{Laidler:2007iy}, \citet{2012ApJ...752...66L} and the citations within for further details of the TFIT process and the improvements gained on mixed wavelength photometry.

\section{Photometric Redshifts and Sample Selection}\label{sec:redshift}
Photometric redshifts for the entire source catalog were calculated using the EAZY photometric redshift software \citep{Brammer:2008gn}. The fitting was done to all available bands using the default reduced template set based on the PEGASE spectral models of \citet{1997A&A...326..950F} with an additional template based on the spectrum of \citet{2010ApJ...719.1168E}. The additional template exhibits features expected in young galaxy populations such as strong optical emission lines and a high Lyman-$\alpha$ equivalent width.

For each galaxy we construct the full redshift probability distribution function (PDF), $P(z) \propto exp(-\chi_{z}^2/2)$, using the $\chi^2$-distribution returned by EAZY. 
Although EAZY allows the inclusion of a magnitude based prior when calculating redshifts, none was included in the fitting due to the large uncertainties still present in the H-band (our photometry selection band) luminosity function at high-redshifts \citep{Henriques:2012gs}.

\subsection{Selection Criteria} \label{sec:sample}
To investigate how the SMF evolves from $z =$ 4-7, we wish to construct a sample of galaxies in the redshift range $3.5 < z < 7.5$. To select a robust sample suitable for SED fitting, we apply a set of additional criteria based on the full redshift probability distribution for each galaxy to construct the different redshift samples, similar to those used in previous high-redshift sample selections \citep{2011MNRAS.418.2074M,Finkelstein:2012hr}. We then apply the following criteria:


\begin{equation}\label{eq:crit1}
\int_{z_{\text{sample}}-0.5}^{z_{\text{sample}}+0.5} P(z)~dz > 0.4
\end{equation}

\begin{equation}\label{eq:crit2}
\int_{z_{\text{peak}}-0.5}^{z_{\text{peak}}+0.5} P(z)~dz > 0.6
\end{equation}
 
\begin{equation}\label{eq:crit3}
(\chi^{2}_{min} /N_{\text{filters}}-1) < 3
\end{equation}

\noindent where $z_{sample}$ = 4, 5, 6 and 7 for the respective bins and $z_{peak}$ is the redshift at the peak of the probability distribution (i.e. minimum $\chi^2$). 

The first criterion (Equation~\ref{eq:crit1}) requires that a significant amount of the probability distribution lies within the redshift range we are examining. The second criterion (Equation~\ref{eq:crit2}) requires that the bulk of the PDF lies close to the peak of the distribution, i.e. that the primary solution is a dominant one. Finally, we require that EAZY provides a reasonable fit to the data (Equation~\ref{eq:crit3}). 

A signal to noise ($\textup{S/N}$) cut is placed on the J and H bands, requiring $\textup{S/N}(J_{125}) > 3.5$ and $\textup{S/N}(H_{160}) > 5$. Known AGN, stars and sources with photometry flagged as effected by artefacts are removed. We also visually inspect each galaxy across all the HST bands, excluding sources which were caused or strongly affected by artefacts such as diffraction spikes, bright stars and image edges which were not excluded by any of the other criteria. 

Of the initial 34930 objects in the CANDELS GOODS South catalog, 3164 objects satisfy our first criterion. Of those objects, 256 are excluded by the second criterion and a further 167 are rejected based on their $\chi^{2}$. The signal to noise criteria exclude a further 274 sources and the remaining criteria exclude a further 204 sources. The resulting final sample comprises 2263 galaxies.

\begin{figure}
\includegraphics[width=84mm]{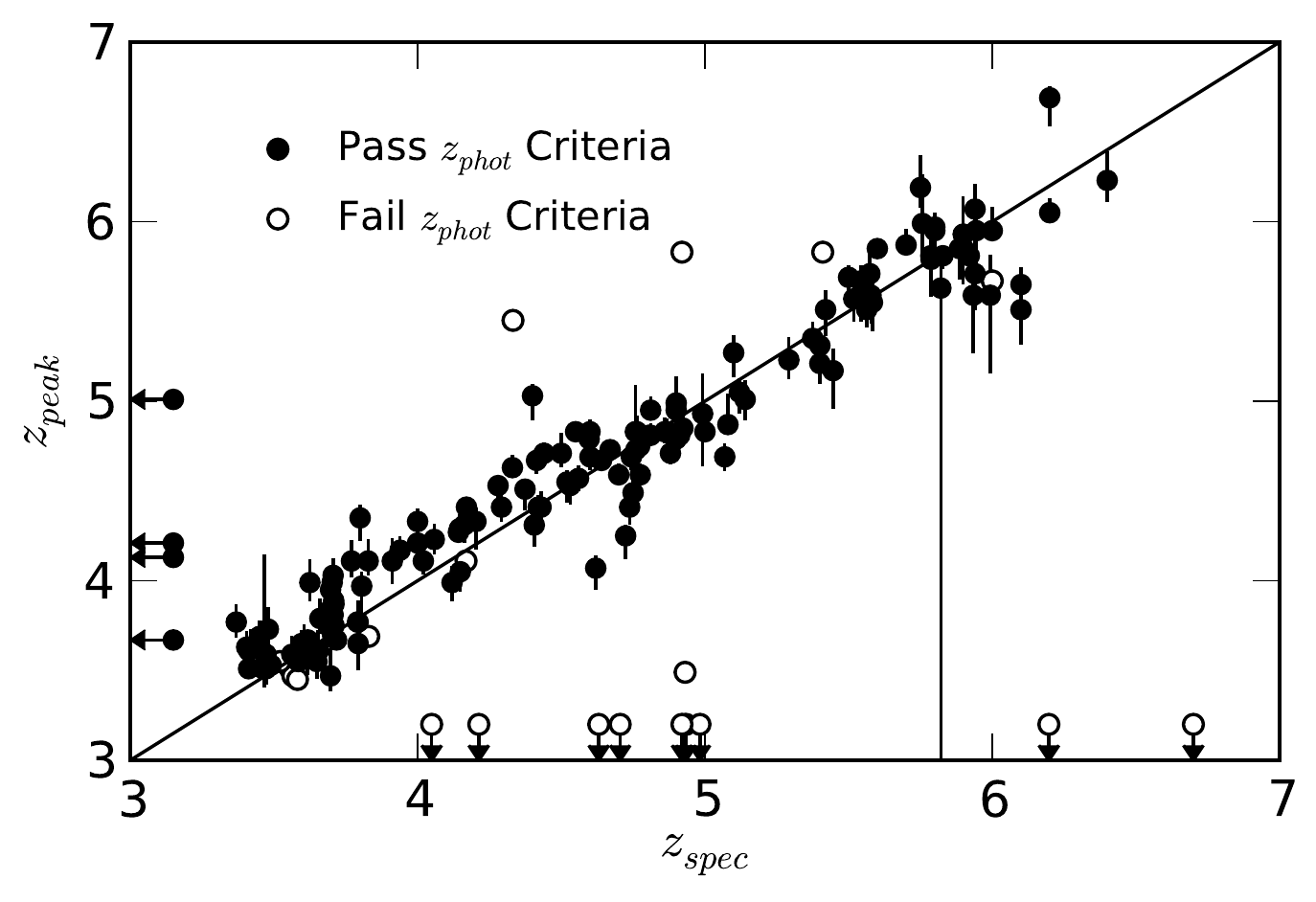}
\caption{Comparison between spectroscopic and photometric redshift for the galaxies in our sample with available spectroscopy and spectroscopic redshift quality of `Good" or better. Filled circles show sources which pass our selection criteria (including interlopers), empty circles show spectroscopically confirmed high-redshift sources which do not pass the selection criteria. The photometric redshift shown is the peak of the probability distribution ($\chi^2$ minimum) with 1-$\sigma$ lower and upper limits.}
\label{fig:specz}
\end{figure}

Figure~\ref{fig:specz} compares the available spectroscopic redshifts for the galaxies in our sample with the corresponding best-fit photometric redshift (minimum $\chi^2$) as found by EAZY. In total, there are 151 spectroscopic redshift matches for galaxies which pass our selection criteria and are therefore included in our samples. In addition, there are a further 21 galaxies with spectroscopic redshifts of $z > 3.5$ which pass the signal to noise, AGN criteria but do not pass the photometric redshift criteria. Of these 21 galaxies, 12 are correctly identified as high redshift galaxies but are excluded due to poor fits (11) or have a redshift very close to the $z = 3.5$ limit but photometric scatter pushes the photometric redshift to below the criteria (1). The remaining 9 spectroscopically confirmed high-redshift sources have best-fit photometric redshifts of $z_{peak} < 3$. The simulations undertaken to correct for selection completeness are outlined in Section~\ref{sec:completeness}.

For the matched galaxies which pass our selection criteria, we find that our redshift accuracy is very good, with a scatter of just $\sigma_{z,O} = rms(\Delta z /1+z_{\text{spec}}) = 0.037$ when outliers are excluded, where $\Delta z =  (z_{\text{spec}}-z_{\text{phot}})$ \citep{Dahlen:2013eu}. We define outliers as $\left | \Delta z \right |/(1+z_{\text{spec}}) > 0.15$, and find an outlier fraction of 2.65\% (4 galaxies). This compares with \citet{2012ApJ...756..164F} who find a scatter of $\sigma_{z}/(1+z) = 0.044$  at $z > 3$ after excluding outliers (defined more strictly as $\left | \Delta z \right | > 0.5$) in the same CANDELS field. We also find that there is very little bias in our photometric redshifts, with $median(\Delta z) = -0.04$. Of the 4 galaxies classed as outliers, all lie at redshifts of $z < 3$ and are low-redshift interlopers which our selection criteria have not been able to exclude.

\citet{Dahlen:2013eu} have recently shown that by combining the results from multiple photometric redshift codes, the scatter and outlier fraction in photometric redshifts can be significantly reduced compared to the results of any single code. For the same set of 151 spectroscopic redshift sources, the photometric redshifts produced through the Bayesian combination outlined in \citet{Dahlen:2013eu} have $\sigma_{z,O} = 0.033$ with an outlier fraction of 3.98\% and $median(\Delta z) = 0.01$. 

Although utilising the photometric redshifts for the CANDELS GOODS-S field produced by this method would result in a small gain in photometric accuracy, we would no longer be able to reproduce the full selection method when running simulations. Given this small improvement, we are confident that the use of photometric redshifts produced by a single code will not adversely affect the overall accuracy of the results.

The matched spectroscopic redshifts are from the following surveys: \citet{LeFvre:2004ge,Stanway:2004gu,Vanzella:2008hp,Hathi:2008ca,Popesso:2009ht,Wuyts:2009gv,Rhoads:2009eb,Vanzella:2009ez,Balestra:2010bt,Kurk:2012ej}. The high-redshift spectroscopic sources within these surveys all derive from initial target selections of predominantly bright Lyman Break galaxy candidates. The measured photometric redshift accuracies are therefore likely biased to a better scatter than the full high-redshift galaxy population. However, examining the redshift accuracy of the mock galaxy catalog used for our selection comparisons in Appendix~\ref{app:selection}, we find that the photometric redshift accuracy remains good down to the lowest masses probed in this survey for galaxies which pass our criteria. For example, for galaxies of $\text{M}_{*} \approx 10^{8.5} \text{M}_{\odot}$, we find a scatter excluding outliers of $\sigma_{z,O} = 0.053$ and $median(\Delta z) = 0.025$ before any $P(z)$ criteria are applied.

To investigate how the SMF evolves from z = 4-7, we then constructed four redshift samples in bins across this redshift range: $z \sim 4 ~(3.5 \leq z < 4.5)$, $z \sim 5~ (4.5 \leq z < 5.5)$, $z \sim 6~(5.5 \leq z < 6.5)$ and $z \sim 7 ~(6.5 \leq z < 7.5)$.

\subsection{Monte Carlo Samples}\label{sec:MC}
Although we find that our photometric redshifts do well when compared with the matched spectroscopic redshifts, the group of outliers are indicative of the difficulties that exist in correctly distinguishing between the Lyman break of high-redshift galaxies at $z > 3$ and strong Balmer break galaxies at more moderate redshifts $z \approx 0.5-2.5$ in low S/N data. \citet{2012ApJ...748..122P,Pirzkal:2013ug} have shown that it is very difficult to categorically classify sources as high-redshift galaxies and not low-redshift interlopers using photo-z's or S/N criteria on the dropout bands.

Previous work using photometric redshifts has dealt with this problem by making use of the full redshift PDF when calculating luminosity functions \citep{2005ApJ...631..126D,2009MNRAS.395.2196M,McLure:2013hh}, thereby incorporating the uncertainty in the analysis. Due to the nature of the SED fitting code used for this work (described in Section~\ref{sec:masses}), the computational effort required to fit the mass at each redshift in order to integrate over the full PDF becomes impractical. As such, we chose to account for these problems in a different manner whilst still dealing with them in a straight-forward way. 

Rather than using only the best-fit redshift from our photometric analysis when selecting our sample, we instead draw the redshift for each galaxy randomly from its full PDF before placing it in the appropriate redshift sample. Where secure spectroscopic redshifts are available, we fix the redshift to that value for all samples (known interlopers are therefore excluded in all samples). This process was repeated 500 times to produce a set of samples to which we then apply the rest of the analysis described in the paper separately. We then average over the results from each sample, using the mean of this full set as our `true' value along with the 1-$\sigma$ upper and lower limits around this mean. 

\begin{table}
\caption{Average sample size and variance for each redshift bin for the 500 Monte Carlo samples generated, see text for details.}
\label{tab:samplesizes}
\begin{tabular}{@{}ccc}
\hline
 Redshift Bin & Mean Sample Size & Variance on sample size  \\
  \hline
 4 & 1235 & 180 \\
 5 & 416 & 63 \\
 6 & 169 & 25 \\
 7 & 42 & 9 \\
 \hline
\end{tabular}
\end{table}

The resulting sample sizes for each redshift bin are shown in Table~\ref{tab:samplesizes}. The varying samples account for both scattering between redshift bins for objects at the boundaries as well as objects moved out of the sample into secondary low-redshift solutions. The effect of this scattering into and out of the samples can be seen when comparing the combined mean samples sizes (1862) to our full high-redshift sample of 2263.

\begin{figure*}
\includegraphics[width=140mm]{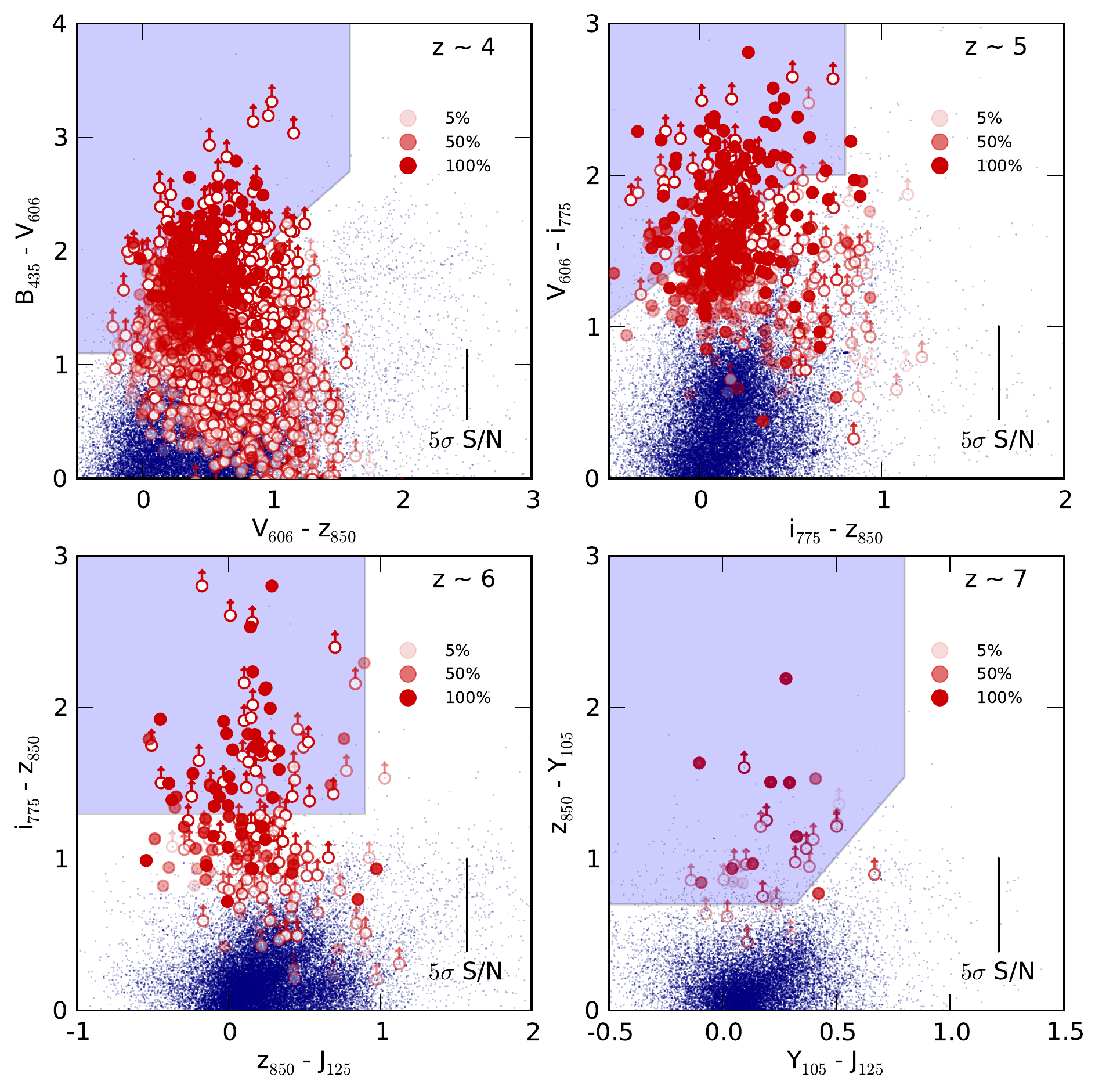}
\caption{The colours of our photometric redshift selected samples in relation to the two-colour cuts typically used to select Lyman break galaxies. Non-detections in a filter are converted to 2-$\sigma$ upper limits when calculating the colours. The shaded blue regions show the region in colour space used to select dropout galaxies in that redshift bin as described in  \citet{2007ApJ...670..928B} and \citet{2012ApJ...754...83B}. The blue points show the colours for the full GOODS-S photometric catalog from which we select our high redshift samples. Red symbols represent galaxies selected in our sample by our selection criteria, where the transparency of the symbols is determined by the number of Monte Carlo samples in which it is selected i.e. the fainter the symbol, the smaller the fraction of MC samples that galaxy is selected in. The legend in each plot illustrates this transparency for galaxies selected in 5, 50 and $100\%$ of Monte Carlo samples. Colours which make use of 2-$\sigma$ upper limits are plotted with open circles, while objects with 2-$\sigma$ detections or better in all bands are plotted with filled circles. Example error bars corresponding to 5-$\sigma$ detections (for both filters in a given colour) are shown for each of the corresponding drop-out colours.}
\label{fig:colours}
\end{figure*}

The strength of using photometric redshifts for sample selection over colour cuts (especially when redshifts would still need to be calculated for a colour-cut sample in order to do SED fitting) is that the method can automatically make use of all available photometry. This is important because although photometric redshifts are still fitting primarily to the characteristic break at Lyman-$\alpha$ targeted by the colour selections, the filters long-ward of the break are useful in excluding low-redshift interlopers \citep{2011MNRAS.418.2074M}. Additionally, the large errors in colour possible due to low signal to noise and possible non-detections in the filter just short of the Lyman break means that likely high-redshift candidates can be scattered well outside the selection region when using colour cuts. 

Figure~\ref{fig:colours} shows the positions of our galaxy samples on the colour-colour planes commonly used to select dropout samples. It is obvious that many of the galaxies selected with photometric redshifts lie outside the selection regions (as taken from \citeauthor{2007ApJ...670..928B}~\citeyear{2007ApJ...670..928B}), especially those galaxies where colours must be calculated using an upper limit. The agreement (or lack of) between dropout selections and photometric redshifts has also been investigated for GOODS-S specifically by \citet{2010ApJ...724..425D}.

To test whether the discrepancy between the observed colours and those required for Lyman break selection can be explained solely by photometric scatter, we performed a range of tests on a mock catalog generated from the semi-analytic models described in \citet{Somerville:2008ed} and \citet{Somerville:2012cq}. Full details of these tests are outlined in Appendix~\ref{app:selection}, however our main finding is that the observed colours can be reproduced from intrinsic Lyman break-like colours and scattering proportional to the observed photometric errors. 

\section{Mass Fitting}\label{sec:masses}
Stellar masses were estimated for our samples using a custom template fitting code with SEDs derived from the synthetic stellar population models of \citet{Bruzual:2003ckb} (BC03 hereafter). Due to the relatively young ages of the stellar populations (as constrained by the age of the Universe at the redshifts involved), the effects of thermally-pulsating asymptotic giant branch stars (TP-AGB) on resulting SEDs are minimal, e.g. \citet{2009ApJ...697.1493S}. As such, we chose not to use the updated SSP models which incorporate stronger contributions for these effects. The use of Charlot \& Bruzual (2007) or \citet{Maraston:2005er} in place of BC03 should have no result on the conclusions found in this work.


\subsection{Model SEDs}\label{subsec:seds}
First, model SEDs are generated from the single stellar populations of Bruzual \& Charlot for a range of population ages, metallicities and star-formation histories (SFHs). For our models and throughout this work we use the initial mass function (IMF) of \citet{Chabrier:2003ki}. Each template is then normalised such that the total stellar mass equals 1 M$_{\odot}$. Nebular emission lines are added to the pure stellar component following the method outlined in Section~\ref{subsection:nebular}. Internal dust extinction is applied following the extinction law of \citet{2000ApJ...533..682C} for the desired range of extinction magnitude $A_{V}$.

When applying dust extinction to the nebular emission, we assume the differential dust extinction between stellar and nebular emissions to be fixed as $E(B-V)_{\text{stellar}} = E(B-V)_{\text{nebular}}$, in contrast to the ratio of $E(B-V)_{\text{stellar}} = 0.44 E(B-V)_{\text{nebular}}$ derived by \citet{2000ApJ...533..682C}. This choice was motivated by the conflicting evidence for the relative extinction of the two emission sources at $z \sim 2$ \citep{Erb:2006ke,ForsterSchreiber:2009hm} and that in the context of these models specifically, the assumed differential extinction ratio and escape fraction are degenerate.

The absolute magnitude at 1500 \AA ~($M_{UV}$) is measured for each template by integrating the flux within a 100 \AA-wide flat bandpass centered on 1500 \AA. Similarly, the UV-continuum slope $\beta$ is calculated by fitting a simple power-law to the integrated template fluxes in the windows described in \citet{1994ApJ...429..582C}. The accuracy of this method in calculating $\beta$ is explored in \citet{Finkelstein:2012hr} and \citet{2013MNRAS.429.2456R}.

Next, each model SED is redshifted in the range $0 \le z < 9$ in steps of $\Delta z = 0.02$, and attenuation by intergalactic neutral hydrogen is applied following the prescription of \cite{1995ApJ...441...18M}. The resulting grid of SEDs is then integrated through the response filter for each of the observed bands, dimmed by the corresponding distance modulus for each redshift.

\subsection{Nebular Emission}\label{subsection:nebular}
Nebular emission lines and continuum are added to the templates following a method similar to previous high-redshift fitting methods, e.g.  \citet{Ono:2010ed,2010A&A...515A..73S,2011MNRAS.418.2074M}, \citet{Salmon:2014tm}. Line strength ratios relative to H$\beta$ for the major Balmer, Paschen and Brackett recombination lines are taken from \cite{Osterbrock:2006ul}, with the total H$\beta$ line luminosity (in erg s$^{-1}$) given by
\begin{equation}
L(\text{H}\beta) = 4.78 \times10^{-13} (1-f_{\text{esc}}) N_{\text{LyC}}
\end{equation}
from \citet{1995A&A...303...41K}, where $f_{\text{esc}}$ is the continuum escape fraction and the number of Lyman continuum photons $N_{\text{LyC}}$ is calculated from each template. The strength of the nebular emission is therefore directly proportional to the number of ionizing photons (Lyman continuum) in the HII region. Since the Lyman continuum emission is dominated by young massive stars, the relative contribution of nebular emission to the total SED is highly dependent on the age of the stellar population and the amount of recent star-formation.  
 
Line ratios for the common metal lines relative to H$\beta$ were taken from the empirical measurements of \cite{Anders:2003ci} for each of the input template metallicities, assuming gas metallicity is equal to the stellar metallicity.
Similarly, the nebular continuum emission luminosity is given by
\begin{equation}
\label{eq:continuum}
L_{\nu} = \frac{\gamma^{(total)}_{\nu}} {\alpha_{B}}(1-f_{\text{esc}}) N_{\text{LyC}}
\end{equation}
where $\alpha_{B}$ is the case B recombination coefficient for hydrogen and $\gamma^{(total)}_{\nu}$ is the continuum emission coefficient given by
\begin{equation}\label{eq:cont_sep}
\gamma^{(total)}_{\nu} = \gamma^{(HI)}_{\nu} + \gamma^{(2q)}_{\nu} +  \gamma^{(HeI)}_{\nu}\frac{n(He^{+})} {n(H^{+})} + \gamma^{(HeII)}_{\nu}\frac{n(He^{++})} {n(H^{+})}
.\end{equation}
$\gamma^{(HI)}_{\nu}$, $\gamma^{(HeI)}_{\nu}$, $\gamma^{(HeII)}_{\nu}$ and $\gamma^{(2q)}_{\nu}$ are the continuum emission coefficients for free-free and free-bound emission by Hydrogen, neutral Helium, singly ionized Helium and two-photon emission for Hydrogen respectively, where the values are taken from \cite{Osterbrock:2006ul}. The coefficients and constants used assume an electron temperature $T=10^4$ K, electron density $n_{e}=10^2$ cm$^{-3}$ and the abundance ratios are set to be $\frac{n(He^{+})} {n(H^{+})} = 0.1$ and $\frac{n(He^{++})} {n(H^{+})} = 0$ \citep{1995A&A...303...41K}.

\subsection{SED Fitting}
The fitting of our SEDs to the observed photometry is done using a Bayesian-like approach, whereby the normalised likelihood $\mathcal{L}(M,t)$ for a given stellar mass, $M$, and template type, $t$, is given by
\begin{equation}
 \mathcal{L}(M,t) = \frac{\exp(-\tfrac{1}{2}\chi^{2}(M,t))}
          {\sum_{t'} \int dM' ~ \exp(-\tfrac{1}{2}\chi^{2}(M',t'))}.
\end{equation}
The $\chi^2$ value is given by
\begin{equation}
  \chi^{2}(M,t) = \sum_{j}^{N_{filters}} \frac{(M F_{j}(t) - F_{j}^{obs})^2} {\sigma_{j}^{2}}
\end{equation}
where $F_{j}(t)$, $F_{j}^{obs}$ and $\sigma_{j}$ are the template flux, observed flux and the observed flux error in the $j$th filter respectively. The template types, $t$, and their associated fluxes correspond to the full range of galaxy parameters (age, star-formation history, dust extinction and metallicity) at the closest matching redshift in the model SED grid.

Because we fit all templates simultaneously, it is therefore straight-forward to calculate the stellar mass probability distribution function (PDF), i.e. 
\begin{equation}
  P(M) \propto \sum_{t} \mathcal{L}(M,t),
\end{equation}
marginalised over all other template galaxy properties (assuming a flat prior). Similarly, PDFs for other parameters such as $\beta$ or $M_{UV}$ ($M_{1500\rm{\AA}}$) can be constructed by summing the likelihoods at a fixed parameter value. Estimating the galaxy parameters in such a way allows us to fully account for errors due to both degeneracies between galaxies properties and errors in the scaling due to the photometric errors.

For our mass fitting, model ages are allowed to vary from 5 Myr to the age of the Universe at a given redshift, dust attenuation is allowed to vary in the range $0 \le A_{V} \le 2$ and metallicities of 0.02, 0.2 and 1 Z$_{\odot}$. Due to the difficulty in obtaining spectroscopy at $z > 3$, the metallicity at high-redshift is not currently well known. Observations of samples at $z \sim 3$ and above \citep{Shapley:2003wi,2008ASPC..396..409M,2012A&A...539A.136S,Jones:2012kn} show that the average metallicity is likely to be mildly sub-solar, however there is a large scatter. \citet{2012A&A...539A.136S} also find that the gas-phase and stellar metallicities are consistent within errors. As such, we choose to fix the metallicity for the nebular emission equal to stellar metallicity.

The star formation histories follow the exponential form $SFR \propto e^{-t/\tau}$ with characteristic timescales of $\tau = $ 0.05, 0.25, 0.5, 1, 2.5, 5, 10, -0.25, -0.5, -1, -2.5, -5, -10 and 1000 (effectively constant SFR) Gyrs. Negative $\tau$ values represent exponentially increasing histories. Fitting is done to the templates both with and without the inclusion of nebular emission. When nebular emission is included in the templates, we assume a moderate escape fraction $f_{\text{esc}} = 0.2$, consistent with the observational constraints on reionization and with simulations \citep{Yajima:2010fb,Fernandez:2011cw,Finkelstein:2012hr,Robertson:2013ji}.

\subsection{Star Formation Rates}\label{sec:SFR}
In order to calculate UV star-formation rates, the rest frame absolute magnitudes (M$_{1500}$) measured from the SED fitting are first corrected for dust extinction using the \citet{Meurer:1999jm} relation
\begin{equation}
A_{1600} = 4.43 + 1.99\beta
\end{equation}
which links the observed UV continuum slope $\beta$ as measured by the SED fitting code (see Section~\ref{subsec:seds}) and the extinction at 1600\AA, $A_{1600}$. For measured $\beta < - 2.23$, where the above relation would imply a negative extinction, the UV extinction was set to 0. UV star-formation rates are calculated using: 

\begin{equation}\label{eq:sfr}
  \textup{SFR} (M_{\odot} \textup{yr}^{-1}) = 
    \frac{L_{UV} (\textup{erg s}^{-1} \textup{Hz}^{-1})} {13.9 \times 10^{27}} ,
\end{equation}
where the L$_{UV}$ conversion factor of \citet{Madau:1998jd} and \citet{KennicuttJr:1998id} corrected to the Chabrier IMF is used ($-0.24$ dex).

In addition to the the star formation rate obtained by this method (SFR$_{\rm{Madau}}$ hereafter), from our SED fitting code we also obtain the instantaneous star formation rate of the best-fitting template for each galaxy (SFR$_{\rm{Template}}$). We find that the two measures agree well at all SFRs with a median$(\log_{10}(\text{SFR}_{\text{Madau}}) - \log_{10}(\text{SFR}_{\text{Template}})) < 0.1$. With the exception of the few galaxies with the highest SFR$_{\rm{Madau}}$, typically SFR$_{\rm{Madau}} > 100$ M$_{\odot} \textup{yr}^{-1}$. These galaxies are red, such that the best-fitting template is an older quiescent stellar population. The Meurer relation however assumes an actively star-forming population with high dust extinction.

We also find that the scatter around the 1:1 relation correlates strongly with the age of the best-fitting SED template, such that younger populations have higher SFR$_{\rm{Template}}$. As we will show in the next section however, individual stellar population parameters such as age and dust extinction are very degenerate in SED fits of high-redshift galaxies. Because of these factors, and for consistency with previous works, we primarily use SFR$_{\rm{Madau}}$ throughout this work. The net effect of the differences in the two star-formation rate estimates can be seen in our observed SFR functions in Section~\ref{sec:SFR}.

\subsection{Image and Detection Simulations}\label{sec:simulations}
By their nature, high-redshift galaxies are small and extremely faint objects. Lying close to the limiting depth in some (or even all) of the observed filters, noise and systematic effects can have a significant effect on the detection and completeness of high-redshift galaxy samples as well as the accurate estimation of their properties. The completeness of our galaxy sample can be separated into two distinct factors: firstly, the inclusion of an object in the initial catalog as a function of the detection image depth, and secondly, the selection of an object in a given sample (e.g $z\approx4$ based on its estimated redshift), which is a function of the overall SED shape and accompanying errors. In this section, we outline a set of detailed simulations undertaken to measure and correct for these effects.

\subsubsection{Completeness Simulations}\label{sec:completeness}
The detection completeness across the field was estimated by inserting thousands of mock galaxies into the detection image (H-band) used for the photometry and recovering them with the same \textsc{SExtractor} parameters and method used for the original sample catalog. The synthetic galaxies were first convolved with the CANDELS WFC3 H-band PSF before being placed randomly across the field with appropriate Poisson noise. The resulting image was then run through the same \textsc{SExtractor} procedure as the initial source detection and the process repeated until a total of $\approx  10^{5}$ input galaxies had been recovered across the entire field.

Galaxy sizes were drawn from a log-normal distribution of mean = 0.15\arcsec and $\sigma = 0.075$, motivated by existing observations of the size evolution of Lyman break galaxies \citep{2004ApJ...600L.107F,2010ApJ...709L..21O,2011A&A...532A..33G,Huang:2013kb} whilst the galaxies profiles were drawn from a distribution of S\'{e}rsic indices centred around $n=1.5$ in the range $0.5 \le n \le 4.0$. Although the precise distribution of morphological profiles for high-redshift galaxies is not well known, studies of lower redshift analogues and stacked samples of LBGs suggests that they are predominantly disk-like ($n < 2$) \citep{Ravindranath:2006ie,Hathi:2007fh}. Our chosen distribution reflects this, with $\sim 80\%$ of input galaxies with $n \leq 2$.

\begin{figure}
\includegraphics[width=84mm]{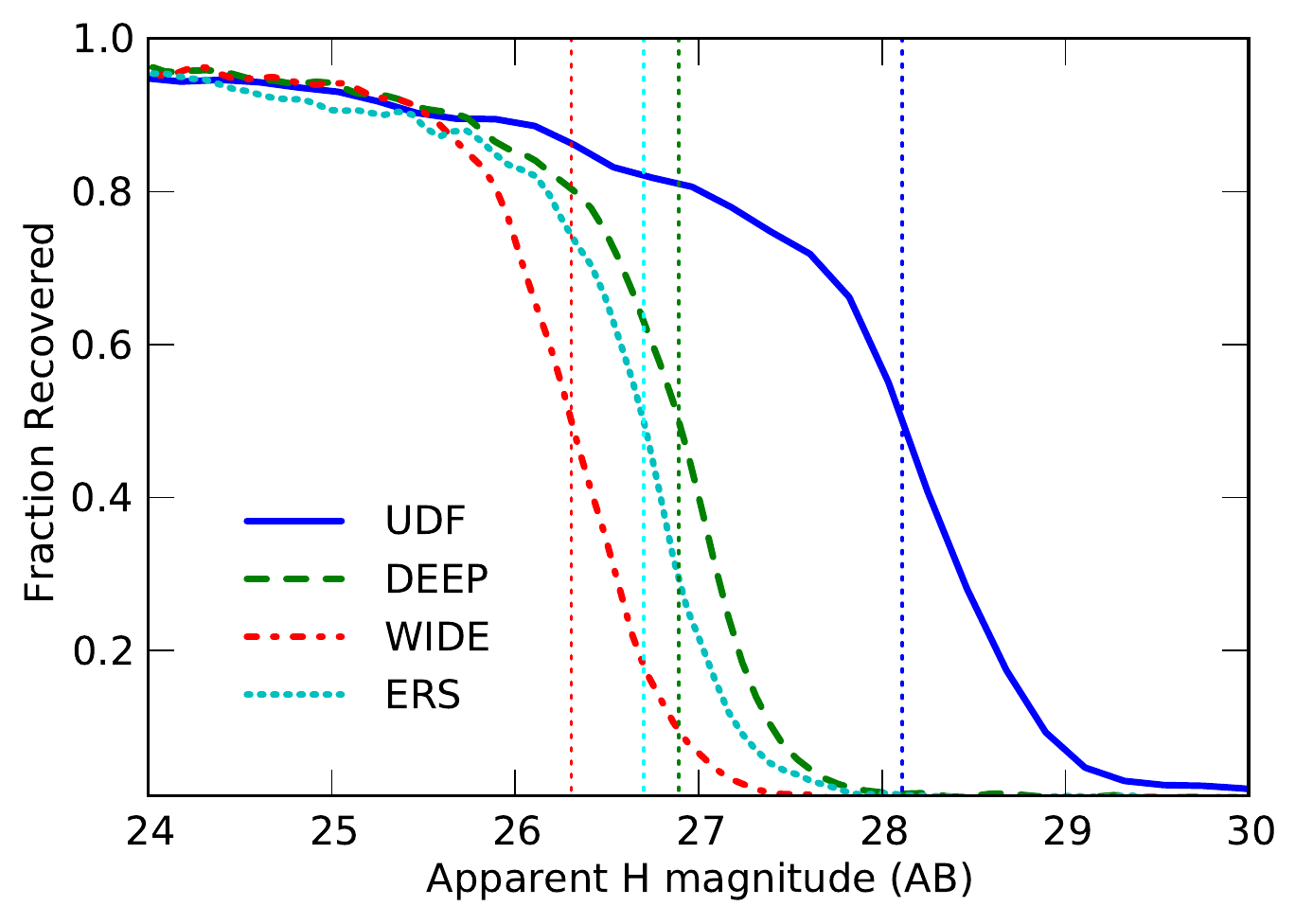}

\caption{Completeness as a function of $H_{160}$ magnitude for each region of the GOODS South field. The vertical dashed lines show the magnitude at which the recovery fraction equals 0.5 for each region of the field.}
\label{fig:completeness}
\end{figure}

Figure~\ref{fig:completeness} shows the resulting completeness curves for each of the image regions. In \citet{Guo:2013ig}, the $H_{160}$ 50\% completeness limit is estimated using the differential number density to be 25.9, 26.6 and 28.1 for the WIDE, DEEP+ERS and UDF regions respectively. When compared to the results of a set of detection simulations similar to those undertaken in our work, \citet{Guo:2013ig} find a good agreement between the two estimates. For the UDF and DEEP+ERS regions, our 50\% completeness limits are in good agreement with those of \citet{Guo:2013ig}. However, in the WIDE region we find a 50\% completeness limit $\approx 0.5$ mag deeper for our input galaxy population.

\citet{2011A&A...532A..33G} demonstrated the significant effect that sizes and morphologies can have on the completeness simulations of high-redshift galaxies. Differences due to the distribution of sizes and slightly differing galaxy profiles used are to be expected. For all regions, the effects of confusion and blending with nearby sources results in a small fraction of input galaxies which are not recovered by the photometry, even at brighter magnitudes.

\subsubsection{Sample Selection}\label{sec:selection_sims}
To estimate the selection functions for each of the redshift bins, a mock photometry catalog of high-redshift galaxies was created and put through the same photometric redshift and sample selection criteria as our real sample. This catalog was constructed by first creating a sample of SEDs drawn randomly from the template sets used for fitting (both with and without nebular emission) with a distribution of $\beta$ centred at $\approx -1.8$ to $-2$, but extending out to $\beta > 1$. Redshifts were allowed to vary in the range $2.5 < z < 9$ and the templates were scaled to $H_{160}$ band magnitudes in the range $22 < H_{160} < 30$, with the corresponding magnitudes in the other filters determined by the shape of each SED.

We produced a catalog in this way, rather than using the mock photometry of semi analytic models as used in Appendix~\ref{app:selection} in order to allow the inclusion of nebular emission in subsequent tests on the stellar mass fitting and ensure good number statistics across all input magnitudes. 

In order to assign photometric errors to the mock photometry (or non-observations where appropriate), each simulated galaxy was assigned a position in the field drawn from the same set of input coordinates as used in the completeness simulations. Photometric errors were then assigned to each photometric band based on the observed flux errors of objects in the original catalog, specific to the region in which it resides (e.g. CANDELS Deep). The flux values for each SED were then perturbed by a gaussian of width equal to the photometric error.

This process does not precisely mirror the method used to produce the observed photometry as it does not include the source extraction for each band individually. However, the resulting catalog is a very close approximation with a catalog of SEDs that have realistic photometric errors and filter coverage across the field, e.g. $Y_{098}$ observations in the ERS region alone. 

To measure the selection efficiency for our high-redshift samples, 100 simulated Monte Carlo samples were created from the template based mock galaxy catalog (as described in this section) using the method outlined in Section~\ref{sec:sample}. From these samples, we measured the fraction of simulated galaxies which pass the selection criteria for any of the high-redshift samples as a function of input redshift and magnitude.

\begin{figure*}
\includegraphics[width=180mm]{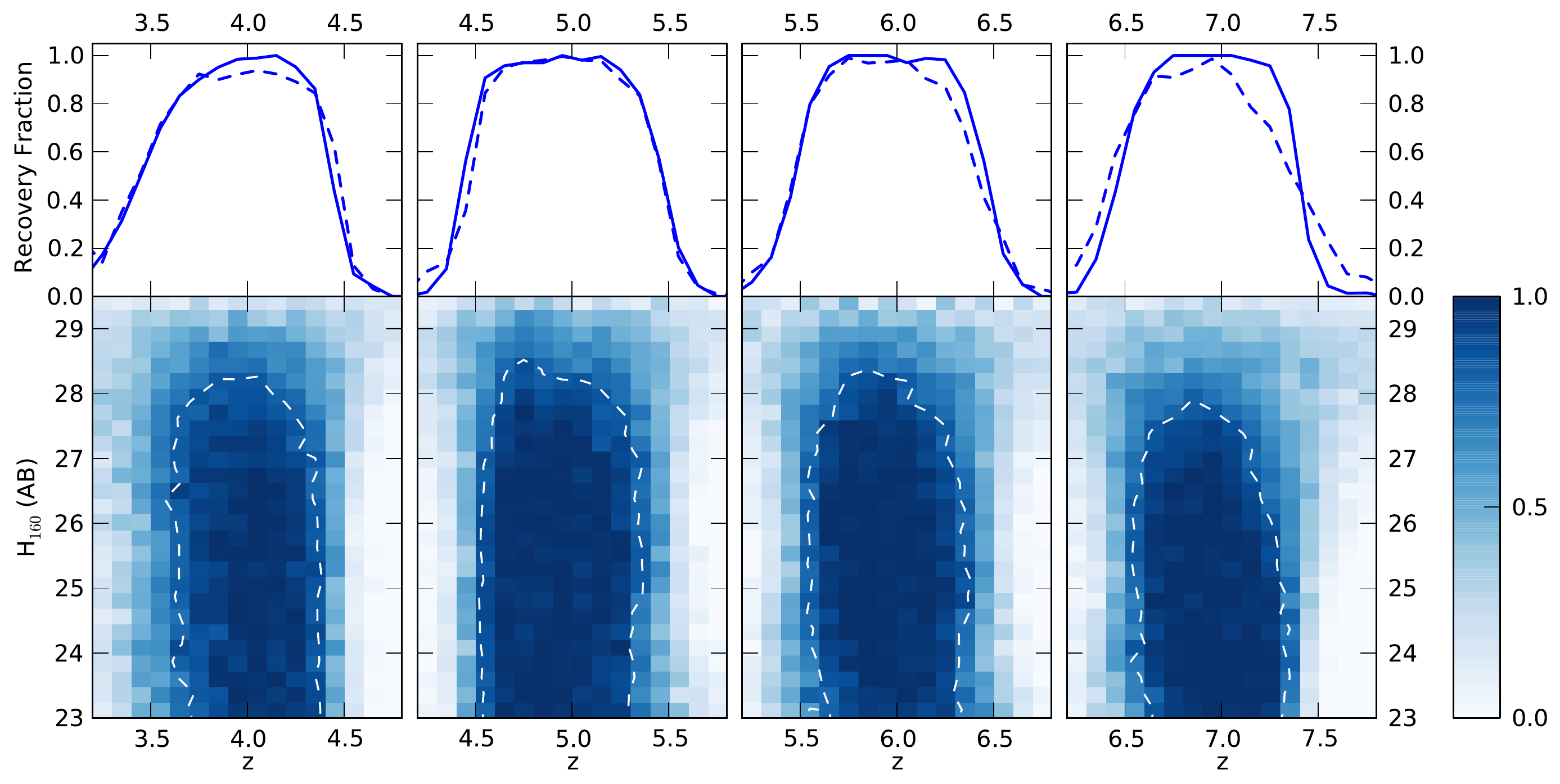}
\caption{Example selection efficiencies for the Ultra Deep Field region of the CANDELS field. The colour scale represents the fraction of input galaxies which pass the $P(z)$ criteria for a given redshift bin as a function of input redshift and apparent magnitude. The dashed white line in the lower sections of the figure shows the 80\% contour in the fraction of recovered galaxies. The upper panels show the recovery fraction as a function of redshift at a fixed input magnitude, $H_{160} = 25$ (continuous) and $H_{160} = 27$ (dashed).}
\label{fig:selection}
\end{figure*}

Figure~\ref{fig:selection} shows the measured selection efficiencies for  the deepest region of the field, the UDF. The selection probabilities (as indicated by the colour scale) do not include the effects of completeness as measured in Section~\ref{sec:completeness}, therefore the lower probabilities measured at faint magnitude are a result of photometric redshift errors due to poor constraints from faint photometry.

For $z\sim 4$ and 5, where the semi-analytic mock catalog used in Appendix~\ref{app:selection} has good number statistics across a wide magnitude range, we reproduce the selection function in the same manner as above and find that the shape of the resulting selection functions are unchanged. We are therefore confident that the photometric selection of our samples is robust to variations in the exact shape of the input SEDs and the limiting factor in selection is the photometric noise.

\subsubsection{Uncertainties in measuring galaxy parameters}
The ability of SED fitting codes to recover the properties of dropout galaxies was well explored by \citet{2010ApJ...725.1644L} who found that stellar mass was the most reliably measured parameter (in comparison to star formation rate and age) and the most robust to assumptions in star-formation history. However this analysis was limited to input and fitted SED models which did not include the effects of nebular emission. The degeneracies in measuring age, dust extinction and star-formation histories from SED fitting have also been well examined e.g. \citet{2010A&A...515A..73S}. Despite these degeneracies, it has been shown that one can reliably measure the UV continuum slope, $\beta$ \citep{2012ApJ...756..164F,2013MNRAS.429.2456R}. Given assumptions about the age and metallicity, i.e. the underlying intrinsic $\beta$, it is then possible to estimate the dust extinction using observations of $\beta$.

For the $\sim 10^5$ galaxies in our simulated catalog which pass the selection criteria, we ran them through the SED fitting code using the same fitting parameters as for our observed data. From these results we are able to test how well the input stellar masses are recovered for our simulated galaxies. In addition, we can also test the accuracy in recovering the other properties of the input stellar populations.

\begin{figure*}
\includegraphics[width=\textwidth]{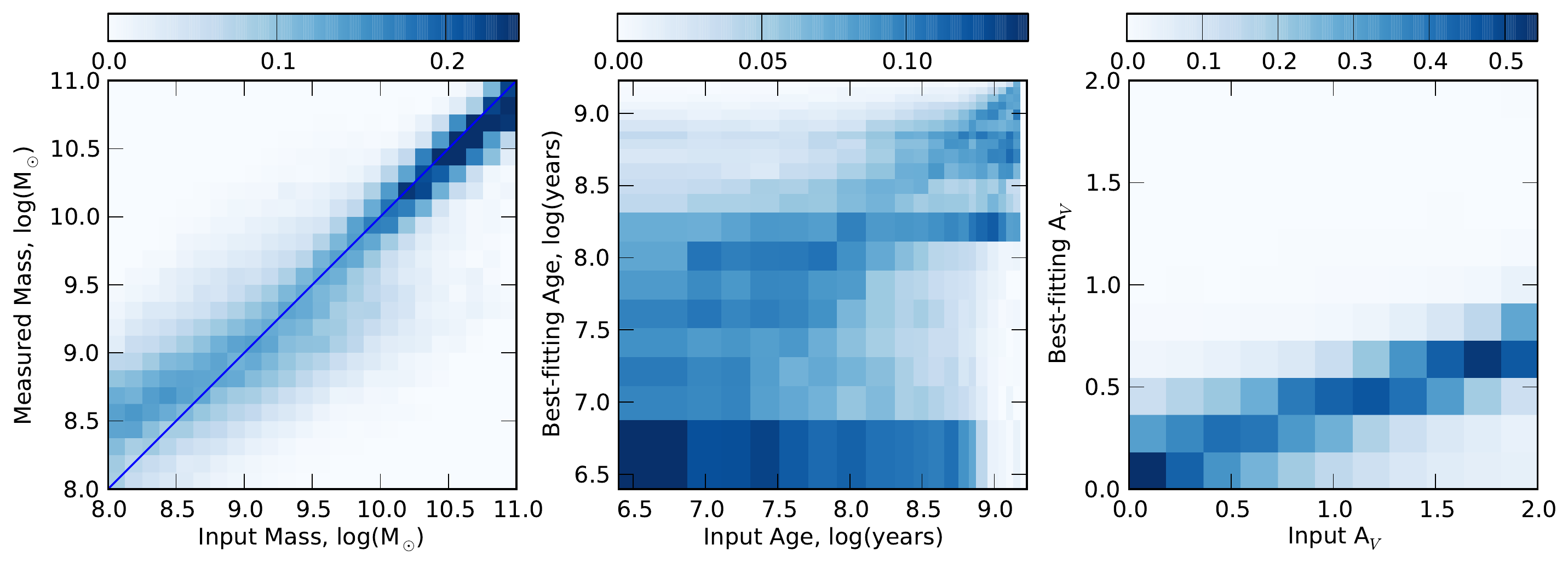}
\caption{2D histograms showing the recovered SED parameters for a set of input SEDs incorporating nebular emission when fitted with nebular emission. The values for the age (centre) and dust (right) are those corresponding to the single best-fitting model whilst the measured mass (left) is taken as $\int M P(M)dM$ for the mass likelihood distribution  marginalised over all other parameters. Each histogram is normalised by the number of input galaxies in each bin and the colour scale corresponds to the fraction of input galaxies at the observed mass (/age/dust extinction).}
\label{fig:masscomparison}
\end{figure*}

Figure~\ref{fig:masscomparison} illustrates how well the SED code is able to recover the stellar masses, ages and dust extinction. As expected, stellar mass is the most robust of the parameters with age and dust extinction showing a very large scatter and bias due to the degeneracy in fitting. Despite these degeneracies in the single best-fitting templates, when calculating the marginalised $\beta$ over all template likelihoods the resulting estimate of $\beta$ is un-biased and well constrained. We show the estimated accuracy of our $\beta$ measurements from these simulations in Appendix~\ref{app:beta}.

For input galaxies with masses $\approx 10^9 \rm{M}_{\odot}$, the median($\log_{10}(\text{M}_{\text{out}}) - \log_{10}(\text{M}_{\text{in}})$) = 0.02, with a standard deviation of 0.4 when input SEDs including nebular emission are fitted with comparable templates. For input masses $\approx 10^{8.5} \rm{M}_{\odot}$ and below, both the bias ($+0.22$ dex at $\approx 10^{8.5} \rm{M}_{\odot}$) and scatter increase.
When mock galaxies with pure stellar SEDs are fitted with pure stellar templates, both the scatter and bias are reduced at all mass ranges. The increased bias and scatter for galaxies with nebular emission is a result of confusion between an older stellar population with a $4000\rm{\AA}$ break and a young star-forming galaxy with strong nebular emission \citep{2009A&A...502..423S,2013MNRAS.429..302C}. 

\begin{figure}
\includegraphics[width=84mm]{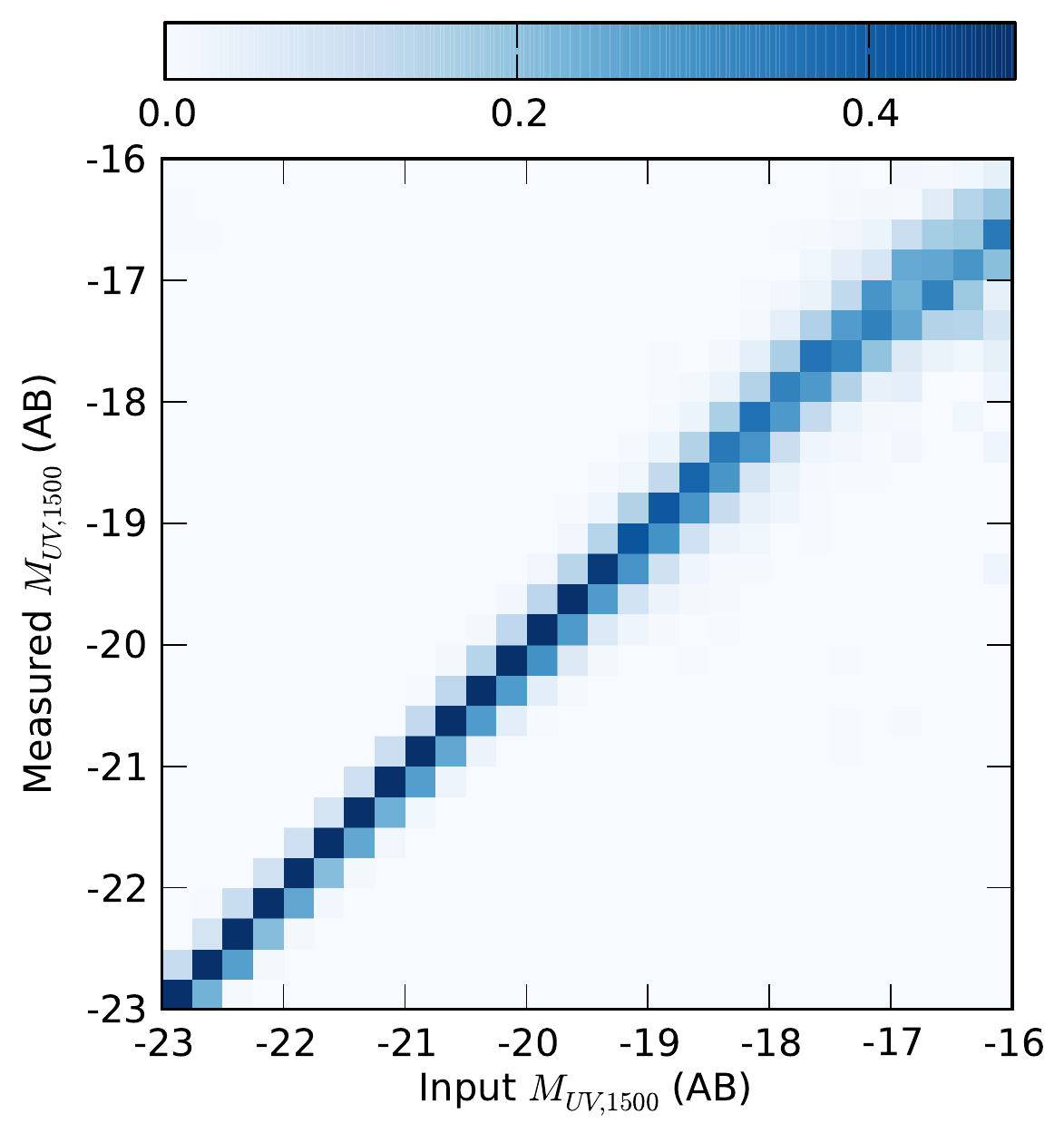}
\caption{Comparison of the recovered versus input $M_{UV}$ for the full mock galaxy sample. As in Figure~\ref{fig:masscomparison}, the histogram is normalised by the number of input galaxies in each bin and the colour scale corresponds to the fraction of input galaxies at the observed $M_{UV}$.}
\label{fig:muvcomparison}
\end{figure}

Finally, we find that the recovered value for $M_{UV}$ is extremely robust across the full dynamic range of our data, with a scatter of $< 0.2$ dex and negligible bias across all redshift out to the limits of our completeness as shown in Figure~\ref{fig:muvcomparison}. From these simulations, we determine that M$_{UV}$ is robust to M$_{UV} \approx$ -17 at $z \approx 4$, reducing to -18 at $z \approx 7$.

\section{Results}\label{sec:results}
\subsection{The $1/V_{\rm{max}}$ estimator}\label{subsec:vmax}
To compute our luminosity and mass functions we use an extension of the $1/V_{\rm{max}}$ method of \citet{Schmidt:1968wj}, treating each of our high-redshift samples as a ``coherent' sample comprised of the individual GOODS South regions with corresponding depths as outlined by \citet{1980ApJ...235..694A,1993ApJ...404...51E,2005A&A...439..863I}. The maximum comoving volume in which a galaxy can be observed and remain in the sample is given by
\begin{equation}
V_{\rm{max},i} =  \sum_{k}^{N_{regions}} \int_{z_{1,k}}^{z_{2,k}} \frac{dV}{dz} dz~d\Omega_{k}
\end{equation}
where the sum, $k$, is over each of the sub-regions in the field with their corresponding solid angle, $d\Omega_{k}$, integration limits $z_{1,k}$, $z_{2,k}$ and $dV/dz$ is the comoving volume element in Mpc$^{3}$ at redshift $z$.
The integration limits are given by $z_{1,k} = z_{min}$ and $z_{2,k} = min \left \{z_{max},z(z_{j},m_{j},m_{max,k}) \right \}$ where $z_{min}$ and $z_{max}$ are the lower and upper boundaries respectively for the given redshift bin, e.g. $z_{min} = 4.5$ and $z_{max} = 5.5$ for the $z\approx5$ sample. The function, $z(z_{j},m_{j},m_{max,k})$, returns the maximum redshift at which an object of apparent magnitude $m_{j}$, observed redshift $z_{j}$, could still be observed given the magnitude limit $m_{max,k}$ of the region.

The mass (or luminosity) function $\phi_{k}$ for discrete bins of mass (/magnitude) $k$ is then:

\begin{equation}\label{eq:vmaxmethod}
\phi_{k}dM = \sum_{i}^{N_{gal}} \frac{w_{i}}{V_{\textup{max},i}}W(M_{k}-M_{i}),
\end{equation}
where the weighting term, $w_{i}$, incorporates corrections for incompleteness and the selection function of the redshift bin as calculated in Section~\ref{sec:simulations}. The window function $W$ is defined as

\begin{equation}
W(x) = 
\begin{cases}
 1 \text{ if } -dM/2 \leq x < dM/2 \\ 
 0 \text{ otherwise }
\end{cases} 
\end{equation}
and $N_{gal}$ is the number of galaxies in the sample.

To incorporate the large error in the stellar masses, where the mass likelihood function for an individual galaxy can span a range much larger than the desired bin widths, we make amendments to Equation~\ref{eq:vmaxmethod}, such that the mass function evaluated for the mass bin $M_{1} < M_{k} < M_{2}$ is given by

\begin{equation}\label{eq:vmaxpdf}
\phi(M) dM = \sum_{i}^{N_{gal}}  \frac{w_{i}}{V_{max,i}} \int_{M_{1}}^{M_{2}} P_{i}(M)dM
\end{equation}
where $P_{i}(M)$ is the probability of galaxy $i$ having stellar mass, M, as calculated from the SED fitting at the fixed redshift for that specific Monte Carlo sample.

\subsection{UV Luminosity functions}
\begin{figure*}
\includegraphics[width=180mm]{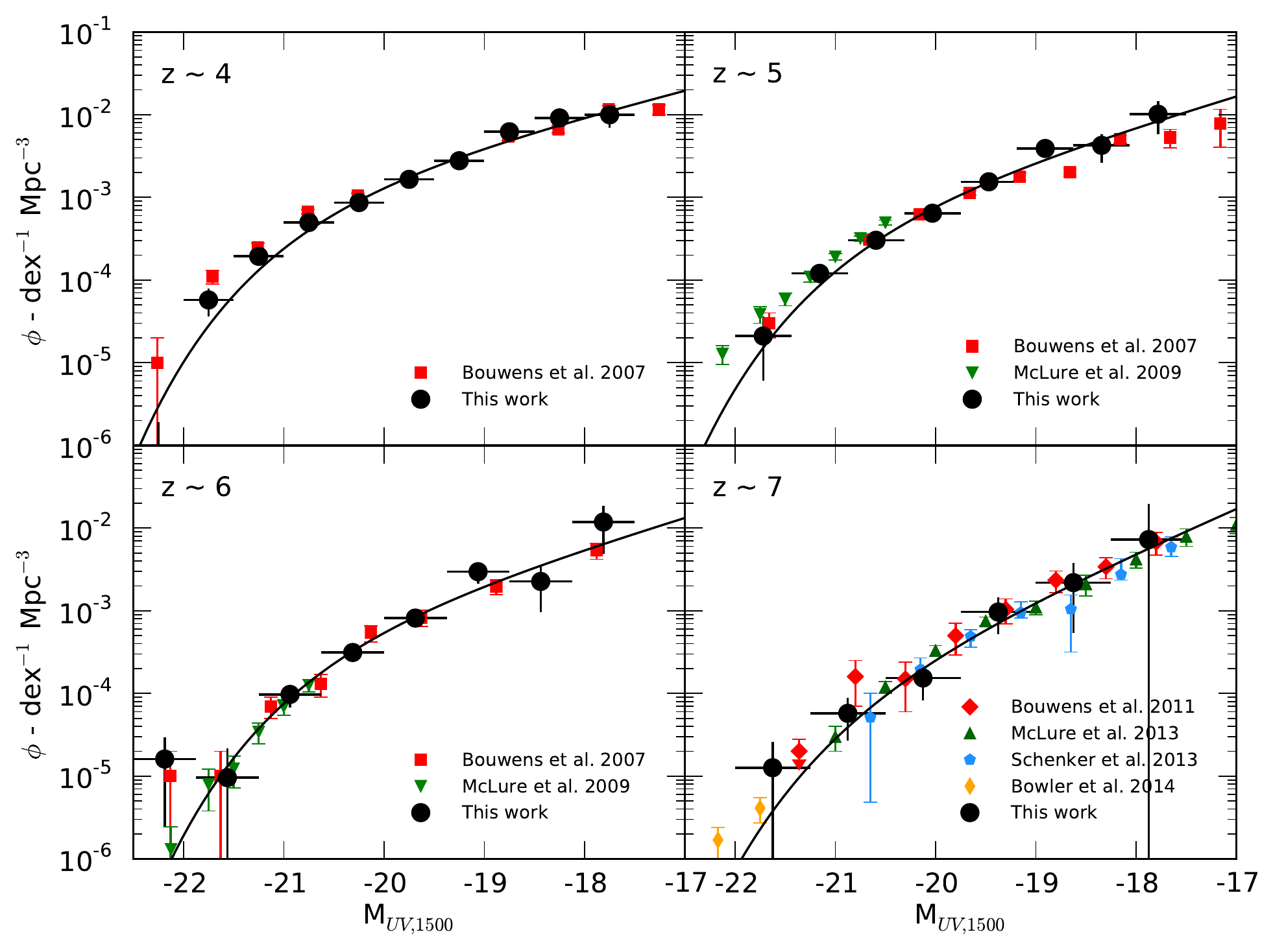}
\caption{A comparison of our $1/V_{\rm{max}}$ luminosity function estimates with those in the literature. We show the results of \citet{2007ApJ...670..928B} at $z \approx 4, 5~\&~6$ (red squares) derived from deep HST observations as well as the ground-based estimates of the bright end of the $z \approx 5$ and 6 luminosity functions by \citet{2009MNRAS.395.2196M}(green downward triangles). For the $z \approx 7$ LF, we show the estimate of \citet{2011ApJ...737...90B} as well as the recent results of \citet{McLure:2013hh} and \citet{Schenker:2013cl} which make use of the deeper UDF12 observations \citep{Koekemoer:2013db} to probe fainter $M_{UV}$ than we are able to.}
\label{fig:luminosityfunctions}
\end{figure*}

As a more robust observable (relative to the stellar mass, Section~\ref{sec:simulations}) with many previous observations, the rest-frame UV luminosity function provides a useful comparison for the method and completeness corrections used in this paper. To ensure that the shapes of our observed mass functions are not affected by biases in our $1/V_{\rm{max}}$ or completeness correction methods we reproduce the luminosity function (LF) for each of our redshift bins, which we can compare to previous work. Figure~\ref{fig:luminosityfunctions} shows the discretized luminosity functions calculated using the method outlined in Equation~\ref{eq:vmaxmethod}. In comparison with previous measurements of the luminosity function at high-redshift we find overall a very good agreement in the general shape of the luminosity functions for our $1/V_{\rm{max}}$ data points.

The data points for this work were also fit through $\chi^2$-minimisation with the \citet{Schechter:1976gl} parameterization

\begin{equation}
  \phi(M) = \phi^{*} \tfrac{\log(10)}{2.5} 10^{-0.4(M-M^{*})(\alpha+1)} 
              \exp{(-10^{-0.4(M-M^{*})})},
\end{equation}

\noindent where $M$ is the rest-frame UV magnitude and $\phi^{*}, \alpha$ and $M^{*}$ are the normalisation, faint-end slope and characteristic UV magnitude as standard. The resulting best-fitting parameters shown in Table~\ref{tab:lum_schecter} along with the corresponding parameters from the selected literature observations shown in Figure~\ref{fig:luminosityfunctions}. At all redshifts, the faint end slope is steep ($\alpha < -1.6$) and shows a tentative steepening towards $z\sim7$. However, the poor constraints on the faint end slope towards the highest redshifts make it difficult to comment on any evolution of the slope that might occur over this redshift range. At all redshifts, the slope is consistent with a fixed slope of $\alpha = -1.7$ \citep{2011ApJ...737...90B}.

At $z\sim4$, our measured $M_{UV}^{*}$ of -20.47$\pm 0.21$ is significantly fainter than that observed by \citet{2007ApJ...670..928B}, who find $M_{UV}^{*} = -21.06 \pm 0.1$. We find a closer agreement with the fainter $M_{UV}^{*}$ observed by \citet{Huang:2013kb} of -20.60$^{+0.13}_{-0.17}$.
At the bright end of the $z \sim 4$ luminosity function, our fit is strongly affected by the very low number density at $M_{UV} \approx -22.25$. Given the relatively small area of our survey field, the numbers of galaxies contributing to the brightest bins is very small ($\approx 1-3$). Differences in the measured (or assumed) redshift when calculating the rest-frame magnitude can therefore have a large effect, e.g. the difference in the distance modulus between $z = 3.5$ and $z = 4$ is $\approx 0.35$. As such, the characteristic cut-off in our Schecter function parameterisation is not well constrained, and the discrepancy is not significant.

\begin{table}
\caption{\citet{Schechter:1976gl} function parameters for $\chi^2$ fits to the $1/V_{\rm{max}}$ luminosity functions. The quoted errors represent the 1-$\sigma$ limits, but do not account for systematic error due to cosmic variance.}
\begin{tabular}{@{}lccc}
\hline
Redshift & $M_{UV}^{*}$ & $\alpha$ & $\phi^{*}$ ($10^{-3}$ Mpc$^{-3}$) \\
\hline
z $\sim 4$ & & & \\
This work &  -20.47 $\pm 0.21$ & -1.77 $\pm 0.18$ & 1.90$^{+0.79}_{-0.65}$ \\
Bouwens et al. 2007 & -21.06 $\pm$ 0.1 & -1.76 $\pm$ 0.05 & 1.1 $\pm$ 0.2 \\
\hline
z $\sim 5$ & & & \\
This work & -20.47 $^{+0.26}_{}$ & -1.90$^{+0.21}_{-0.16}$ & 1.07$^{+0.59}_{-0.14}$\\
Bouwens et al. 2007 & -20.69 $\pm$ 0.13 & -1.66 $\pm$ 0.09 & 0.9 $^{+0.3}_{-0.2}$ \\
McLure et al. 2009 & -20.73 $\pm$ 0.11 & -1.66 $\pm$ 0.06 & 0.94 $\pm 0.19$\\
Bouwens et al. 2012 & -20.60 $\pm$ 0.23 & -1.79 $\pm$ 0.12 & 1.4 $^{+0.7}_{-0.5}$ \\
\hline
z $\sim 6$ & & & \\
This work & -20.31$^{+0.84}_{-1.59}$ & -1.91$^{+0.91}_{-0.59}$ & 0.95$^{+2.21}_{-0.91}$ \\
Bouwens et al. 2007 & -20.29 $\pm$ 0.19 & -1.77 $\pm$ 0.16 & 1.2 $^{+0.6}_{-0.4}$ \\
McLure et al. 2009 & -20.04 $\pm$ 0.12 & -1.71 $\pm$ 0.11 & 1.8 $\pm 0.5$\\
Bouwens et al. 2012 & -20.37 $\pm$ 0.3 & -1.73 $\pm$ 0.20 & 1.4 $^{+1.1}_{-0.6}$ \\
\hline
z $\sim 7$ & & & \\
This work & -20.47$^{+1.43}_{}$ & -2.31$^{+1.31}_{-0.19}$ & 0.29$^{+2.87}_{-0.13}$ \\
Bouwens et al. 2011 & -20.14 $\pm$ 0.26 & -2.01 $^{+0.14}_{-0.15}$ & 0.86$^{+0.70}_{-0.39}$ \\
McLure et al. 2013 & -19.90 $^{+0.23}_{-0.28}$ & -1.9 $^{+0.14}_{-0.15}$ & 1.10$^{+0.56}_{-0.45}$ \\
Schenker et al. 2013 & -20.14 $^{+0.36}_{-0.48}$ & -1.87 $^{+0.18}_{-0.17}$ & 0.64$^{+0.56}_{-0.27}$ \\
\hline
\end{tabular}
\label{tab:lum_schecter}
\end{table}

\subsection{Observed mass-to-light ratios}\label{subsec:masslightratios}
In the past, the relationship between a galaxy's stellar mass and its UV luminosity (or $\log_{10} (\text{M}_{*})$ and $M_{UV}$) has been used as both a diagnostic of galaxy formation histories \citep{2009ApJ...697.1493S} and as a tool for estimating the galaxy stellar mass function at high-redshift \citep{Gonzalez:2011dn}. In a scenario where galaxies form their stars continuously, a strong $\log_{10} (\text{M}_{*})$-$M_{UV}$ relation should form, whereas more stochastic bursty star formation modes could result in a relation with wider scatter and a weaker trend.

\begin{figure*}
\includegraphics[width=130mm]{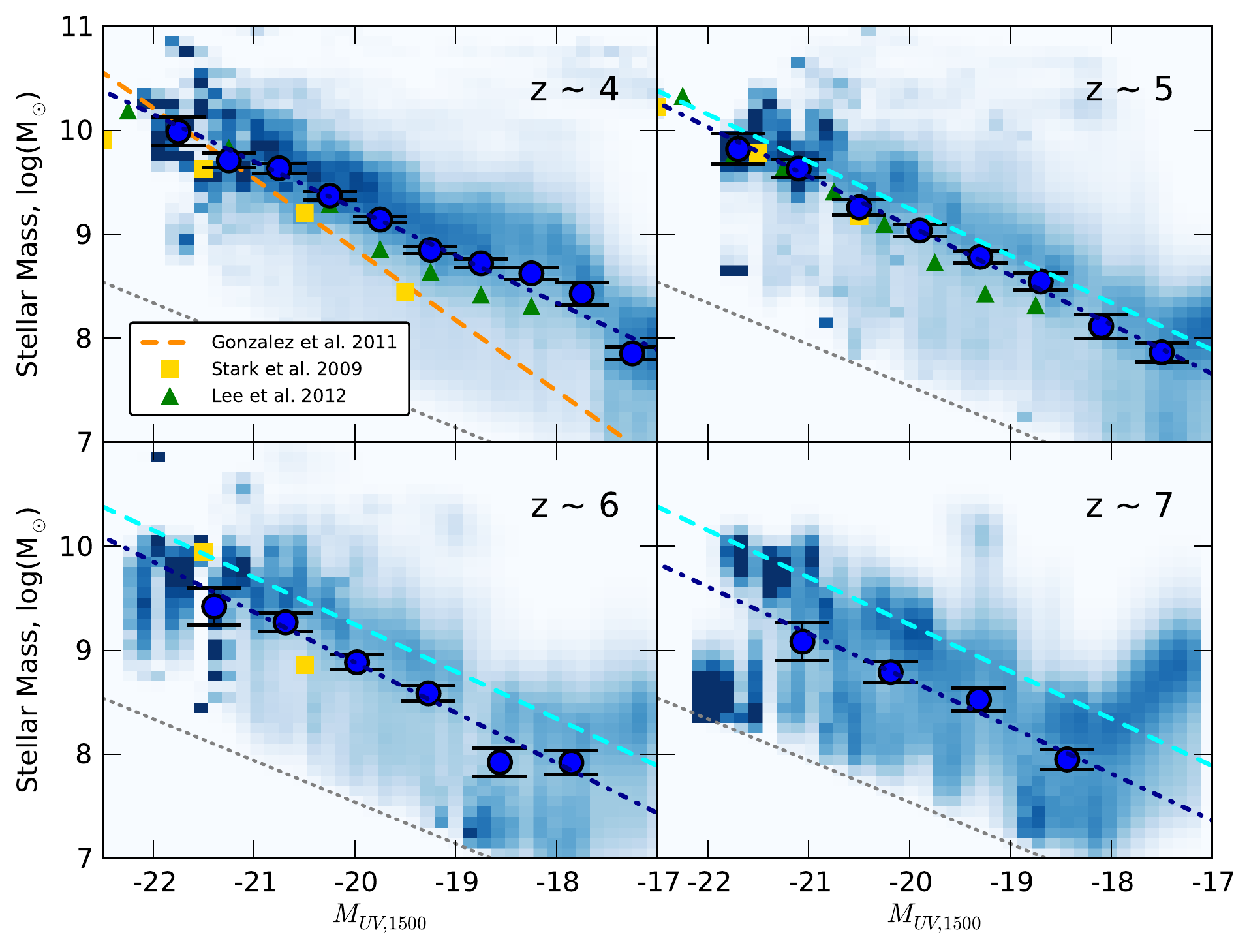}
\includegraphics[width=130mm]{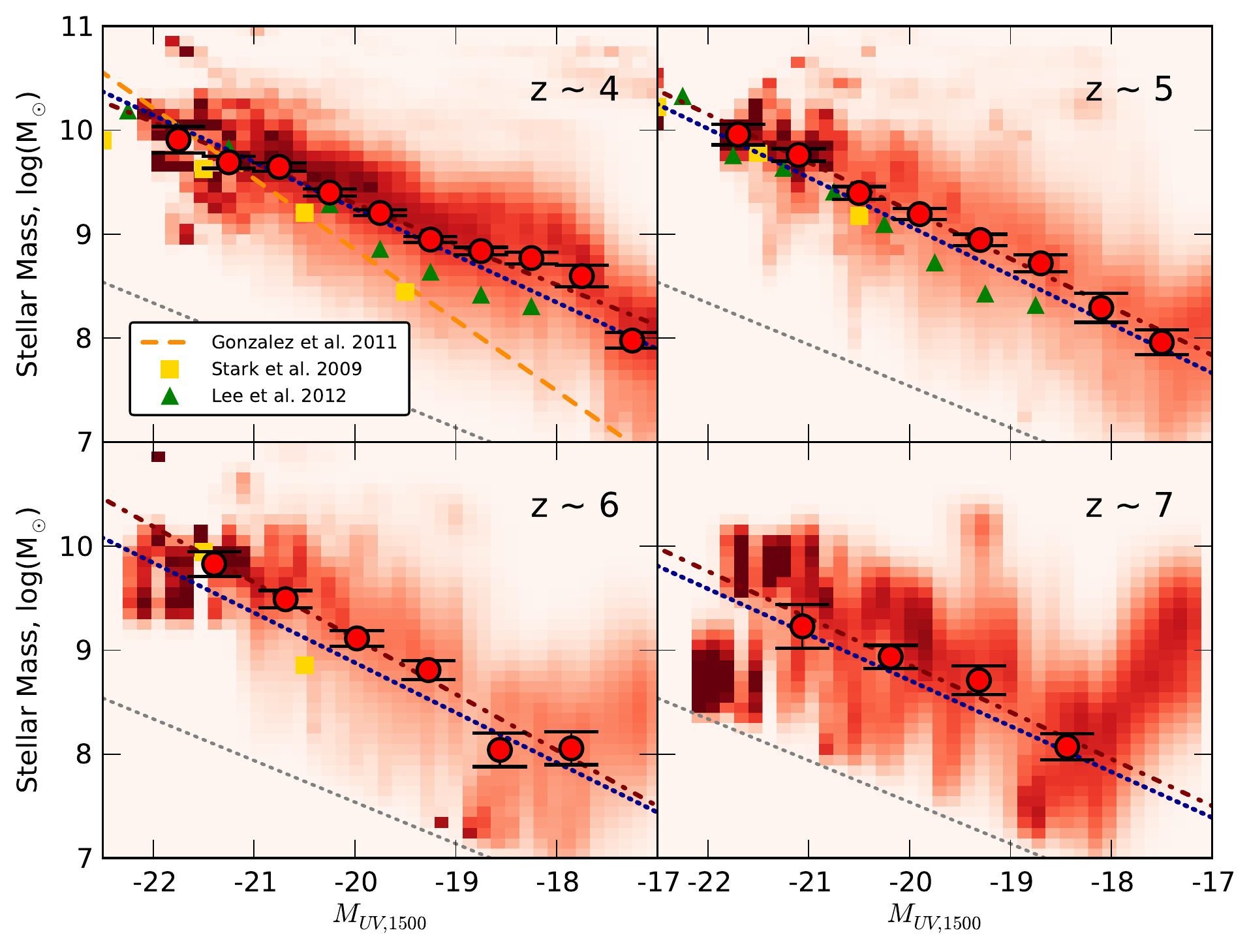}
\caption{Top: Probability distribution of the mass-to-light ratios observed when nebular emission is included in the fitting, stacked across all of the Monte Carlo samples. The values are normalised such that the probability at each value of $M_{UV}$ integrates to unity The blue dot-dashed line represents the average of the best-fitting line to robust means in each of the Monte Carlo samples, with the corresponding average means and their errors shown by the blue circles. The $z \sim 4$ relation is shown for reference at high-redshifts (cyan dotted line). Bottom: The corresponding probability distributions, bi-weight means and best-fitting relation (red dot-dashed line) when nebular emission is excluded from the SED fitting.  The blue dotted line shows the best-fitting relation from the top panel (including nebular emission) for each sample. In both panels, the orange dashed line shows the mass-to-light ratio observed by \citet{Gonzalez:2011dn}, measured for their $z \approx 4$ sample and applied across all bins. The green triangles and yellow squares show the average stellar mass in $M_{UV}$ bins as calculated by \citet{2012ApJ...752...66L} and \citet{2009ApJ...697.1493S} respectively, all stellar masses have been converted to the same Chabrier IMF. The grey dotted line represents the template in our SED fitting parameters with the lowest mass to light ratio.}
\label{fig:MLratios}
\end{figure*}

Using the stellar mass and $M_{UV}$ probability distributions produced by our SED fitting code, we plot the observed mass-to-light ratios for each of our redshift bins in Figure~\ref{fig:MLratios}. For a given  Monte Carlo sample (see Section~\ref{sec:sample}), the 2D $\log_{10} (\text{M}_{*})$-$M_{UV}$ probability distributions of each galaxy in the redshift bin are summed. The resulting PDFs of each sample are then summed to create a combined PDF in each redshift bin across all Monte Carlo samples. Finally, we normalise such that the probability at each value of $M_{UV}$ integrates to unity. By plotting the observed $\log_{10} (\text{M}_{*})$-$M_{UV}$ in this way, we take into account the full redshift and fitting errors. However, this representation is still subject to the effects of small number statistics for the brightest and faintest galaxies. 

As such, we also show the biweight mean $\log_{10}(\text{M}_{*})$ for bins of $M_{UV}$ within the range of reliably measurable $M_{UV}$. For $z \geq 5$, each bin contains a minimum of 5 galaxies by design, whilst for the $z \sim 4$ sample each bin contains a minimum of 10 galaxies. To these means, we fit linear functions with intercept $\log_{10}\text{M}_{*(M_{UV}=-19.5)}$ and slope $d\log_{10}\text{M}_{*}/dM_{UV}$. The best-fit values for each redshift sample are shown in Table~\ref{tab:ML} for stellar mass estimates both with and without the inclusion of nebular emission. 

As has been seen in many previous studies, we observe a clear `main-sequence' trend of increasing mass with increasing UV luminosity, and a large scatter about this trend. For the bright galaxies (M$_{UV} < -21$), our results agree well with those of \citet{2009ApJ...697.1493S}, \citet{Gonzalez:2011dn} and \citet{2012ApJ...752...66L}. Over the full range of UV luminosity, we find a shallower trend with $M_{UV}$ than \citet{Gonzalez:2011dn}. We also find that this trend evolves in normalisation between redshift $z \sim 4$ and 7. 

The change in the observed normalisation of the $\log_{10} (\text{M}_{*})$-$M_{UV}$ relation with redshift as a consequence of the inclusion of nebular emission has been examined before \citep{Shim:2011cw,deBarros:2012wa,Schenker:2013ep,Stark:2013ix}. However, we find that although this trend for decreasing normalisation with redshift is enhanced when nebular emission is included in the mass fitting, the trend still exists when fitting with pure stellar templates (see Table~\ref{tab:ML}). 

In Figure~\ref{fig:MLratios}, the effect of including nebular emission in the stellar mass estimate can be seen clearly in the bottom panel. At all redshifts, the average stellar mass for a given $M_{UV}$ is lower when nebular emission is included. \citet{Salmon:2014tm} also consider the effects of adding nebular emission lines to the SED models for galaxies in the same redshift range, and they find similar changes to the derived masses as we do here. In addition, the median stellar masses they observe for UV faint galaxies are higher than those of \citet{Gonzalez:2011dn} and \citet{2012ApJ...752...66L}, consistent with our observations.

Because in our SED fitting on the mass, we restrict the age of the templates to be less than the age of the universe at that redshift, the range of $\log_{10} (\text{M}_{*})$-$M_{UV}$ ratios available in the fitting does vary with redshift, i.e. a galaxy at $z = 7$ can never have as old as a stellar population as a galaxy at $z = 4$. If the fits to galaxies at $z \sim 6$ and 7 were being restricted by this upper limit, the limits set by the template set could create an artificial evolution in the scaling of the $\log_{10} (\text{M}_{*})$-$M_{UV}$ relation with redshift. However, examining the best-fitting SED parameters across all of the Monte Carlo samples, we find that at all redshifts and $M_{UV}$ values the highest best-fitting mass lies well below the maximum mass allowed by the template set. From this, we conclude that the observed scaling is therefore physical and not a result of systematics in our analysis. 

The slopes of our fitted $\log_{10} (\text{M}_{*})$-$M_{UV}$ relations are all close to that of a constant mass to light ratio (M$_{*} \propto -0.4$ $M_{UV}$) across the full range in luminosity. This implies there is no strong evolution of the mass to light ratio with luminosity. \citet{2012ApJ...752...66L} suggested the source of the change in their observed $\log_{10} (\text{M}_{*})$-$M_{UV}$ ratio could be due to a luminosity dependent extinction, a result which had also been implied by the evolution of $\beta$ with $M_{UV}$ seen by \citet{2012ApJ...754...83B}. Subsequent observations by \citet{Dunlop:2011jl} and \citet{2012ApJ...756..164F} have found no obvious luminosity dependence. However, recent studies by \citet{Bouwens:2013vf} and \citep{Rogers:2014bn} with greatly increased sample sizes and greater dynamic range confirm an unambiguous colour-magnitude relation. While measurements of $\beta$ for our sample do not exhibit strong evidence for such a strong luminosity dependent extinction at any redshift (see Appendix~\ref{app:beta}), our sample does not contain a statistically significant number of the brightest and faintest galaxies to rule out such evolution given the large scatter and error on $\beta$. 

Due to the increasing uncertainty in stellar mass measurements for galaxies below $10^{9} \text{M}_{\odot}$, the average mass-to-light ratios for the faintest galaxies could become increasingly biased towards fainter UV luminosities. As such, we cannot rule out a change in the $\log_{10} (\text{M}_{*})$-$M_{UV}$ slope at faint luminosities like that inferred by \citet{2012ApJ...752...66L}. To better constrain the average mass-to-light ratio for faint galaxies, detailed stacking across the full SEDs as a function of $M_{UV}$ would be required. Restricting our analysis to the brightest galaxies ($M_{UV} < -19.5$) at $z\sim4$ and 5 where the potential biases are minimised, we find no significant change in the fitted $\log_{10} (\text{M}_{*})$-$M_{UV}$ slopes.

\begin{table}
\caption{The best-fitting slope and intercepts of the $\log_{10} (\text{M}_{*})$-$M_{UV}$ mass-to-light relation, averaged across all Monte Carlo samples. At $z\sim 4$ and 5, we also show in parentheses the best-fitting values when the fits are restricted to only the brightest galaxies ($M_{UV} < -19.5$).}
\centering
\begin{tabular}{c c c}
\hline
$z$ & $\log_{10}\text{M}_{*~(M_{UV}=-19.5)}$ & $d\log_{10}\text{M}_{*}/dM_{UV}$ \\ 
\hline
 & \emph{With Nebular Em.} & \\
4 & $9.02 \pm 0.02$ & $-0.45\pm 0.02$ \\
  & ($9.06 \pm 0.05$) &  ($-0.42 \pm 0.06$) \\
5 & $8.84 \pm 0.04$ & $-0.47\pm 0.04$ \\
  & ($8.85 \pm 0.12$) &  ($-0.46 \pm 0.11$) \\
6 & $8.64 \pm 0.06$ & $-0.48\pm 0.07$ \\
7 & $8.49 \pm 0.09$ & $-0.44\pm 0.12$ \\
 & & \\
 & \emph{Without Nebular Em.} & \\
4 & $9.10 \pm 0.02$ & $-0.39\pm 0.02$ \\
  & ($9.12 \pm 0.05$) &  ($-0.37 \pm 0.05$) \\
5 & $9.00 \pm 0.04$ & $-0.46\pm 0.04$ \\
  & ($9.00 \pm 0.09$) &  ($-0.45 \pm 0.08$) \\
6 & $8.84 \pm 0.07$ & $-0.54\pm 0.07$ \\
7 & $8.63 \pm 0.11$ & $-0.45\pm 0.13$ \\ 

\hline
\end{tabular}
\label{tab:ML}
\end{table}

In hydrodynamical simulations of galaxies at $z > 5$, \citet{Wilkins:2013kh} found a relationship between the intrinsic $L_{1500}$ (excluding dust absorption) and $\text{M}/L_{1500}$ which is roughly constant. This relationship is also seen to evolve, with the normalisation decreasing with increasing redshift. When dust extinction was applied to the intrinsic model luminosities based on the $\beta$ observations of \citet{2012ApJ...754...83B}, the observed $\log_{10} (\text{M}_{*})$-$M_{UV}$ exhibited a much stronger correlation comparable to that observed by \citet{Gonzalez:2011dn}.

\subsection{Stellar mass functions at high-redshift}\label{sec:smf}
\begin{figure*}
\includegraphics[width=180mm]{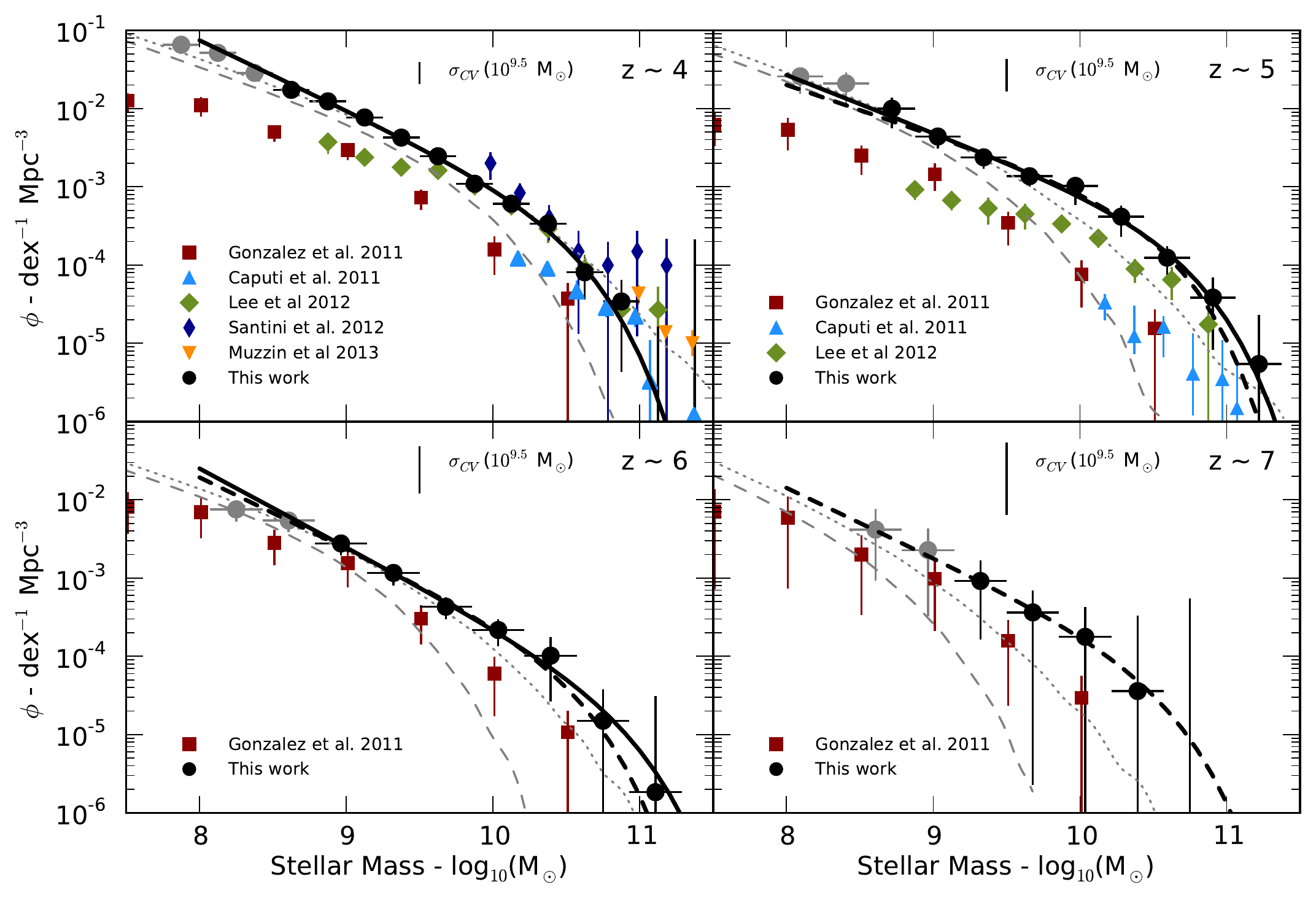}
\caption{The $1/V_{\rm{max}}$ stellar mass functions for the high-redshift samples. Error bars take into account random Poisson noise as well as the scatter between the Monte Carlo samples due to photometric redshift uncertainty. The black circles show the mass bins included in the $\chi^2$ fitting to the \citet{Schechter:1976gl} functions based on the stellar mass limits described in the text. The dashed and dotted lines show the stellar mass functions calculated by applying the best-fitting mass-to-light ratio (including nebular emission, see Table~\ref{tab:ML}) to the literature luminosity functions at each redshift with a scatter of 0.2 (dashed) and 0.5 (dotted) dex. For the $z\sim4$ bin, the Schecter fit of \citet{2007ApJ...670..928B} was used, whilst at $z\sim5~\&~6$ and $z\sim7$ the fits of \citet{Anonymous:96uKWdy6} and \citet{McLure:2013hh} respectively were used to generate the luminosity distribution. We also show using the error bars at the top of each panel the cosmic variance expected for galaxies of stellar mass $\approx 10^{9.5}$ M$_{\odot}$, as predicted by the method outlined in \citet{Moster:2011ip}.}
\label{fig:massfunctions}
\end{figure*}

Following the method outlined in Section~\ref{subsec:vmax}, specifically using the $1/V_{\rm{max}}$ method in Equation~\ref{eq:vmaxpdf}, we construct the stellar mass function for each of our high-redshift samples. The resulting stellar mass functions are shown in Figure~\ref{fig:massfunctions}. Our data points and errors take into account the stellar mass errors, Poisson errors and the errors due to the photometric redshift uncertainty.

The black lines in Figure~\ref{fig:massfunctions} show the best-fitting \citet{Schechter:1976gl} functions from $\chi^2$ minimisation to the the $1/V_{\rm{max}}$ data above our chosen mass completeness limits (black points). We perform two sets of fits to our data. Firstly, we allow all 3 parameters to vary (solid line, $z\sim 4, 5$ and 6) and secondly, we fix the characteristic mass such that $\text{M}^{*} = \text{M}_{z\sim4}^{*}$ (dashed line, $z\sim 5, 6$ and 7). The parameters for these fits are shown in Table~\ref{tab:mass_schecter}. 

Because there exists such a large scatter in the observed mass-to-light ratios for the high-redshift galaxies, accurately estimating the mass completeness limit is non-trivial. For a given mass near the completeness limit, there could exist a significant contribution from galaxies below the luminosity limit proportional to how large the actual intrinsic scatter is. However, it is not known how much of the observed scatter is due to photometric error (and therefore photometric redshift estimates and parameters from SED fitting) and how much is intrinsic.
Rather than trying to correct for galaxies lost through incompleteness down to the lowest observed masses, we instead restrict our analysis to masses unaffected by this scatter. We calculate this mass limit by taking the 95\% mass percentile of the observed galaxies within 0.5 dex of our $H_{160}$ magnitude limit, finding it to be $\approx 10^{8.5}$ M$_{\odot}$. Under the reasonable assumption that the intrinsic scatter in the mass-to-light ratio does not rapidly increase below our detection limit, the contribution to masses above $\approx 10^{8.5}$ from galaxies below the limit will be negligible.

Additionally, as shown in Section~\ref{sec:simulations}, the accuracy of stellar mass estimates begins to deteriorate at lower masses with an increasing bias which could lead to biased slopes. Taking these factors into account, the limits chosen when fitting the stellar mass functions were $\log_{10} (\text{M}_{\odot}) =$ 8.55, 8.85, 8.85 and 9.15 for $z \sim$ 4, 5, 6 and 7 respectively.

Inspecting the observed mass functions in Figure~\ref{fig:massfunctions}, it can be seen that the exact limit should have little effect on the measured slope within reasonable bounds at $z \sim 4-6$. Choosing limits $\pm 0.5$ dex would not affect our conclusion that the mass function is steep.

\begin{table}
\centering
\label{tab:mass_schecter}
\caption{\citet{Schechter:1976gl} function parameters for $\chi^2$ fits to the $1/V_{\rm{max}}$ mass functions. For the $z \sim 5$, 6 and 7 samples, we do two fits, one in which $\log_{10}(\text{M}^{*})$ is allowed to vary, and one in which it is fixed to the best-fitting value for the $z\sim 4$ sample. The quoted errors represent the 1-$\sigma$ errors from fitting marginalised over the remaining parameters but do not account for any systematic errors due to cosmic variance.}
\begin{tabular}{@{}cccc}
\hline
$z$ & $\log_{10}(\text{M}^{*})$ & $\alpha$ & $\phi^{*}$ ($10^{-4}$ Mpc$^{-3}$) \\
\hline
4 & 10.51$^{+0.36}_{-0.32}$ & -1.89$^{+0.15}_{-0.13}$ & 1.89$^{+3.46}_{-1.32}$ \\
& & & \\
\hline
5 & 10.68$^{+0.98}_{-0.46}$ & -1.74$^{+0.41}_{-0.29}$ & 1.24$^{+4.77}_{-1.19}$\\
 & 10.51 & -1.64$^{+0.15}_{-0.17}$ & 2.21$^{+0.80}_{-0.76}$\\
& & & \\
\hline
6 & 10.87$^{+1.13}_{-1.06}$ & -2.00$^{+0.57}_{-0.40}$ & 0.14$^{+4.11}_{-0.14}$ \\
 & 10.51 & -1.90$^{+0.27}_{-0.31}$ & 0.46$^{+0.36}_{-0.26}$ \\
& & & \\
\hline
7 & 10.51 & -1.89$^{+1.39}_{-0.61}$ & 0.36$^{+3.01}_{-0.35}$ \\
& & & \\
\hline
\end{tabular}
\end{table}

Due to the small sample size at $z \sim 7$ and the large errors in estimating the stellar mass (from both the photometric redshift and fitting errors), the mass function is very poorly constrained. The range of acceptable values for $\alpha$ cover an extremely wide range but are consistent with the slope of $\alpha \approx -1.9$ found for the lower redshift bins and the slope of the corresponding luminosity function. Over the redshift range examined by this work, the errors in $\alpha$ are too large to infer any evolution in slope with redshift.

As we are observing only a single field, we are unable to estimate the cosmic variance in the number densities by comparing the field to field variation. We use the updated QuickCV code of \citet{Moster:2011ip} (see also \citeauthor{Newman:2002fa}~\citeyear{Newman:2002fa}) to estimate the cosmic variance as a function of mass in each of our redshift bins for a survey field with the dimensions of CANDELS GOODS South. In Figure~\ref{fig:massfunctions} we show the estimated error on the counts for galaxies of mass $\approx 10^{9.5}$ M$_{\odot}$. For stellar mass $\approx 10^{10}$ M$_{\odot}$ and above, the cosmic variance predicted by this method exceeds $>100\%$ at $z \sim 6$ and 7. However, due to the lack of constraints on the galaxy bias at high-redshift, there is a large uncertainty on these estimates. When compared to the field-to-field variation observed by \citet{2012ApJ...752...66L}, our estimates represent a conservative assessment of the likely cosmic variance. With the full CANDELS imaging now complete, future analysis incorporating all five of the separate survey fields should allow much more robust measures on the true cosmic variance at high-redshift.

In addition to the $1/V_{\rm{max}}$ estimates, we also estimate the stellar mass function using a method analogous to that of \citet{Gonzalez:2011dn}. For each sample, $10^6$ UV magnitudes in the range $-23 < M_{UV,1500} < -13$ are drawn from the observed luminosity functions from the literature at each redshift \citep{2007ApJ...670..928B,Anonymous:96uKWdy6,McLure:2013hh}. The $M_{UV,1500}$ are then converted to stellar masses using the best-fitting relations from Table~\ref{tab:ML} for each redshift sample and a scatter of 0.2 (dashed lines) or 0.5 dex (dotted lines).

At $z \sim 4$, the $1/V_{\rm{max}}$ data shows a good agreement with the mass functions generated by this method. At higher redshifts however, luminosity based mass functions increasingly underpredict the number density at high masses for the same fixed scatter in the $\log_{10} (\text{M}_{*})$-$M_{UV}$ relation. To increase the number densities at high mass to match those observed, either a more strongly evolving $\log_{10} (\text{M}_{*})$-$M_{UV}$ relation (in direct disagreement with that observed) or a significantly greater scatter in the intrinsic $\log_{10} (\text{M}_{*})$-$M_{UV}$ ratios is required. 

\subsubsection{Comparison with the literature}
\citet{2011MNRAS.413..162C} studied the massive end of the stellar mass function at $3 \leq z \leq 5$ over a wide area in the UKIDSS Ultra Deep Survey (UDS), using photometric redshifts for a sample of $4.5\mu m$ selected galaxies. Our observed number densities show a broad agreement at $z \sim 4$ ($3.5 \leq z < 4.25$ for \citeauthor{2011MNRAS.413..162C}) for $\log_{10}\text{M}_{*} > 10.5$ but are significantly higher at lower masses. The same is also true across all masses at $z > 4$ ($4.25 \leq z < 5.0$). However for both redshift samples, \citet{2011MNRAS.413..162C} find a very steep low-mass slope when parameterised with a \citet{Schechter:1976gl} fit, in agreement with our results. At the massive end of the $z\sim4$ galaxy SMF, our results agree with those of \citet{Muzzin:2013bl} over the limited mass range covered by both works.

Covering a significantly smaller area than those observations but probing to lower masses are the $1/V_{\rm{max}}$ observations of \citet{Santini:2012jq}. We find a good agreement with these results, however due to the small number statistics at the high mass end of the SMF, the errors on both sets of observations are large.

Another measurement of the stellar mass function at $z \sim 4 - 5$ is that of \citet{2012ApJ...752...66L}, who study the SMF for a Lyman break selected sample in the GOODS North and South fields \citep{2004ApJ...600L..93G}. At $z \sim 4$, for stellar masses $\log_{10}\text{M}_{*} > 9.5$ the two results are in excellent agreement. Below this mass range, the significantly steeper low-mass slope measured in this work results in a higher number densities than those found by \citet{2012ApJ...752...66L}. At $z \sim 5$, we again find higher number densities although there is some agreement at the highest masses. 

\citet{Gonzalez:2011dn} provides the only previous observation that covers the full redshift range and is also the only work which does not construct the galaxy stellar mass function from the galaxy masses directly. Instead, \citet{Gonzalez:2011dn} measure the $\log_{10} (\text{M}_{*})$-$M_{UV}$ relation at $z \sim 4$ (testing for consistency with the smaller samples at $z \sim 5$ and 6). The mass-to-light ratio is then applied to the observed UV luminosity function for each redshift bin, allowing the estimation of the SMF to lower masses and higher redshift than would otherwise have been possible. Because their $\log_{10} (\text{M}_{*})$-$M_{UV}$ ratio is fixed at all redshifts, there is less evolution in the SMF than we observe. The slope of the $\log_{10} (\text{M}_{*})$-$M_{UV}$ relation observed by \citet{Gonzalez:2011dn} also results in a low-mass slope which is shallow across all redshift bins.

\begin{figure}
\includegraphics[width=80mm]{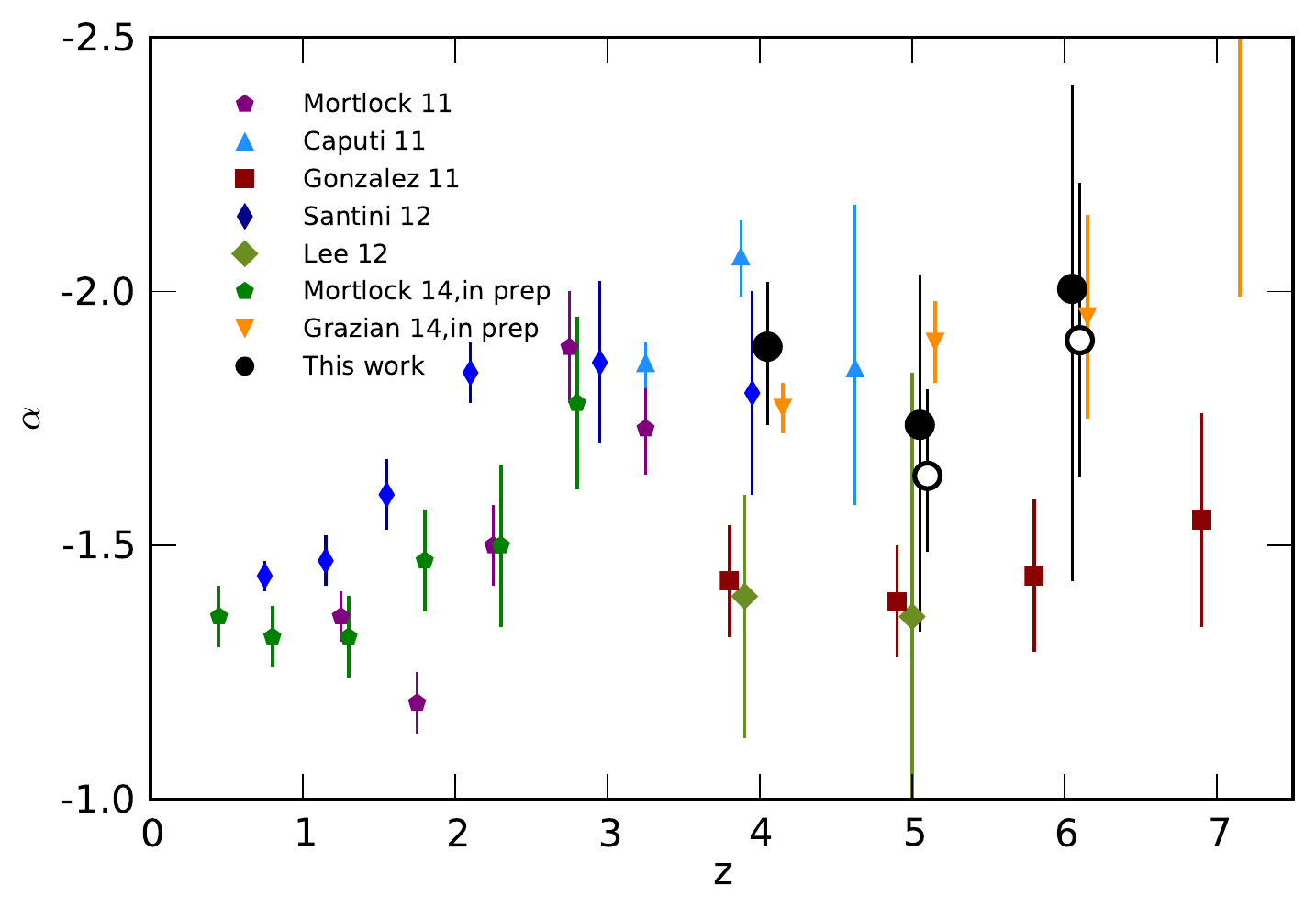}
\caption{Evolution of the low-mass slope from $z = 0$ to $z \sim 7$. We show the best-fitting $\alpha$ for both the freely varying (filled circles) and fixed $\text{M}_{*}$ (empty circles) fits. The fits for $z\sim7$ were excluded due to the poor constraints. We show results from the recent literature for the stellar mass function at lower redshifts and at $z > 3$. Shown are the $\alpha$ quoted for single \citet{Schechter:1976gl} fits to the observed data where $\alpha$ has been left as a free parameter in the fitting or has been estimated analytically \citep{Gonzalez:2011dn}.}
\label{fig:alpha_evolution}
\end{figure}

When placed in the context of low-redshift observations of the stellar mass function, our results at $z \sim 4$ are a continuation of the trend of increasing $\alpha$ with redshift. In Figure~\ref{fig:alpha_evolution}, we show recent observations of the low-mass slope of the SMF from $z = 0$ out to $z \sim 7$. We include only results where $\alpha$ has been fitted as a free parameter and the values of $\alpha$ quoted are from the single Schecter function parameterisations of the SMF.

At $z \sim 3$, there is a broad agreement in the estimations of the low-mass slope at $\alpha \approx -1.8$. By $z \sim 4$, there is a much larger disagreement between observed value spread across $\alpha \sim -2$ to $-1.4$. It is important to note that the observations with a shallower low-mass slope, \citet{Gonzalez:2011dn,2012ApJ...752...66L}, are those with galaxies selected using the Lyman break technique colour cuts and source detection using optical bands (typically $z_{850}$). In contrast, those with steep slopes, \citet{2011MNRAS.413..162C}, \citet{Santini:2012jq} and this work, use photometric redshift selection as well as near- or mid-infrared band for source detection. The best-fitting $\alpha$ of this work are also in good agreement with the maximum-likelihood estimates of Grazian et al. \emph{in prep}, an independent analysis of the combined CANDELS GOODS and UDS (UKIDSS Ultra Deep Survey) fields.

\subsubsection{Comparison with theory}\label{sec:theory}
\begin{figure*}
\centering
\includegraphics[width=175mm]{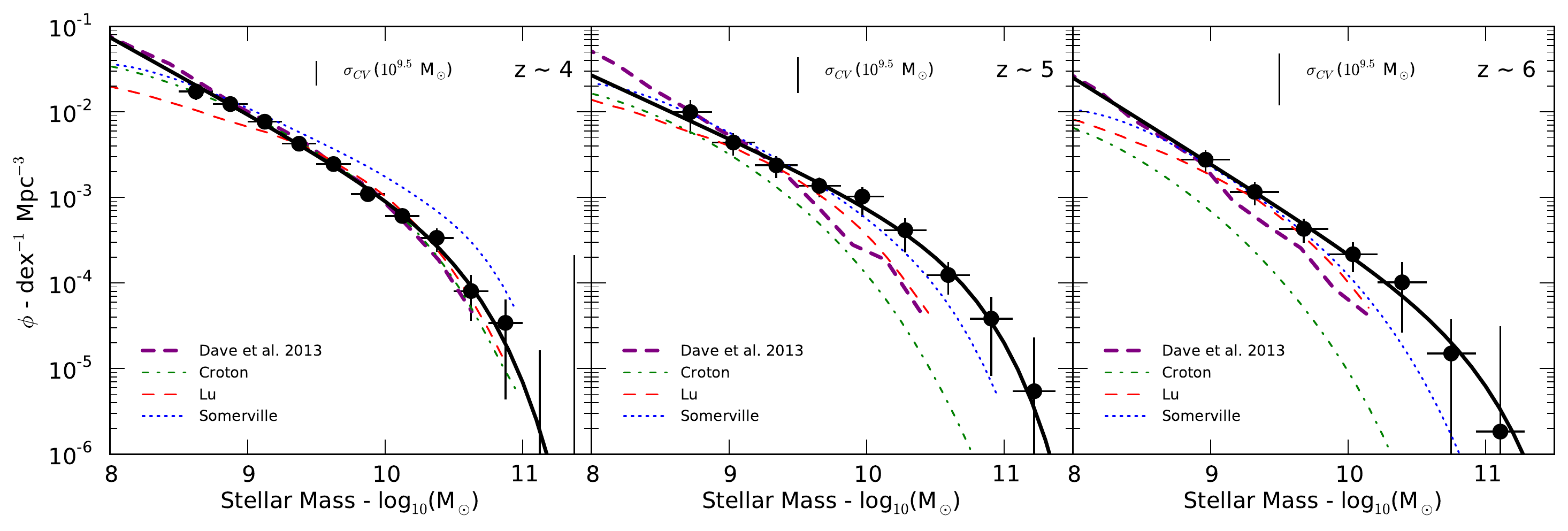}
\caption{Comparison of the observed galaxy stellar mass functions in this work with theoretical model predictions at $z \sim 4$, 5 and 6. We show the semi-analytic models of \citet{Croton:2006ew}, \citet{Somerville:2008ed} and \citet{Lu:2011hj}, using the error convolved stellar mass functions as outlined in \citet{Lu:2013ui}. The dashed purple line shows the results from the hydrodynamical simulations of \citet{Dave:2013bf}.}
\label{fig:smf_theory}
\end{figure*}

In Figure~\ref{fig:smf_theory}, we compare our measurements of the observed stellar mass function with the predictions of both smoothed particle hydrodynamic (SPH) and semi-analytic models. The SPH predictions are taken from the hybrid energy/momentum-driven wind (ezw) model of \citet{Dave:2013bf}. We also show the predictions of three semi-analytic models (SAM), from \citet{Croton:2006ew}, \citet{Lu:2011hj} and \citet{Somerville:2008ed} (see also \citeauthor{Somerville:2012cq}~\citeyear{Somerville:2012cq}). Details of the three models and an in-depth comparison between the model predictions across all redshifts can be found in \citet{Lu:2013ui}. The number densities have been renormalised to the comoving volume of the cosmology used throughout our paper ($H_{0} = 70$ kms$^{-1}$Mpc$^{-1}$, $\Omega_{0}=0.3$ and $\Omega_{\Lambda}=0.7$). Analysis is restricted to $z \leq 6$ due to limits on the robustness of simulations at higher redshifts from the numerical resolution of the simulations.

Inspecting the mass functions at $z \sim 4$, there is excellent agreement between the observations and the models of \citet{Dave:2013bf}, \citet{Croton:2006ew} and to a lesser extent \citet{Lu:2011hj}. Of the three SAM predictions, \citet{Somerville:2008ed} shows the least agreement at $z\sim 4$ due to the over-abundance of higher mass galaxies. However, at higher redshifts the reverse is true, with the \citet{Somerville:2008ed} models providing the best match to the observed $z\sim5$ and 6 mass functions.

Of the four model predictions presented here, the SPH simulations of \citet{Dave:2013bf} exhibit the steepest low-mass slopes and the closest agreement with our observations. The steepening of the low-mass slope in this model (from $z \sim 0$ to $z > 3$) is a result of decreasing contribution from wind recycling at high-redshifts. The resulting feedback at high-redshift has a smaller mass dependence than other models. This can be seen when compared to the SAM model of \citet{Lu:2011hj} which has feedback with a much stronger mass dependence owing to increasingly strong (or efficient) feedback in low-mass haloes. 

The SPH predictions, along with those of the \citet{Lu:2011hj} SAM, most closely match the evolution in the overall normalisation of the number densities across the observed redshift range. The other semi-analytic models undergo a much stronger evolution in the number density of the most massive galaxies. It is important to take into account the fact that all 3 of the semi-analytic models are tuned to match only the $z = 0$ stellar mass function. The range of acceptable parameters at $z = 0$ found by \citet{Lu:2011hj} results in a broad distribution of predicted stellar mass functions at high-redshift. Nevertheless, it is clear that our new observations of the high-redshift SMF can be used to further constrain our best models of galaxy evolution.

\subsection{Stellar Mass Density}
\begin{figure}
\centering
\includegraphics[width=80mm]{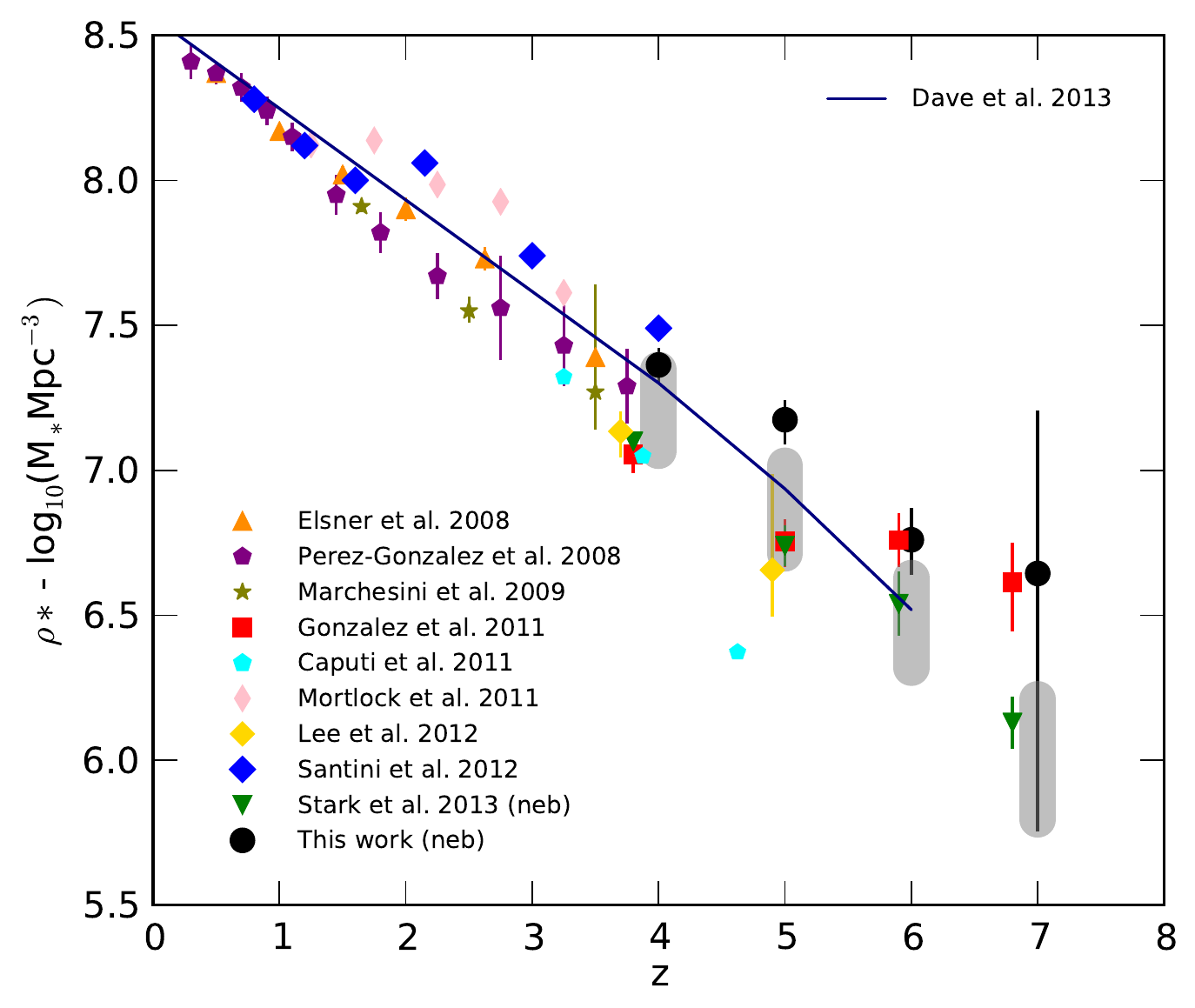}
\caption{Observed stellar mass densities (for M $>10^8$ M$_{\odot}$). All literature values have been converted to a Chabrier/Kroupa IMF as appropriate. The grey regions at $z \sim 4$, 5, 6 and 7 show the range in stellar mass density traced by the luminosity function-based mass functions described in Section~\ref{sec:smf}, the lower and upper limits correspond to 0.2 and 0.5 dex of scatter in the applied mass-to-light ratios respectively.}
\label{fig:smd_evolution}
\end{figure}

We compute the total stellar mass density (SMD) by integrating the fitted \citet{Schechter:1976gl} function from $\text{M}_{*} = 10^8~\text{to}~10^{13} \text{M}_{\odot}$, with 1-$\sigma$ errors estimated from the minimum and maximum SMD within the 1-$\sigma$ contours for the fit parameters (see Table~\ref{tab:smd}). For the $z\sim 5$, 6 and 7 samples, we use the best-fitting parameters with $\text{M}^{*} = \text{M}_{z\sim4}^{*}$. The results are shown in Figure~\ref{fig:smd_evolution} as the solid black points. We also show results from the literature across all redshift ranges, converted to the same cosmology and IMF (Chabrier/Kroupa).

\begin{table}
\centering
\label{tab:smd}
\caption{Stellar mass densities integrated from the Schecter parameters in Table~\ref{tab:mass_schecter} for $\text{M} > 10^8 \text{M}_{\odot}$,. Error bars correspond to the minimum and maximum stellar mass densities within the $1~\sigma$ contours of the mass function fits.}
\begin{tabular}{c c}
\hline
z & $\rho_{*} (\log_{10}~\text{M}_{\odot} \text{Mpc}^{-3}$)\\
\hline
4 & $7.36\pm0.06$ \\
5 & $7.17^{+0.07}_{-0.08}$ \\
6 & $6.76^{+0.11}_{-0.12}$ \\
7 & $6.64^{+0.56}_{-0.89}$ \\
\hline
\end{tabular}
\end{table}

Our observations show the continuation of the rapid decline in global stellar mass density towards high-redshifts, falling by a factor of between $\sim 4$ and 40 in the $\sim 1$ Gyr between $z \sim 4$ and 7. This rate of stellar mass growth observed is higher than observed by \citet{Gonzalez:2011dn} over the same time period but comparable to that found by \citet{Stark:2013ix} when the large uncertainty in the $z\sim7$ SMD is taken into account.

At $z \sim 4$, our results lie within the range of past stellar mass density measurements at this redshift. Although larger than the results of the Lyman break selected samples \citep{Gonzalez:2011dn,2012ApJ...752...66L,Stark:2013ix}, we find a SMD less than that of \citet{Santini:2012jq} and comparable to that of some of the other photometric redshift selected samples \citep{PerezGonzalez:2008cq,Marchesini:2009ef}. As could be inferred from the stellar mass functions, the stellar mass densities of \citet{Dave:2013bf} underpredict observed SMD at $z \sim 5$ and $z \sim 6$ but shows a good agreement at $z \sim 4$. Similarly, the range of densities covered by our luminosity based mass functions (grey regions) are significantly lower than the directly observed SMD in all redshift bins apart from $z\sim4$. 

\subsection{Star Formation Rates}
\subsubsection{Specific star formation rates}
Earlier observations of the sSFR evolution at $z > 3$, with mass estimates excluding the effects of nebular emission, showed the sSFR at a fixed mass remained roughly constant at $\sim 2$ Gyr$^{-1}$ with increasing redshift \citep{2009ApJ...697.1493S,Gonzalez:2010hm,2012ApJ...754...83B}. Such a plateau in the sSFR evolution was at odds with most plausible models of galaxy evolution (as explored by \citeauthor{Weinmann:2011hh}~\citeyear{Weinmann:2011hh}).

However, it has since been shown that the inclusion of nebular emission in stellar mass estimates at high-redshift has a significant effect on the redshift evolution of the specific star formation rate (sSFR) \citep{2009A&A...502..423S,2010A&A...515A..73S,Stark:2013ix,Gonzalez:2014do}. By lowering the measured mass for a fixed star formation rate, the inclusion of nebular emission results in a higher sSFR proportional to the strength (or effect on the estimated stellar mass) of the emission lines.

\begin{figure}
\centering
\includegraphics[width=80mm]{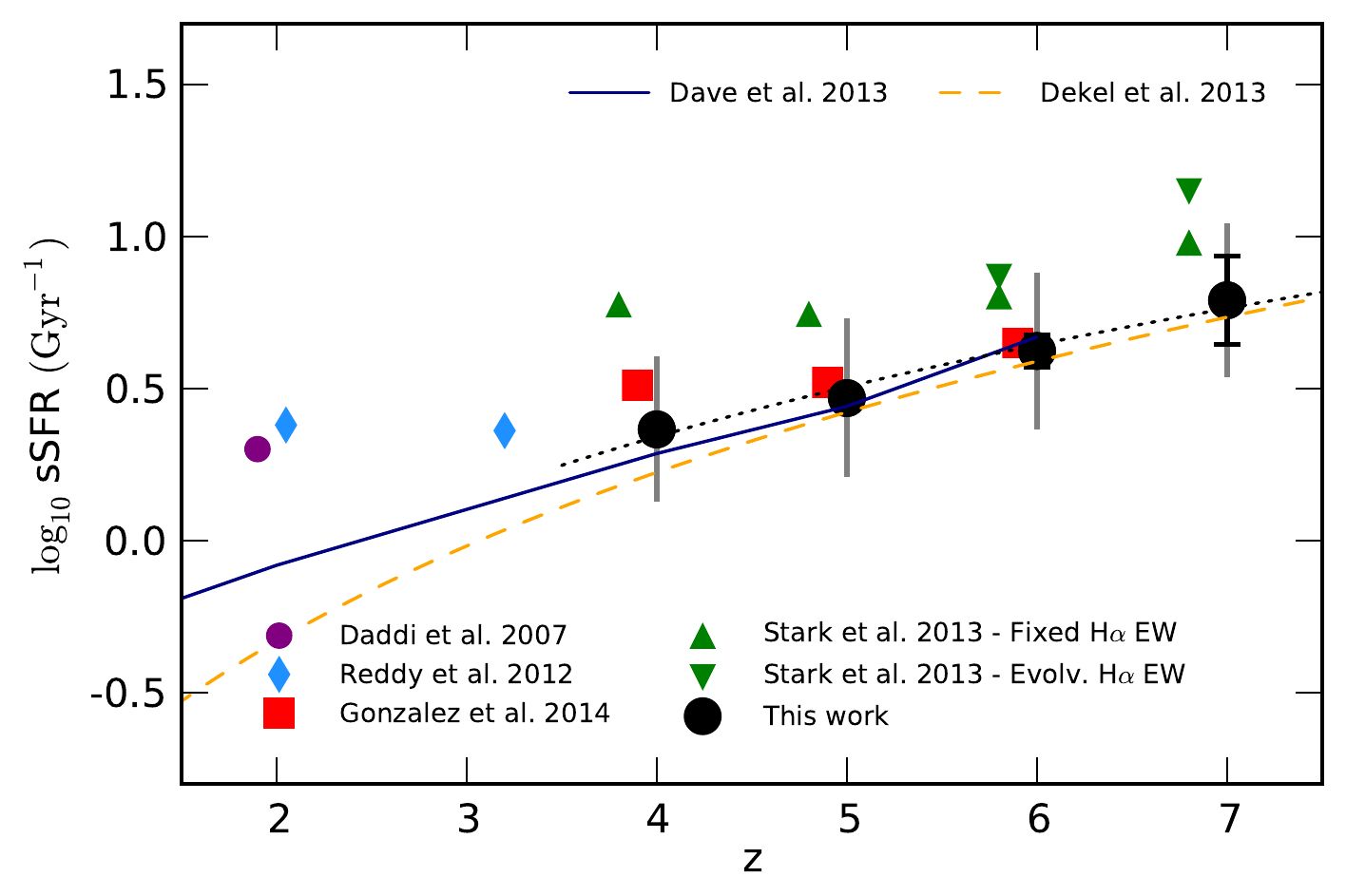}
\caption{Biweight mean specific star-formation rates (sSFR) and error on the mean for galaxies of mass $M_{*} = 5 \times 10^9 M_{\odot}$ as a function of redshift for this work (black circles). We find the scatter in sSFR, taken as the biweight scale of the distributions (grey error bars), to be $\approx 0.25$ dex across all redshifts. The dashed yellow line shows the evolution of the specific accretion rate, $\propto (1 + z)^{2.5}$, as outlined in \citet{Dekel:2013id}. The blue line shows the sSFR predicted by \citet{Dave:2013bf}, the model which most closely matches the observed stellar mass function (see Section~\ref{sec:theory}). The black dotted line shows the best-fitting power law to the results in this paper from $z \sim 4$ to 7 ($\propto (1 + z)^{2.06 \pm 0.25}$).}
\label{fig:ssfr_evolution}
\end{figure}

In Figure~\ref{fig:ssfr_evolution}, we show our results for the sSFR (when using $\text{SFR}_{\text{Madau}}$) in a stellar mass bin at $\log_{10}(\text{M} / \text{M}_{\odot}) = 9.7 \pm 0.3$ alongside previous observations at $z > 2$. We find an average sSFR of $2.32\pm0.08$, $2.94\pm0.20$, $4.21\pm0.54$ and $6.2\pm2.5$ $\text{Gyr}^{-1}$ for $z\sim 4$, 5, 6 and 7 respectively. Our observations show a clear trend in increasing sSFR with redshift in the redshift range $4 \leq z \leq 7$. The observed sSFR are in very good agreement with those of \citet{Gonzalez:2014do} but are systematically lower than those of \citet{Stark:2013ix} over the same redshift range. However, as noted in \citet{Stark:2013ix}, the introduction of 0.5 dex of intrinsic scatter to the $log_{10}\text{M}_{*}-M_{UV}$ used when estimating their sSFR would result in a reduction of $2.8\times$ at $z\sim4$. Such a large intrinsic scatter would be fully consistent with the $log_{10}\text{M}_{*}-M_{UV}$ relations and stellar mass functions observed in this paper. Taking this offset into account, the increasing consensus in the observed sSFR at high redshift is encouraging.
 
Performing a simple best fit to our observed sSFR across all four redshift bins, weighted to the measured scatter, gives sSFR $\propto (1 + z)^{2.06 \pm 0.25}$ (black dotted line). This trend is much more consistent with theoretical expectations of the sSFR evolution than a plateau at $\sim 2$ Gyr$^{-1}$, whereby the increased accretion of cold gas onto haloes results in higher specific star formation rates in the early universe. This can be seen in the evolution of the specific accretion rate \citep{Neistein:2007fg,Dekel:2013id}, $\propto (1 + z)^{2.5}$, shown as the orange dashed line in Figure~\ref{fig:ssfr_evolution}. Also shown in Figure~\ref{fig:ssfr_evolution} are the simulation predictions of \citet{Dave:2013bf} which are in good agreement with both our observations and the the specific accretion rate model at $z > 3$. 

Although we find a strong agreement with the zero-th order specific accretion rate model over the redshift range covered in this work, such a comparison should not be made in isolation from the previous observations of the sSFR at lower redshift. By $z \sim 4$ and below, the observed sSFR begins to diverge strongly from that predicted by the specific accretion rate and the SPH models of \citet{Dave:2013bf}. If we include the additional observations of \citet{2007ApJ...670..156D} and \citet{2012ApJ...754...25R} we find a best-fit of sSFR $\propto (1 + z)^{0.9 \pm 0.2}$, consistent with that found by \citet{Gonzalez:2014do}. However, restricting the power law fit previously calculated to only $z \sim 5$, 6 and 7, we find sSFR $\propto (1 + z)^{2.6 \pm 0.4}$.

Given the large degeneracies and assumptions inherent to the stellar mass and star formation estimates \citep{deBarros:2012wa,2013A&A...549A...4S}, there may still be significant systematic errors in the observed sSFR at high redshift. For example, \citet{Salmon:2014tm} find that the use of an SMC-like extinction curve instead of \citet{2000ApJ...533..682C} results in systematically lower sSFR for their sample at $z \sim 4$ to $z \sim 6$ using the same CANDELS data as our work. Similarly, the poor constraints on the properties of nebular emission and the escape fraction at high redshift allows for a wide range of plausible scenarios which could affect the measured stellar mass and SFR significantly. This is especially important at $z \sim 6$ and 7 where the impact of nebular emission is strongest \citep{Stark:2013ix,Smit:2013ud,Gonzalez:2014do}. Although the previous tension between theory and observations at $z > 4$ has been largely resolved, improved constraints on the stellar populations and star formation rates of high-redshift galaxies are still required before robust comparisons can be made. 

\subsubsection{Star formation rate functions and the cosmic star formation rate density}
\begin{figure*}
\centering
\includegraphics[width=140mm]{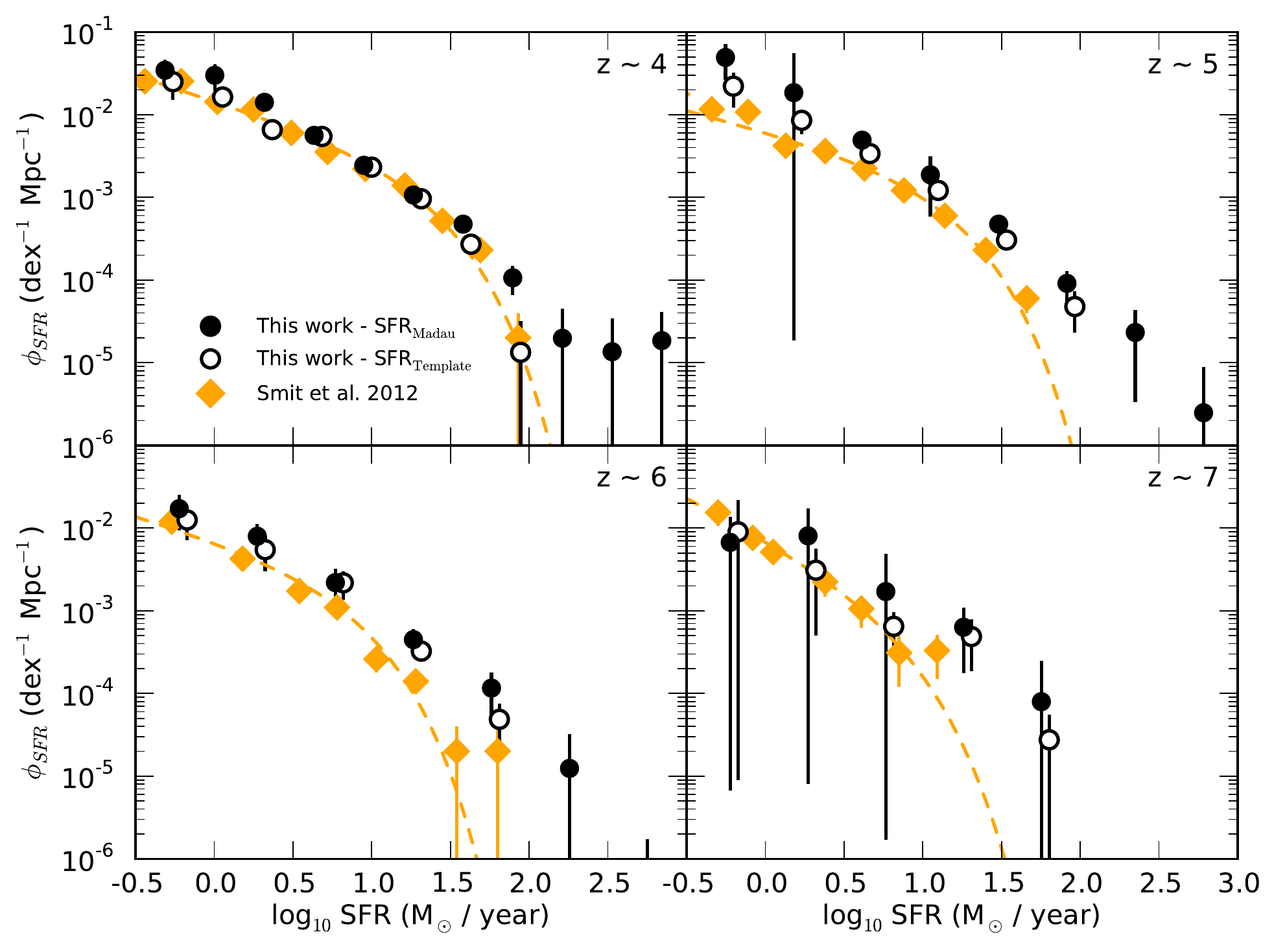}
\caption{Star-formation rate functions calculated using the $1/V_{\rm{max}}$ estimator as outlined in Equation~\ref{eq:vmaxmethod_sfr}. The filled black circles correspond to star-formation rates estimated from the dust corrected UV luminosity whilst the open black circles correspond to the best-fitting star-formation rate from the SED fitting, see Section~\ref{sec:SFR}. The SFR-functions of \citet{Smit:2012is} converted to a Chabrier IMF are shown by the yellow diamonds.}
\label{fig:sfr_function}
\end{figure*}

To measure the evolution of the star-formation rate (SFR) density for across our observed redshift bins, we use the previously calculated $1/V_{\rm{max}}$ values for each galaxy to construct a SFR function analogous to the mass or luminosity function for the same data, such that
\begin{equation}\label{eq:vmaxmethod_sfr}
\phi_{SFR,k}d\epsilon = \sum_{i}^{N_{gal}} \frac{w_{i}}{V_{max,i}}W(\epsilon_{k}-\epsilon_{i}),
\end{equation}
where $\epsilon = \log_{10}(\text{SFR}_{UV})$. The SFR functions for our high-redshift samples are shown in Figure~\ref{fig:sfr_function} for both the dust corrected UV star-formation rates ($\text{SFR}_{\text{Madau}}$) and the SED star-formation rates as outlined in Section~\ref{sec:masses}. At low to moderate star-formation rates ($\log_{10}(SFR_{UV}) \leq 1.5$), the two SFR estimates are in good agreement across all redshifts as was seen when the two estimates for individual galaxies were compared. 

The SFR function estimates of \citet{Smit:2012is} (converted to the same IMF used in this work) exhibit lower star-formation rates than we observe at all redshifts, with the exception of the $\text{SFR}_{\text{template}}$ based estimate at $z\sim4$ which shows excellent agreement across the full SFR range. \citet{Smit:2012is} correct the observed UV luminosity functions \citep{2007ApJ...670..928B,Anonymous:96uKWdy6} to intrinsic magnitudes using the same \citet{Meurer:1999jm} relation as outlined in Section~\ref{sec:masses}. Since the underlying UV luminosity functions for both observations show a good agreement, this discrepancy can be attributed solely to the $\beta$ values and methodology when correcting extinction. The \citet{2012ApJ...754...83B} $M_{UV}-\beta$ relations used to correct for dust extinction in \citet{Smit:2012is} exhibit a stronger UV luminosity dependence as well as a bluer average colour than that observed in our work (see Appendix~\ref{app:beta}). 

\begin{figure}
\centering
\includegraphics[width=80mm]{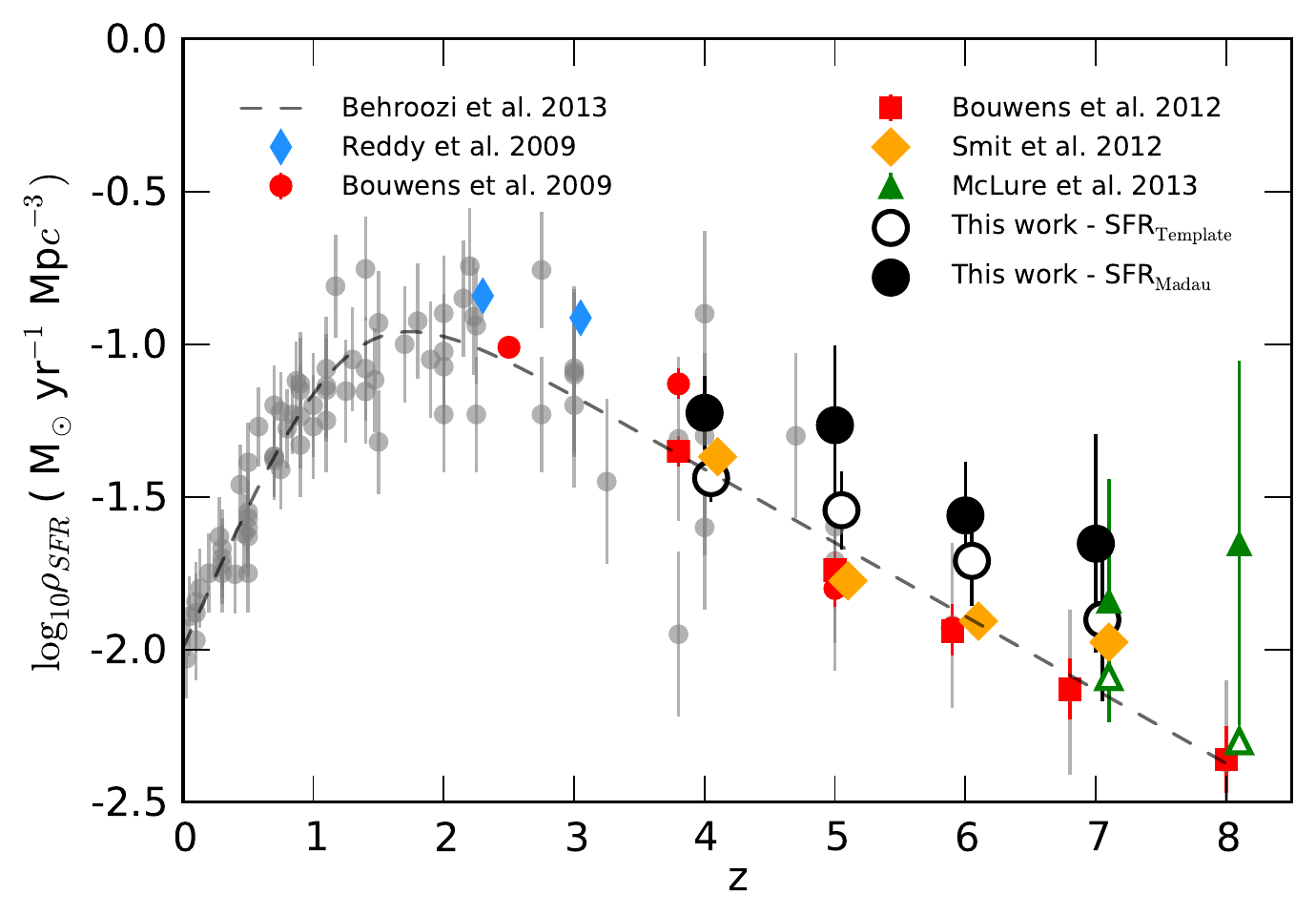}
\caption{Evolution of the SFR density as estimated from the SFR functions in Figure~\ref{fig:sfr_function}. The black filled circles show the SFR density calculated by integrating under the observed data directly using SFR$_{\rm{Madau}}$. The empty black circles show the corresponding estimates using SFR$_{\rm{Template}}$, these points have been offset by $+0.05 z$ for clarity. We show the recent compilation of SFR density observations (grey circles) and the fitted functional form from \citet{Behroozi:2013fg}. Shown separately are the dust-corrected UV SFR estimates of \citet{Anonymous:r_woy2UF}, \citet{Bouwens:2009ik}, \citet{2012ApJ...754...83B},  \citet{Smit:2012is} and \citet{McLure:2013hh}. The \citet{Smit:2012is} values were calculated by integrating the analytic SFR functions presented in their work from $\log_{10}(SFR_{UV}) = -0.47$ and above, in line with the limits used in other observations (see text). For the UV SFR density observations of \citet{McLure:2013hh} (open triangles), we apply a dust correction based on the observed $\beta$ slopes of the same survey as measured by \citet{Dunlop:2013kp} (filled triangles). We assume a fixed average extinction (with respect to $M_{UV}$) consistent with their observations, using $\left \langle \beta \right \rangle = -2.1 \pm 0.2$ for $z \sim 7$ and $\left \langle \beta \right \rangle = -1.9 \pm 0.3$ for $z \sim 8$.}
\label{fig:sfr_density}
\end{figure}

Integrating the SFR function across a suitable range gives the global SFR density. For consistency with other UV SFR density measurements, the lower bound in the integration of the SFR function was chosen to be $\log_{10}(SFR_{UV}) = -0.47$, equivalent to $0.03 L^{*}_{z=3}$, and the data points were integrated directly in steps. The evolution of the cosmic star formation rate density is shown in Figure~\ref{fig:sfr_density}. Alongside our new estimates of the SFR density at high-redshift we show the compilation of observed cosmic SFR from $z=0$ to $z=8$ from \citet{Behroozi:2013fg} and their fitted functional form to the same data. Our observations show a clear rise in the cosmic SFR over the $\sim 1$ Gyr between $z\sim7$ and $z\sim4$ with an increase of $\approx 0.5$ dex over this period.

As expected from the inspection of the SFR functions, the integrated SFR densities predicted by \citet{Smit:2012is} are lower than those observed in this work. The same is true for the cosmic SFR observed by \citet{2012ApJ...754...83B} which makes use of the same $M_{UV}-\beta$ relations. Due to the steep UV luminosity functions, the cosmic SFR is dominated by the faint galaxy population. The difference in dust correction resulting from the redder observed UV continua in this work therefore has a large effect on the observed dust corrected SFR density. 

In the recent $\beta$ observations of \citet{Bouwens:2013vf}, the authors find systematically redder $\beta$'s than for \citet{2012ApJ...754...83B}. They find that for a fixed redshift and rest-frame UV luminosity, $\Delta\beta \sim 0.13 - 0.19$ for $z = 4-6$ and $\Delta\beta \sim 0.22$ at $z\sim7$. Following the the \citet{Meurer:1999jm} relation, we estimate that the corresponding difference in dust corrections would be $\Delta A_{1500} \sim 0.26 - 0.38$ for $z = 4-6$ and $\Delta A_{1500} \sim 0.44$ at $z\sim7$. An increase in the \citet{Smit:2012is} star-formation rates of this magnitude would bring the two dust-corrected UV SFR-function estimates into greater agreement (and hence the corresponding SFR densities). However, such a correction is a simplification, and does not take into account the different $M_{UV}-\beta$ slopes observed and the effects of scatter. We can therefore not make a fully quantitative comparison of how SFR functions based on the $\beta$ observations of \citet{Bouwens:2013vf} would compare with those in this work.

Applying a dust correction based on the observations by \citet{Dunlop:2013kp} to the uncorrected SFR density observed by \citet{McLure:2013hh} (for the same sample and photometry), we find the results are in good agreement with our observations at $z \sim 7$. In \citet{Bouwens:2013vf}, the authors claim that the $\beta$ observations of \citet{Dunlop:2013kp} are biased red-ward by $\Delta \beta \sim 0.13$, our estimated dust-corrected SFR-density for \citet{McLure:2013hh} could therefore be a factor of $~0.26 dex$ too high. We note that if we apply a correction to the $\beta$'s observed in this work in order to match the observed $M_{UV}-\beta$ relations of \citet{Bouwens:2013vf}, our SFR density estimates would be reduced by $\Delta \rho_{SFR} \sim 0.05 - 0.1 dex$.

As with our observations of the specific star formation rate, the possible systematic errors resulting from the treatment of dust could have a significant effect on the observed SFR density. The importance of this can be seen in the difference between the UV and SED-fitting SFR functions and their corresponding SFR density estimates. The rarer red objects selected by our photometric redshift samples can contribute a significant fraction of cosmic SFR density if assumed to be dusty star-forming objects, i.e. $\beta$ corrected UV star-formation rates. Although the growing availability of spectroscopic data for high redshift galaxies will help reduce the uncertainty in some of these assumptions, the independent SFR observations at $z > 3$ promised by ALMA and LOFAR will be essential for obtaining robust measures of the cosmic SFR in the early universe.

\section{Summary}\label{sec:summary}
In this paper, we make use of the deep data provided by the CANDELS survey of the GOODS South to study the stellar mass growth of galaxies in the first 2 billion years of galaxy evolution. For a photometric redshift selected sample, we present new measurements of the galaxy stellar mass function across the redshift range $z \sim 4$ to 7 along with observations of the UV star formation rate of this sample. Stellar masses for the sample are measured from SED template fitting incorporating the effects of nebular emission, previously shown to have a significant effect on the observed stellar masses at high-redshift. 

Using the rest-frame UV magnitudes and UV continuum slopes measured by our SED fitting code, we also calculate dust-corrected star formation rates for our sample. From these we derive specific star formation rates and a measure of the cosmic SFR density as a function of redshift. Our primary conclusions are as follows:

\begin{itemize}
  \item Our new observations of the stellar mass functions at $z \sim 4$ to $z \sim 7$ exhibit steep low-mass slopes across the whole redshift range. These slopes are significantly steeper than previous observations in this redshift regime and are much closer to those observed in the UV luminosity functions of these same objects and recent observations at lower redshifts.  
  \item The observed stellar mass to UV luminosity ratio of our sample exhibits minimal evolution with luminosity, with close to a constant M/L$_{UV}$ in all redshift bins. The overall normalisation of the $\log_{10} (\text{M}_{*})$-$M_{UV}$ undergoes a significant increase in the scaling of this relation over time.
  \item From our observations of the stellar mass function, we calculate the stellar mass density at $z \sim 7$ is $6.64^{+0.58}_{-0.89}$ $\log_{10}~\text{M}_{\odot} \text{Mpc}^{-3}$ rising to $7.36\pm0.06$ at $z \sim 4$ for galaxies M$> 10^8 \text{M}_{\odot}$ and a Chabrier IMF.
  \item At a fixed stellar mass ($M = 5 \times 10^9 M_{\odot}$), the mean specific star formation rate rises with redshift. We find sSFR $= 2.32\pm0.08~\text{Gyr}^{-1}$ at $z \sim 4$, rising to $6.2\pm2.5~\text{Gyr}^{-1}$ at $z \sim 7$. These results are in good agreement with other estimates of sSFR which incorporate nebular emission in the stellar mass estimates. 
  \item We observe a rapid decline in the cosmic star formation rate at $z > 4$, but find star formation rate densities up to $\approx 0.5$ dex higher than those of \citet{2012ApJ...754...83B} and \citet{Smit:2012is} at the same redshifts. We conclude that much of this difference can be attributed to the rarest objects with large amounts of inferred dust extinction. Future spectroscopic and long-wavelength observations will be vital in better understanding star-formation rates in this epoch. 
\end{itemize}

\section*{Acknowledgements}
We thank the anonymous referee for their thorough review and help in greatly improving the paper. This work is based on observations taken by the CANDELS Multi-Cycle Treasury Program with the NASA/ESA HST, which is operated by the Association of Universities for Research in Astronomy, Inc., under NASA contract NAS5-26555. This work is based in part on observations made with the Spitzer Space Telescope, which is operated by the Jet Propulsion Laboratory, California Institute of Technology under a contract with NASA. Support for this work was provided by NASA. AM acknowledges funding via an ERC consolidator grant (PI: McLure). We would also like to acknowledge funding from the Science and Technology Facilities Council (STFC) and the Leverhulme Trust. 

\bibliographystyle{mn2e_2}
\bibliography{mass_functions}

\appendix
\section{Selection method comparison}\label{app:selection}
Traditionally, (star-forming) galaxies at high-redshift have been selected using the Lyman break technique, whereby galaxies are selected based on the observed colours across the redshifted Lyman break in their spectra.

When the observed colours of our photometric redshift selected galaxies are plotted in the same way, the selected galaxies span a range of colours far wider than those encompassed by the Lyman break galaxy (LBG) selection criteria. Many of the galaxies have colours which would place them in the locus spanned by low-redshift galaxies, according to the Lyman break criteria. This has been observed before by \citet{2010ApJ...724..425D}, who find a similar range of colours for galaxies with photometric redshifts at $z \sim 4$ and 5.

This raises the question as to whether this discrepancy is solely due to photometric scatter in the relevant colours, if they have stellar populations different from those expected for the Lyman break criteria, or if in fact those galaxies outside the selection criteria are low-redshift interlopers or catastrophic failures in the photometric redshift estimation.

To answer these questions, we have taken a sample of mock galaxies from the CANDELS semi-analytic models of \citet{Somerville:2008ed} and \citet{Somerville:2012cq} across all redshifts. From the full SAM catalog of galaxies from $z = 0$ to $z > 8$, we have included all galaxies at $z > 3$ and a randomly selected sample of a quarter of the galaxies at $z = 3$ and below (subsequent calculations of the interloper fraction fully correct for this reduced number density at $z < 3$). The resulting sample of $\sim 260,000$ galaxies consists of approximately equal numbers of high-redshift galaxies and a fully representative sample of low-redshift galaxies across their corresponding luminosity and colour distributions. For full details of the mock galaxy properties as a function of redshift we refer the reader to \citet{Somerville:2008ed} and \citet{Lu:2013ui}.

We then assign photometric errors to the intrinsic fluxes in each band based on the observed errors in the original catalog and then perturb the flux by those errors. The resulting colours should then indicate the effects of photometric scattering on the intrinsic colours. We have assumed the errors are gaussian (the fluxes are perturbed by a value drawn from a gaussian where $\sigma =$ Flux Error) and have applied the errors based on the measured flux errors of objects with equivalent fluxes in the UDF, DEEP and WIDE regions. 

To further illustrate this process we also select an example galaxy from our mocks which we can follow through the individual steps. The photometric properties of the galaxy and the corresponding photometric errors are outlined in Table~\ref{tab:eg_photom}. The galaxy has a redshift of $z = 5.01$, a stellar mass of $\approx 8 \times 10^{8} \text{M}_{\odot}$ and a UV-continuum slope of $\beta = -2.1$.
 
\begin{table*}
    \centering
    \label{tab:eg_photom}
    \caption{Intrinsic magnitudes, fluxes and typical observation errors for the example galaxy and the CANDELS GOODS South region. The first row outlines the true intrinsic fluxes $F_{true}$ in the key filters at $z\sim5$. In the second row we show the measured 5-$\sigma$ limiting magnitudes (or fluxes) for the DEEP region estimated in \citet{Guo:2013ig} for the photometry used in this paper. The third row shows the average and standard deviation flux error for objects in the photometric catalog \citep{Guo:2013ig} with fluxes within 0.1 dex of the intrinsic flux for our example galaxy (i.e. the distribution from which our assigned photometric error is drawn). The final row shows the 'observed' fluxes ($F_{obs}$) for our example galaxy after assigning a flux error $\sigma_{F}$ and perturbing the intrinsic flux by a value drawn from a gaussian with width $\sigma = \sigma_{F}$.}
    \begin{tabular}{lcccccccc}
    \hline
    ~                      & \multicolumn{2}{c}{$V_{606}$} & \multicolumn{2}{c}{$i_{775}$} & \multicolumn{2}{c}{$z_{850}$} & \multicolumn{2}{c}{$H_{160}$} \\ \hline
    ~                      & AB        & $\mu Jy$ & AB & $\mu Jy$ & AB & $\mu Jy$ & AB   & $\mu Jy$ \\ \hline
    Intrinsic Flux - $F_{true}$           & 29.35         & 0.0066   & 27.14  & 0.0506  & 26.83  & 0.0673 & 26.91 &  0.0625  \\
    DEEP 5-$\sigma$ limit   & 29.35        & 0.0066   & 28.55  & 0.0138  & 28.55  & 0.0138 & 27.36 &  0.0413   \\
    Mean Error $\pm 1$ SD  & -         & $0.0054\pm0.0028$    & -  & $0.0111 \pm  0.0057$  &   & $0.0145 \pm 0.0113$   &   -   & $0.0099 \pm 0.0074$ \\ 
    'Observed' Flux - $F_{obs}$  & $>27.82$     & $0.009 \pm 0.009$  & 27.07  & $0.054\pm0.010$  & 26.85 & $0.066\pm0.013$   & 27.01        & $0.057 \pm 0.006$   \\
    \hline
    \end{tabular}
\end{table*}
 
\begin{figure}
\centering
\includegraphics[width=80mm]{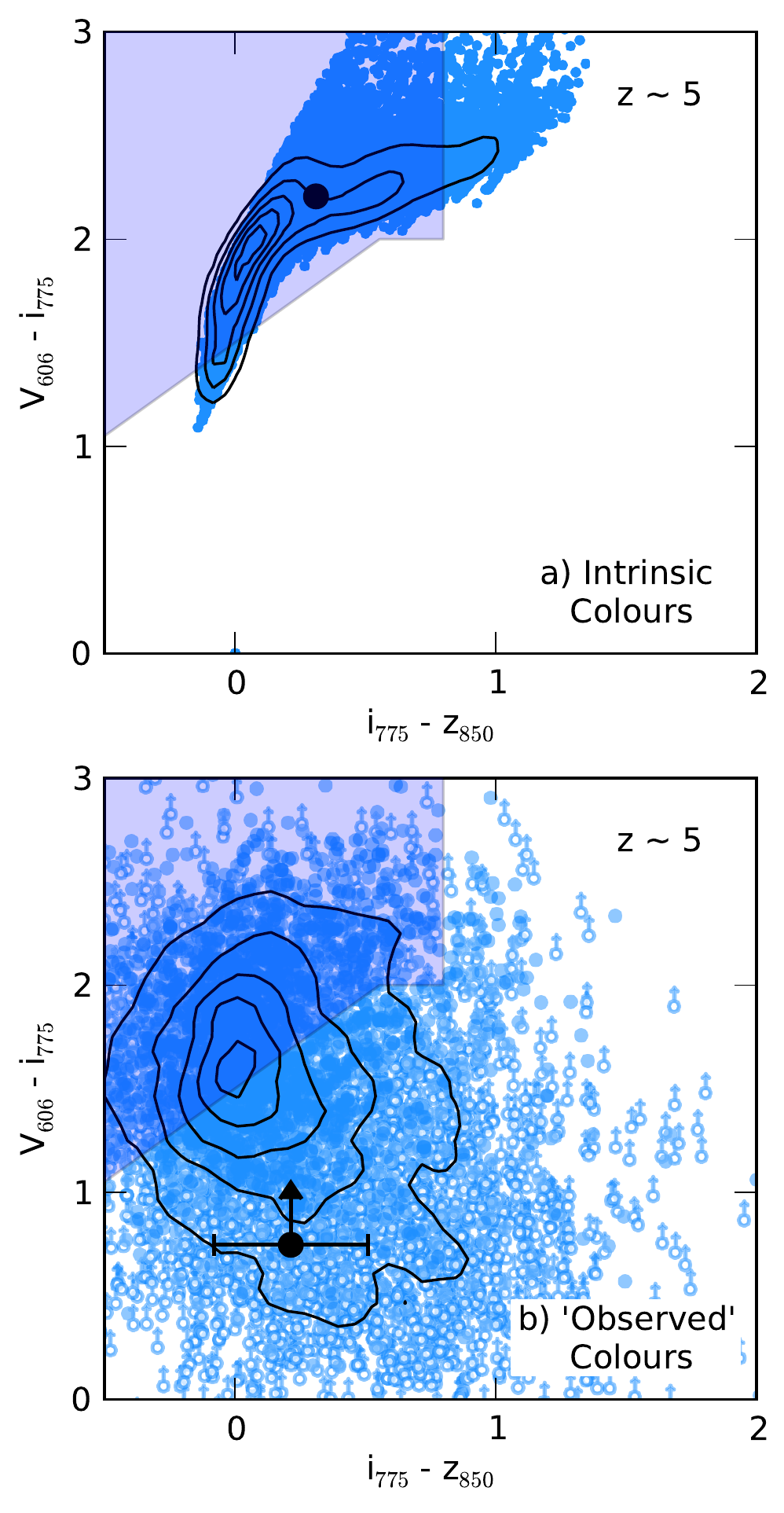}
\caption{a) Intrinsic colours of galaxies at $4.5 < z < 5.5$ from the CANDELS semi-analytic mock catalog. Also shown are contours representing the density of points, with the innermost contour corresponding to five times the density of the outermost contour. The larger black point shows the intrinsic colour of the galaxy from the example outlined in the text. b) Observed colours of galaxies at $4.5 < z < 5.5$ from SAM mock sample after the photometry has been perturbed by flux errors proportional to the observed flux errors in the CANDELS DEEP region of observed photometry. Open circles represent colours constructed from 2-$\sigma$ upper limits. As in a), the innermost contour corresponds to five times the density of points of the outermost contour.}
\label{fig:mock_col}
\end{figure}

The intrinsic colours of the mock galaxies at $z \sim 5$ ($4.5 \leq z < 5.5$) are shown in Figure~\ref{fig:mock_col}. It is clear that the colours spanned by the galaxies lie well within the Lyman break selection criteria of \citet{2007ApJ...670..928B}, with only a small fraction of galaxies redder than the criteria in the colour above the break or bluer across the Lyman break. Our input population therefore closely matches the colours for which the colour-colour criteria have been designed. The same is true across all redshift bins. 

The lower panel of Figure~\ref{fig:mock_col} shows the colours of our mock galaxy catalog after being perturbed by errors drawn from the CANDELS DEEP region. Only galaxies with $S/N(H_{160}) > 5$ are shown, matching the selection criteria for our high-redshift samples, resulting in a sample of 5673 galaxies with $4.5 < z_{\text{true}} < 5.5$. As for our observed objects in Figure~\ref{fig:colours}, 2-$\sigma$ upper-limits are used to derive magnitudes for non-detections ($S/N < 2$) and objects with negative fluxes.  

It is immediately clear that photometric scatter pushes the colour spanning the Lyman break to values much lower in $V_{606} - i_{775}$ than the range covered by the selection criteria. However, the main locus of galaxies still resides either within or within the typical error of the Lyman break selection region. In addition, the majority of mock galaxies with `observed' $V_{606} - i_{775} < 1$ are those with a non-detection in $V_{606}$ and hence represent a lower limit. The same effect occurs across all regions, with the lower photometric errors in deeper GOODS South region resulting only in a fainter magnitude for an equivalent signal to noise in the optical bands. 

The systematic shift towards lower values of $V_{606} - i_{775}$ once photometric scatter is included can be explained by the relative signal-to-noise ratios in the filters above and below the Lyman break. Given the depth of each filter (see Table~\ref{tab:eg_photom}), objects with relatively faint apparent magnitudes above the break and intrinsic colours $> 1-2$ will always have a significantly lower signal-to-noise in the filter below the break. In this scenario, $V_{606}$ fluxes which are scattered to higher values will result in a brighter more robust magnitude. Conversely, objects which are scattered to fainter magnitudes are more likely to result in non-detections, requiring the use of upper limits which will push the observed colours down.

These simulations show that high-redshift galaxies can exhibit colours across the Lyman break well outside the traditional selection criteria. However, it is also important to show that the galaxies selected by photometric redshift with colours outside the colour criteria are indeed these high-redshift galaxies rather than lower redshift galaxies in the same colour space.

\begin{figure}
\centering
\includegraphics[width=80mm]{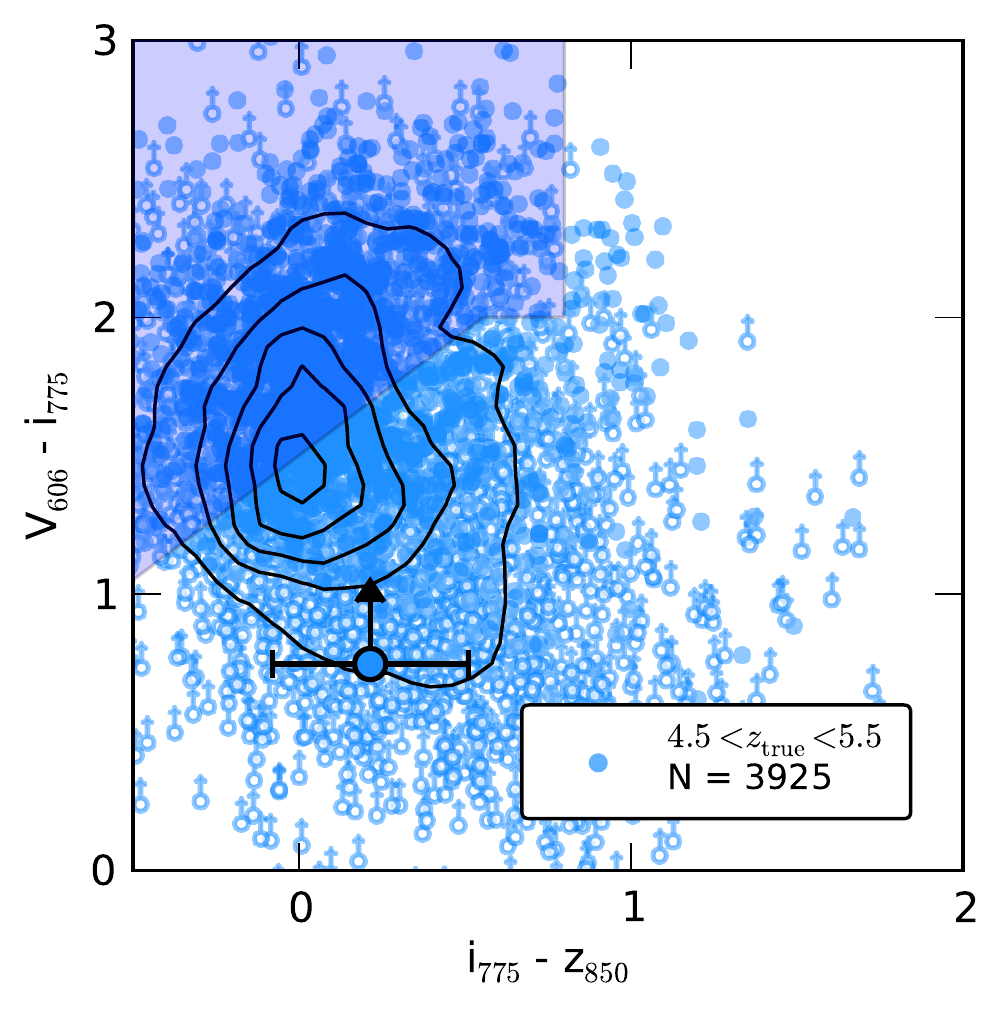}
\includegraphics[width=80mm]{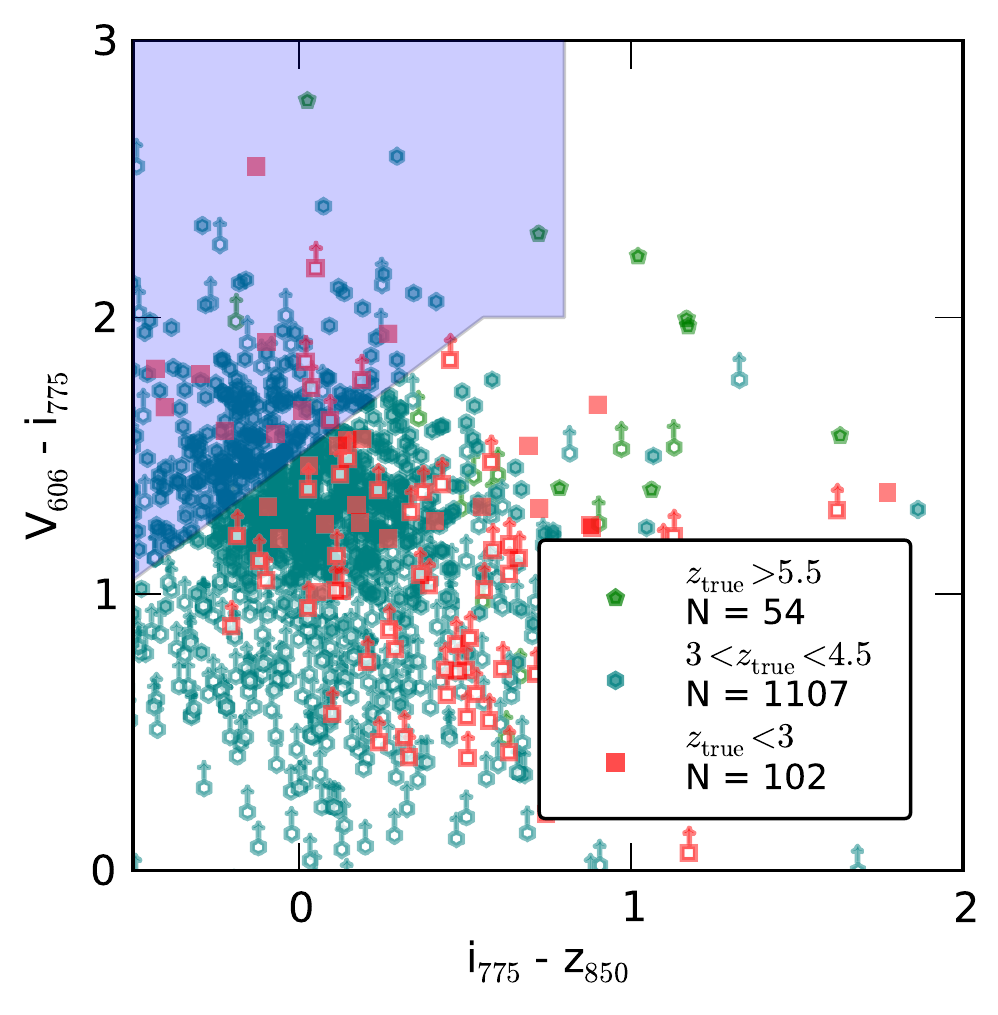}
\caption{\emph{Top:} Observed colours of galaxies from the SAM mock sample which pass our photometric redshift selection criteria, have best-fitting photometric redshifts in the range $4.5 < z_{\text{phot}} < 5.5$ and have true redshifts in the range $4.5 < z_{\text{true}} < 5.5$. As in Figure~\ref{fig:mock_col}, the innermost contour corresponds to five times the density of points of the outermost contour. Open circles with arrows represent colours constructed from 2-$\sigma$ upper-limits. The separately marked large blue circle corresponds to the the example galaxy which is correctly estimated to be $z \sim 5$. \emph{Bottom:} Observed colours of galaxies from the SAM mock sample which pass our photometric redshift selection criteria and have best-fitting photometric redshifts in the range $4.5 < z_{\text{phot}} < 5.5$ but have true redshifts outside of the desired redshift bin. As in the top panel, open symbols with arrows represent colours constructed from 2-$\sigma$ upper-limits. In both panels, N is the number of galaxies in the corresponding sample. As outlined in the text, the number of $z_{\text{true}} < 3$ galaxies shown represents a quarter of those expected in a fully representative sample. Using the best-fitting photometric redshift and our selection criteria, the low-redshift interloper fraction for this sample $= (102 \times 4) / (54 + (102 \times 4) + 1107 + 3925) = 0.07$. This low-redshift interloper fraction is reduced to $\approx 0.06$ when we generate our Monte Carlo samples.}
\label{fig:mock_col_photz}
\end{figure}

To address this, we next calculate photometric redshifts for the SAM mock galaxy sample incorporating the photometric errors using the same method as described in Section 2 and apply our sample selection criteria. By applying additional cuts based on the full $P(z)$ distribution, the number of interlopers can be reduced at the expense of excluding some real sources.  How strict the selection criteria are is a balance between minimising the contamination from interlopers and scatter at the bin edges and maximising the number of real high-redshift galaxies in the sample. In Figure~\ref{fig:mock_col_photz} we show the colours of galaxies which pass our high-redshift selection criteria. The top panel of Figure~\ref{fig:mock_col_photz} shows those galaxies with $4.5 < z_{\text{true}} < 5.5$, these galaxies span the full range of colours traced by the error perturbed colours of input high-redshift galaxies shown in the previous plots. In the case of the Table~\ref{tab:eg_photom} example galaxy, the best-fitting photometric redshift is $z = 5.0^{+0.2}_{-0.5}, 5.2\pm0.1$ and $4.86^{+0.27}_{-0.15}$ for the DEEP, UDF and WIDE errors respectively. 

In the bottom panel of Figure~\ref{fig:mock_col_photz}, we show the selected galaxies which have true redshifts outside of the desired redshift range. At $z\sim 5$, the majority of low-redshift ($z < 3$) interlopers which are selected to be $z\sim5$ by the photometric redshift selection exhibit colours which lie outside of the Lyman break colour criteria. The fraction of low-redshift interlopers is very small compared to the number of 'real' high-redshift galaxies in this colour space. However, as redshift increases, the fraction of low-redshift interlopers increases such that at $z \sim 7$ based on the best-fitting $z_{peak}$ alone, the fraction of outliers equals $\sim 0.60$, 0.51 and 0.72 for DEEP, UDF and WIDE respectively. Clearly, basing high-redshift samples on the best-fitting photometric redshift alone would produce highly biased samples. Further $S/N$ or photometric criteria such as those used in this work are clearly required to produce a reliable sample.

Applying the selection criteria and generating Monte Carlo samples as outlined in Section~\ref{sec:sample}, the low-redshift interloper fractions for our mock samples are reduced to an estimated 0.008, 0.06, 0.15 and 0.22 for $z \sim 4$, 5, 6 and 7 respectively. This was estimated by combining the fractions calculated for each field (assuming the ERS region to have interloper fractions comparable to the DEEP region) proportional to the number of high-redshift galaxies selected from each region of the field. 

The lower panel of Figure~\ref{fig:mock_col_photz} also highlights the importance of fully incorporating the photometric redshift errors when creating high-redshift galaxy samples. Approximately 20\% of the galaxies selected as $z\sim5$ have true redshifts below the desired range. However upon closer inspection, we find that the median true redshift for the $3 < z_{\text{true}} < 4.5$ points (turquoise hexagons) is 4.4 whilst the median best-fitting photometric redshift for the same sample is $4.6$ with average 1-$\sigma$ errors of $^{+0.18}_{-0.35}$.

By making use of the full $P(z)$ distribution estimated by the photometric redshift code as we do in this work (see Section~\ref{sec:MC}), galaxies with $P(z)$ which span the redshift boundaries will be scattered between and contribute to both adjacent redshifts bins between different MC samples (or scattered out of the sample e.g. $z < 3.5$ or $z > 7.5$). Throughout this work, the errors resulting from this photometric redshift uncertainty are incorporated in the analysis and errors presented.

\begin{figure}
\centering
\includegraphics[width=80mm]{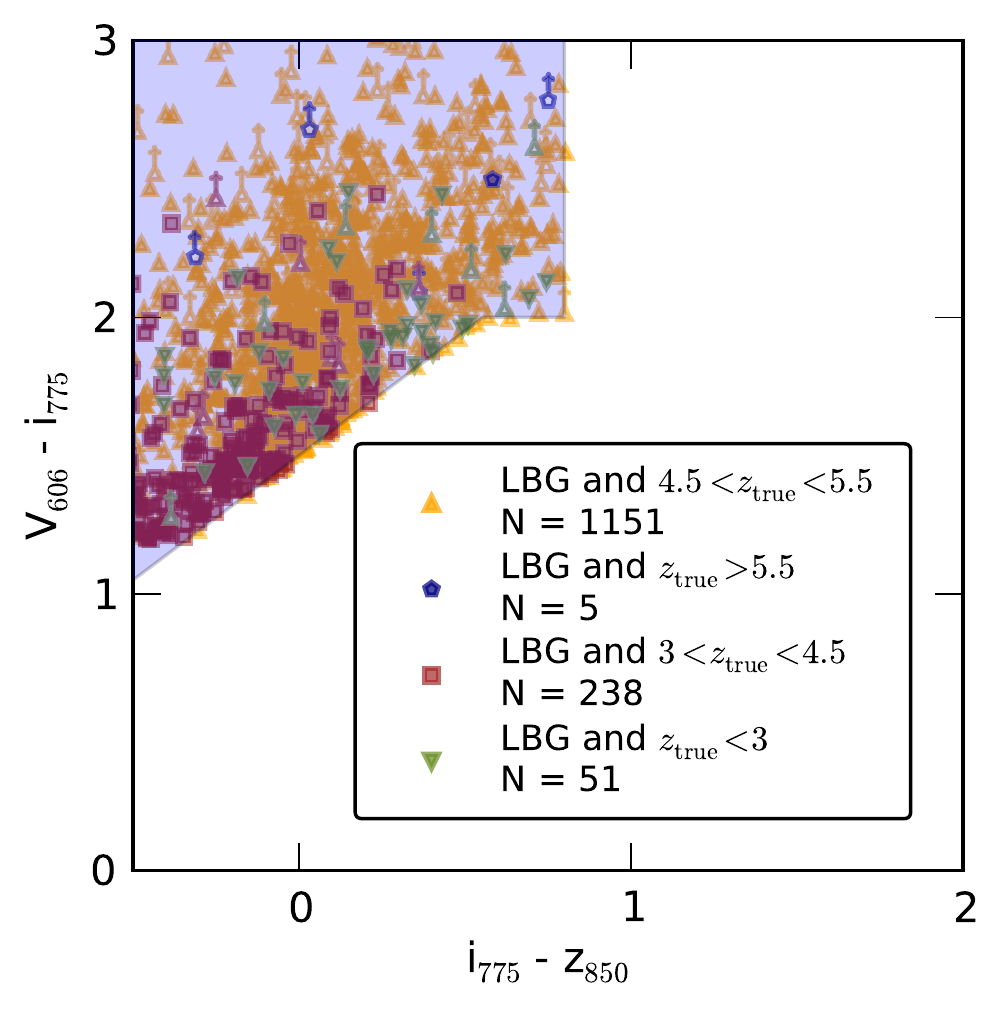}
\includegraphics[width=80mm]{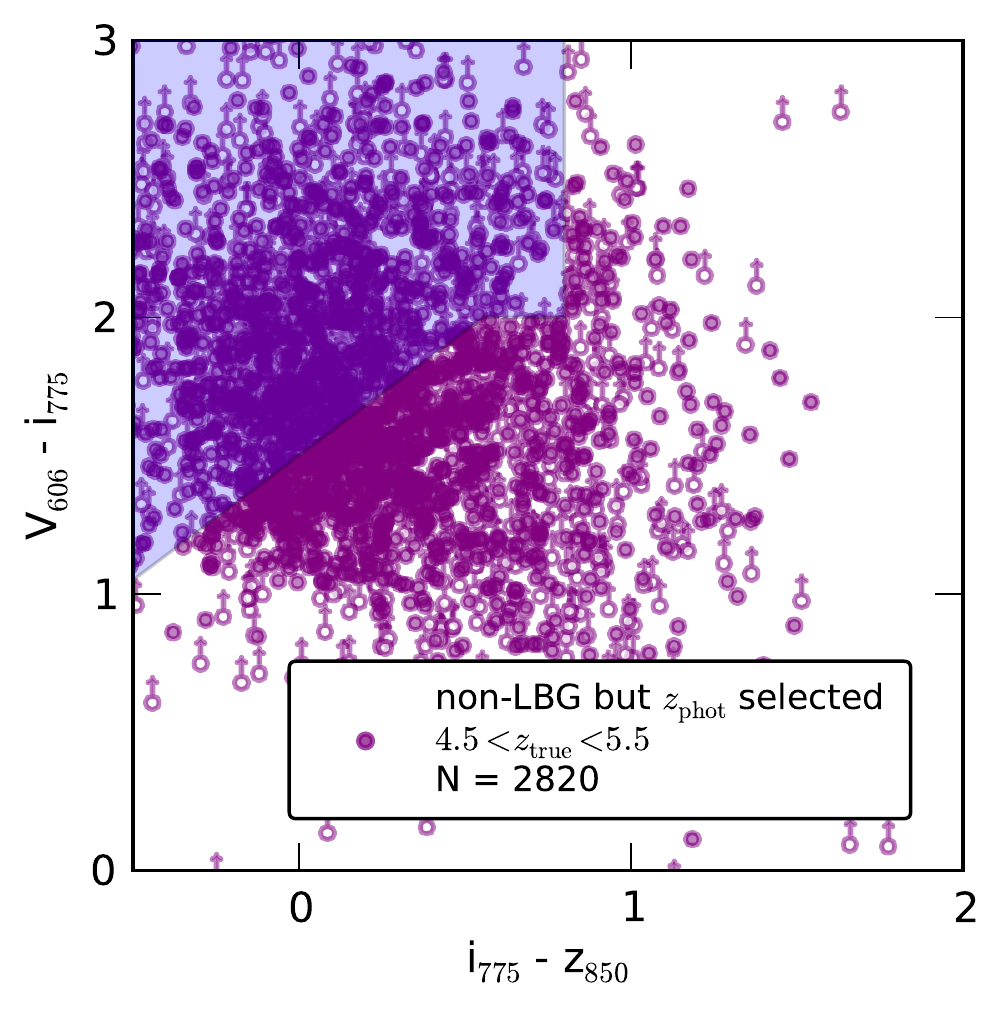}
\caption{\emph{Top:} Observed colours of galaxies from the SAM mock sample which pass the Lyman break selection criteria outlined in the text, separated into bins of intrinsic redshifts. In contrast to previous plots and in keeping with common LBG selection techniques, when calculating colours the measured magnitude is used down to a $S/N = 1$ and the 1-$\sigma$ upper limit is used below this. The same is true for colours plotted in the bottom panel. As outlined in the text, the number of $z_{\text{true}} < 3$ galaxies shown represents a quarter of those expected in a fully representative sample. The low-redshift interloper fraction for this sample $= (4*51) / (5 + (51 \times 4) + 238 + 1151) = 0.13$. \emph{Bottom:} Observed colours of galaxies from the SAM mock sample with $4.5 < z_{\text{true}} < 5.5$ which fail the Lyman break selection criteria outlined in the text but are correctly selected as $z\sim5$ galaxies by the photometric redshift selection used in this work.}
\label{fig:mock_col_LBG}
\end{figure}

To compare the low-redshift interloper fractions and the robustness of our photometric redshift selection, we also run the SAM mock catalog through a Lyman break selection process. Our Lyman break selection criteria are based on the $V_{606}$-dropout criteria of \citet{2012ApJ...754...83B} and we exclude sources with $S/N > 2$ in any of the bands blueward of the dropout bands ($U_{\text{CTIO}}$, $U_{\text{VIMOS}}$ and $B_{435}$). We also require $S/N(i_{775}) > 5.5$, comparable with other Lyman break selections at this redshift, e.g. \citet{Giavalisco:2004et} and \citet{2006AJ....132.1729B}. However we note that by choosing a stricter optical $S/N$ requirement, the purity of the sample can always be improved at the expense of total sample size. For consistency with other LBG selections, when making the colour cuts, we use the observed magnitudes for detections above 1-$\sigma$ and the 1-$\sigma$ upper limit below this.

We caution that since the mock photometric catalog in this section is designed to replicate the $H_{160}$ selection of the observational data used in this paper, the detection criteria for our Lyman break sample will differ from those in the literature based solely on the optical (e.g. \citeauthor{2007ApJ...670..928B}~\citeyear{2007ApJ...670..928B}). Therefore, the selection efficiencies and low-redshift interloper fractions calculated here represent only the Lyman break technique as applied to the CANDELS data in this paper specifically. As such, we do not make any claims regarding the low-redshift interloper fraction of Lyman break selection elsewhere in the literature.

\begin{figure}
\centering
\includegraphics[width=80mm]{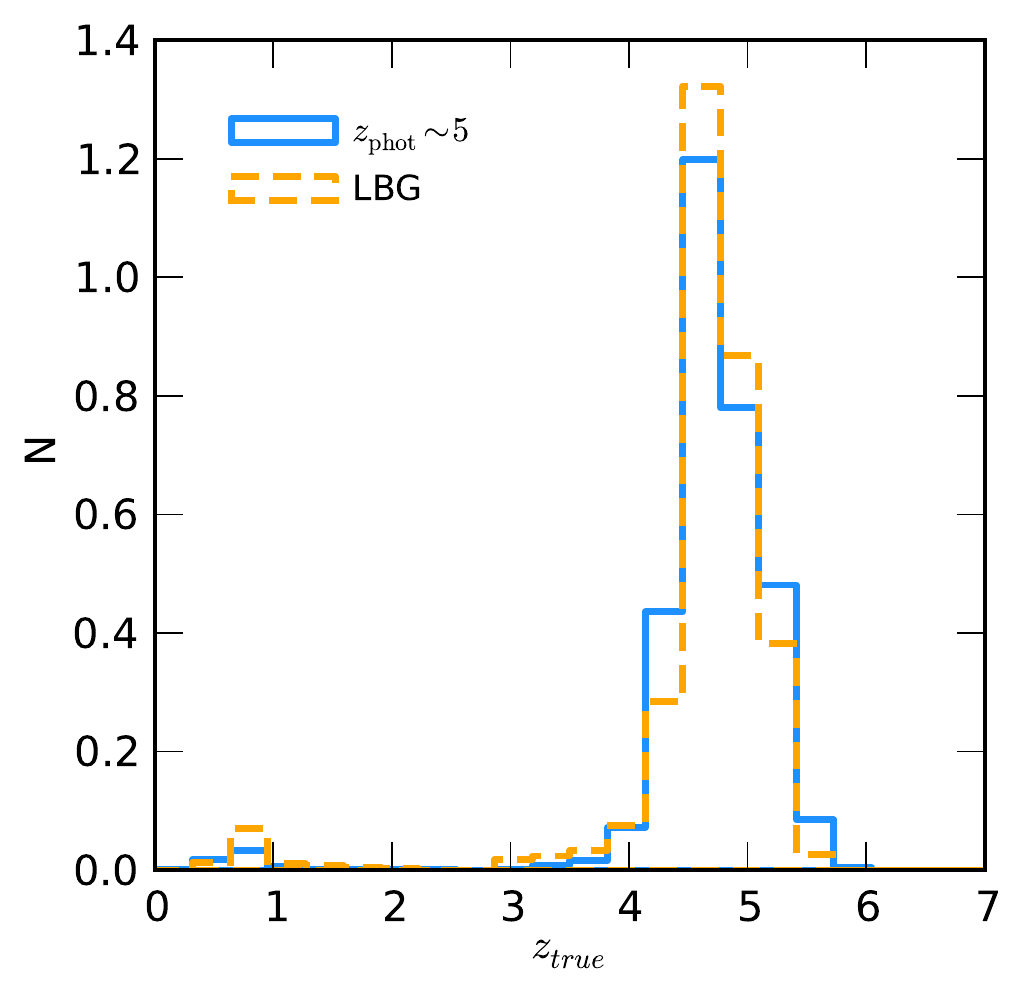}
\caption{Normalised number densities as a function of true redshift for the photometric redshift and Lyman break galaxy samples generated for our SAM mock galaxy catalog.}
\label{fig:mock_z_hist}
\end{figure}

In the upper panel of Figure~\ref{fig:mock_col_LBG}, we show the colour distribution and sample sizes for our LBG sample. For this sample, we find a low-redshift interloper fraction comparable to that of the photometric redshift selection. In the lower panel of Figure~\ref{fig:mock_col_LBG}, we show galaxies which have true redshifts in the range $4.5 < z_{\text{true}} < 5.5$ and do not satisfy all of the LBG criteria but do pass the photometric redshift selection criteria. Although many of these galaxies have colours outside the LBG colour criteria, photometric redshifts are also able to select galaxies which fail the non-detection or optical $S/N$ criteria. In Figure~\ref{fig:mock_z_hist}, we also compare the intrinsic redshift distribution of the two selection methods. Both the low-redshift contaminants and scatter at bin extremities are clearly visible for both samples.

\begin{figure}
\centering
\includegraphics[width=80mm]{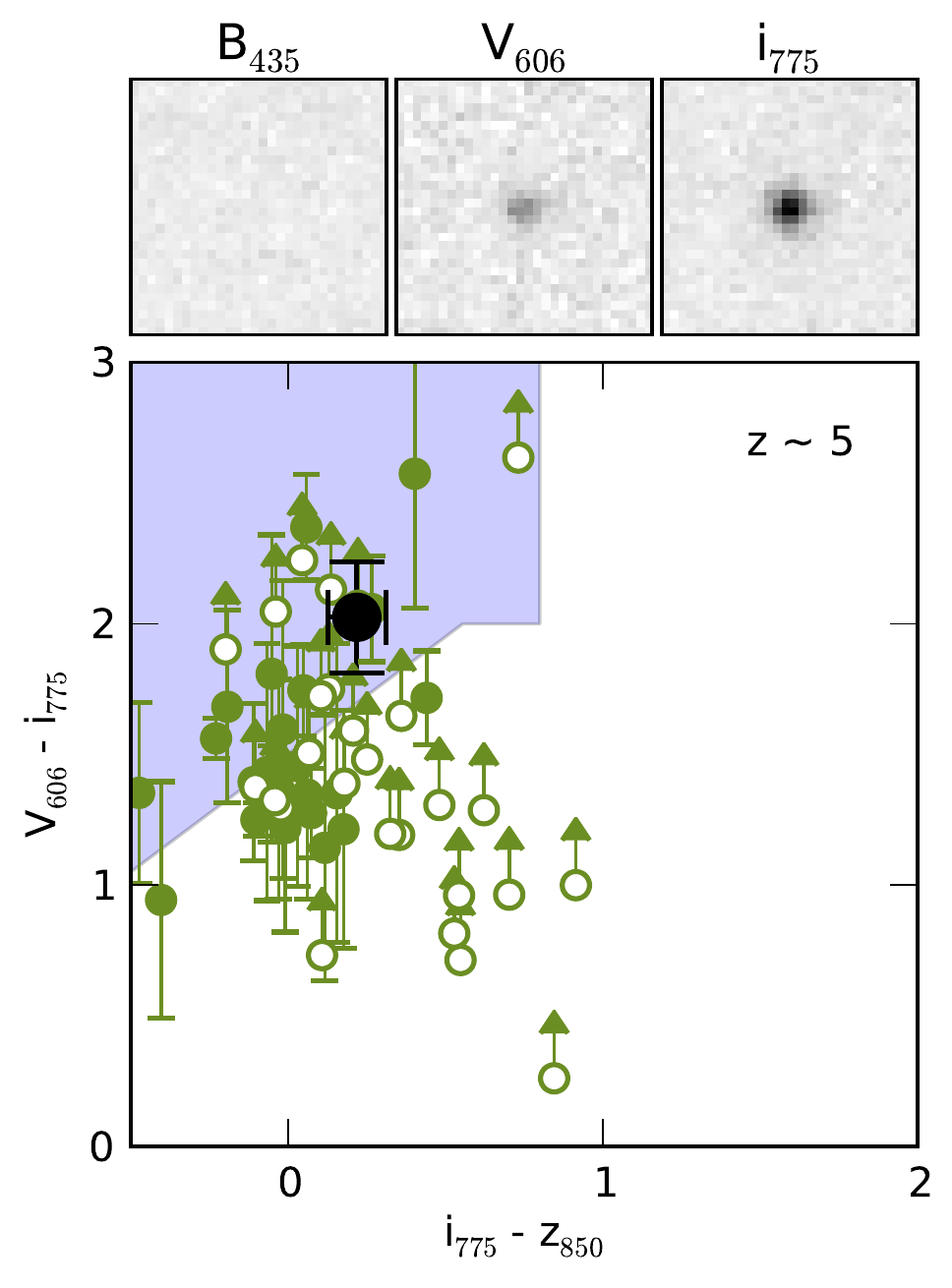}
\caption{\emph{Top panels}: $1.8 \times 1.8$ arcsec$^2$ postage-stamp images of the median stacked faint sources in the $B_{435}$, $V_{606}$ and $i_{775}$ filters. \emph{Main panel}: The observed colours of the individual faint sources are shown by the smaller green circles. Open circles represent sources where the $V_{606}$ has been calculated from the $2\sigma$ flux upper limit. The large black circle shows the measured colour for the stacked images.}
\label{fig:stack_colours}
\end{figure}

As a further step to demonstrate the difference in the observed Lyman break colours can be explained solely by photometric scatter, we examine the photometry for a median stack of the 50 candidate $z \sim 5$ galaxies in the CANDELS DEEP region with the lowest $S/N(H_{160})$. Stacking the photometry of a large enough number of sources should cancel out most of the photometric noise, with the resulting images closely reproducing the average intrinsic colours of the input galaxies. In Figure~\ref{fig:stack_colours}, we show the initial observed colours (or lower limits) for each of the faint $z\sim5$ candidates along with the observed colours of the stacked sources. Although the majority of the input galaxies have colours outside the Lyman break selection, the combined stack has a colour which places it more than 1-$\sigma$ inside the desired region and fully consistent with the expected $z \sim 5$ colours.

Also shown in Figure~\ref{fig:stack_colours} are the median stacked images in the 3 filters from below to above the Lyman break. Crucially, for any high-redshift candidate galaxy, the filters at wavelengths lower than the Lyman selection colours should contain zero flux due to the complete absorption by inter-galactic Hydrogen. The median stack in $B_{435}$ for our faint sample contains no trace of flux, with a 2-$\sigma$ upper limit of $29.48$ within a $0.6"$ diameter aperture.

While the tests presented in this appendix cannot account for objects with peculiar intrinsic colours (i.e. significantly different from those predicted by semi-analytic or synthetic stellar population models), they demonstrate that the observed colour distribution can be fully accounted for by the photometric scattering of the expected intrinsic colours. We also conclude that photometric redshift selection can be much less sensitive to photometric scatter than the Lyman break selection criteria for the same redshift range. Furthermore, that it is also able to correctly select high-redshift galaxies which are not identified by the traditional Lyman break selection techniques. In addition, the samples produced by both methods contain similar fractions of low-redshift galaxy contamination when criteria of comparable strictness are applied.

\section{Observed UV continuum slopes}\label{app:beta}
\begin{figure}
\centering
\includegraphics[width=75mm]{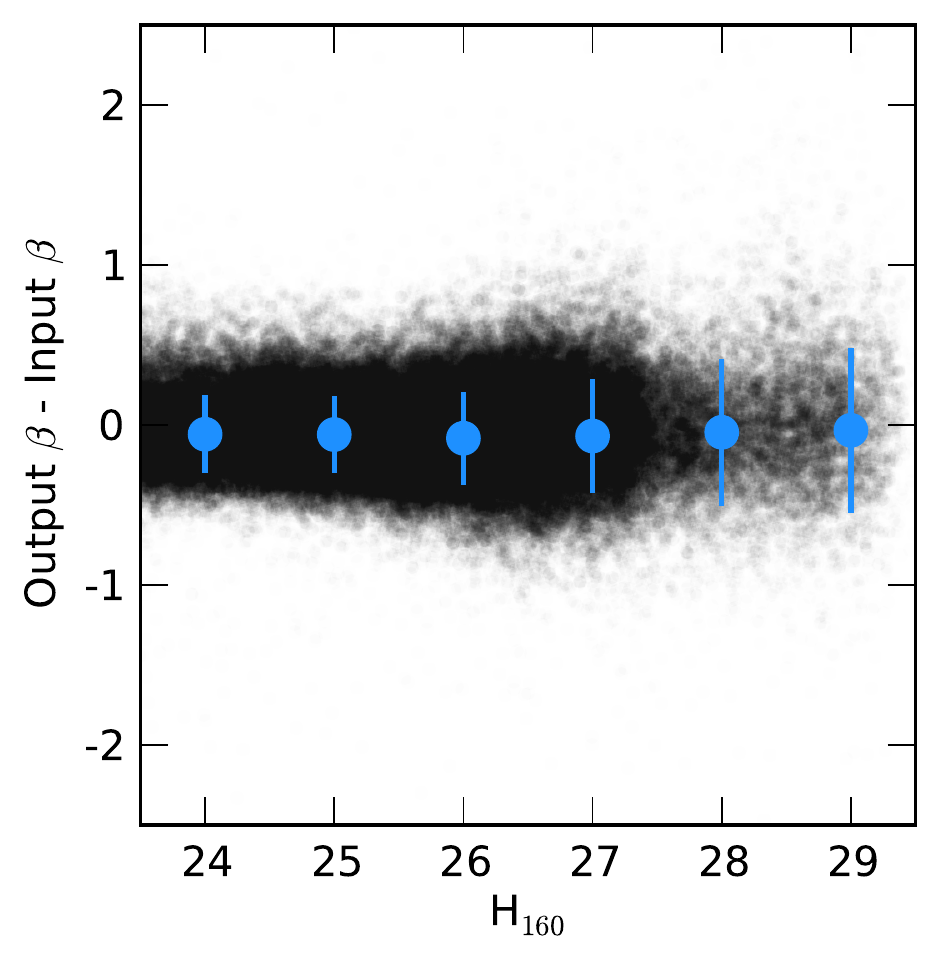}
\caption{Recovered $\beta$ - input $\beta$ as a function of apparent H$_{160}$ magnitude. The blue circles show the mean $\beta_{out}$ - $\beta_{in}$ in bins with width $=1$ magnitude. The bias ($\left | median (\beta_{out}-\beta_{in}) \right |$ is less than 0.1 for all magnitudes, whilst the standard deviation increases from $= 0.24$ at $H_{160} = 24$ to $= 0.44$ at $H_{160} = 28$}
\label{fig:betacomparison}
\end{figure}

As one of the key observables that it is possible to accurately measure for high-redshift galaxies using photometry, the UV continuum slope ($\beta$) has been well studied but with initally conflicting results \citep{Dunlop:2011jl,Wilkins:2011fs,2012ApJ...754...83B,Finkelstein:2012hr,2013MNRAS.429.2456R,Bouwens:2013vf}. The method used in this work to measure $\beta$ follows a similar procedure to that outlined in \citet{Finkelstein:2012hr}. The relative accuracy of the different methods and the effects of differing sample criteria are explored in depth by \citet{2013MNRAS.429.2456R}, however from our simulations (Section~\ref{sec:simulations}) we can test the accuracy of our fitting directly. Figure~\ref{fig:betacomparison} shows the difference between the input and measured $\beta$ as a function of H$_{160}$ magnitude for all regions of the GOODS-S field combined.  

\begin{figure}
\centering
\includegraphics[width=80mm]{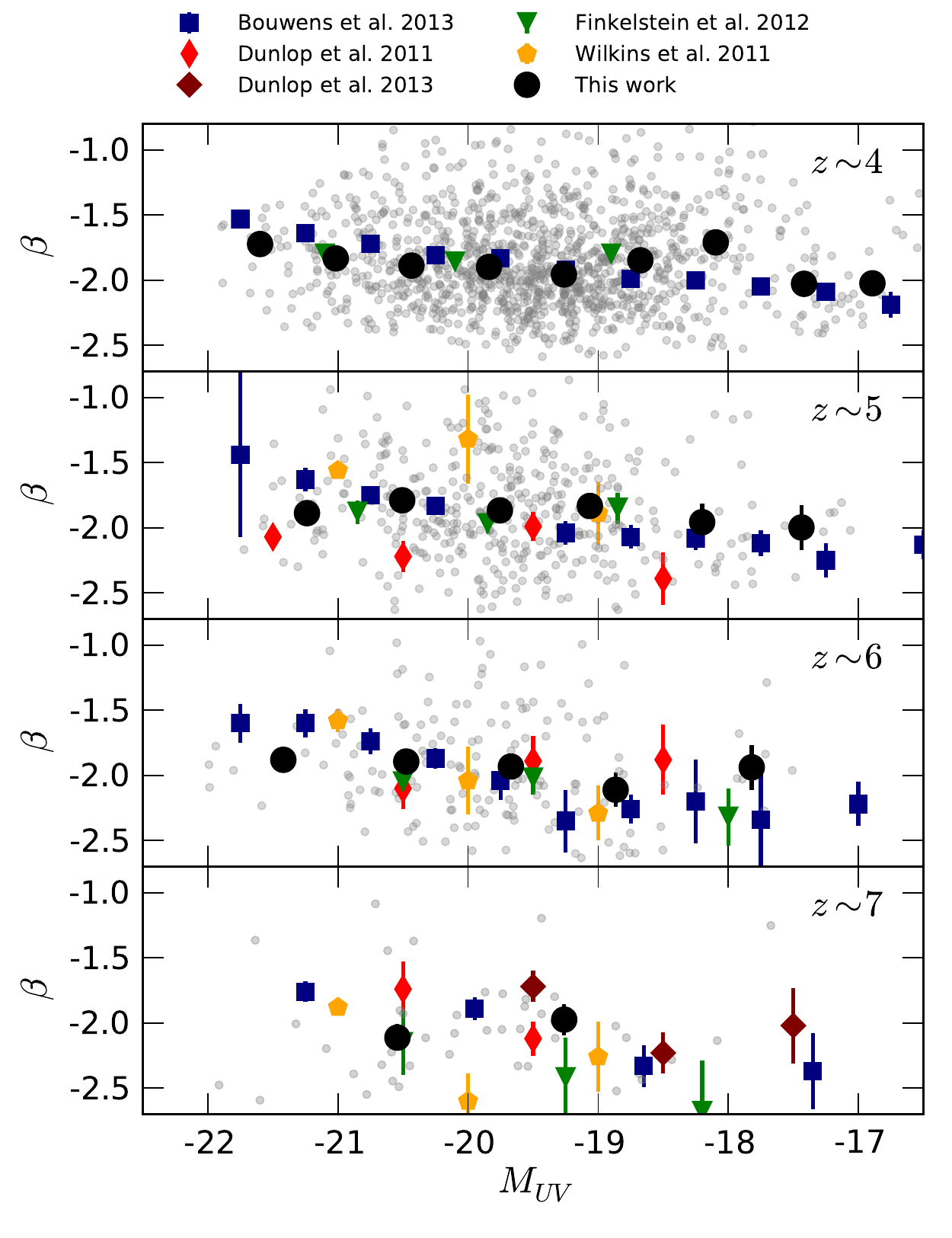}
\caption{Measured UV continuum slope as a function of UV magnitude for this work and previous studies. The background grey circles show the individual points for one of the Monte Carlo samples used in our work. The black circles show the biweight mean $\beta$ and corresponding standard error on the mean as a function of $M_{UV}$ averaged over 100 of our Monte Carlo samples. Also shown are the equivalent $M_{UV}$ binned means available from the literature.}
\label{fig:beta_muv}
\end{figure}

\begin{figure}
\centering
\includegraphics[width=80mm]{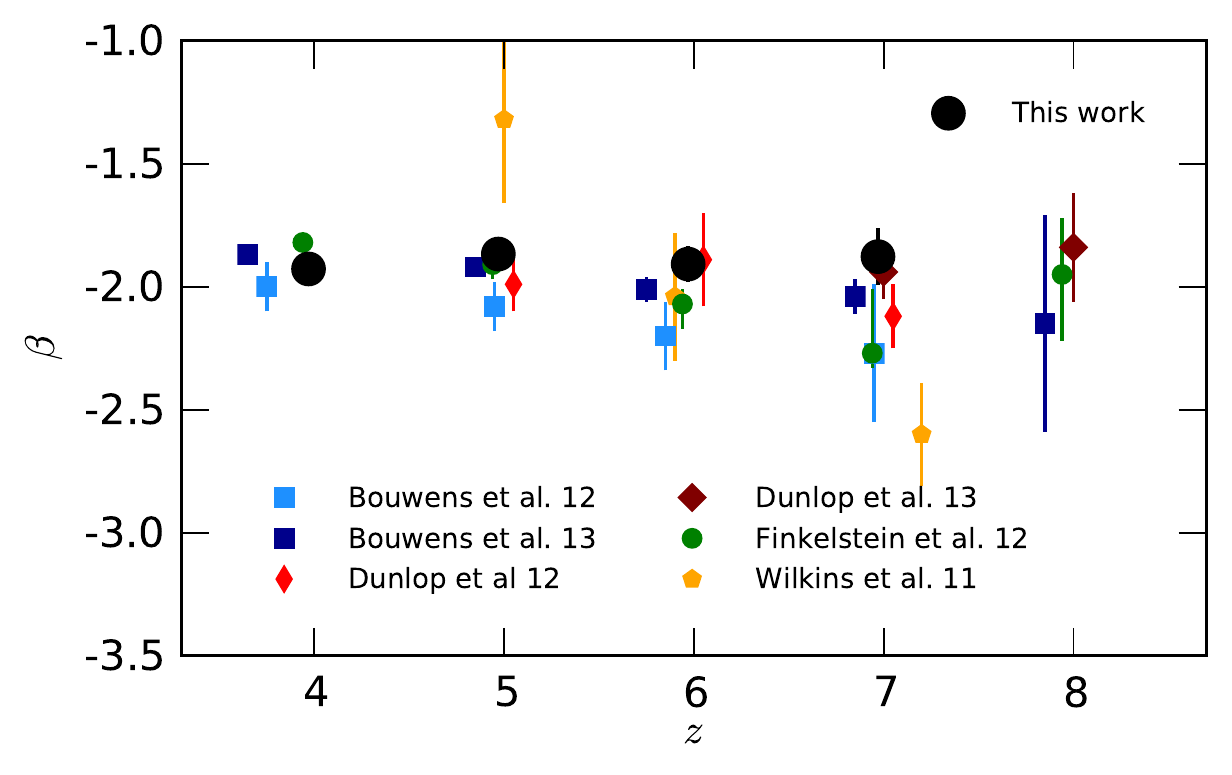}
\caption{UV continuum slope as a function of redshift for a fixed UV magnitude. The points from \citep{2012ApJ...756..164F} are for a fixed $M_{UV} \sim 20$, all other data points are for $M_{UV} \sim 19.5$.}
\label{fig:beta_z}
\end{figure}

In \citet{2013MNRAS.429.2456R}, it was shown that the SED fitting method for measuring $\beta$ suffers from a red bias when measuring the average slope. This is a result of limits placed on the measured $\beta$ by the bluest template available in the fitting, artificially clipping the measurement to values above that limit. The problem is most severe for the faintest galaxies or those which have the least secure photometric redshifts. In Figure~\ref{fig:beta_muv}, we show the measured $\beta$ as a function of UV magnitude for one of our Monte Carlo samples alongside the $M_{UV}$ binned bi-weight means and other values from the literature. We do not see a strong piling up of sources at the bluest templates ($\approx -2.69$ when nebular emission is included) suggesting our observations are not strongly affected by this. However, it may have a small effect on the average $\beta$ for a fixed $M_{UV}$ (see Figure~\ref{fig:beta_z}).

Because we apply the dust correction to each galaxy based on its own measured $\beta$, rather than an observed average, our dust corrections will be unaffected by any such bias if it does exist. For any galaxy bluer than $\beta = -2.23$, the applied extinction based on the relation of \citet{Meurer:1999jm} is 0.

\label{lastpage}

\end{document}